\documentclass[a4paper,11pt]{article}
\usepackage{jheppub} 
\usepackage{url}
\usepackage{lineno}
\usepackage{multirow}
\usepackage{subcaption}
\usepackage{amsthm,amsmath,amssymb}
\usepackage{mathrsfs}
\usepackage{gensymb}
\usepackage{cleveref}
\usepackage{hyperref}
\usepackage{rotating,tabularx}

\newcommand{\kspipi}{K^{0}_{S}\pi^+\pi^-}
\newcommand{\kslpipi}{K^{0}_{S/L}\pi^+\pi^-}
\newcommand{\klpipi}{K^{0}_{L}\pi^+\pi^-}
\newcommand{\Dz}{D^0}
\newcommand{\Dzb}{\bar{D}^{0}}
\newcommand{\kk}{K^+K^-}
\newcommand{\pipi}{\pi^+\pi^-}

\newcommand{\pipipio}{\pi^+\pi^-\pi^0}
\newcommand{\klpio}{K^{0}_{L}\pi^0}
\newcommand{\kspiopio}{K^{0}_{S}\pi^0\pi^0}

\newcommand{\kenu}{K^{+}e^{-}\nu_{e}}
\newcommand{\kthreepi}{K^{+}\pi^{-}\pi^{+}\pi^{-}}
\newcommand{\kpipio}{K^{+}\pi^{-}\pi^{0}}
\newcommand{\kpi}{K^{+}\pi^{-}}

\newcommand{\kspio}{K^{0}_{S}\pi^0}
\newcommand{\kseta}{K^{0}_{S}\eta}

\newcommand{\ksomega}{K^{0}_{S}\omega}
\newcommand{\klpiopio}{K^{0}_{L}\pi^0\pi^0}

\newcommand{\gevcc}{\,{\rm GeV}/{c^{2}}}

\arxivnumber{1234.56789} 

\title{\boldmath Updated model-independent measurement of the strong-phase differences between $D^0$ and $\bar{D}^0 \to \kslpipi$ decays}

\collaboration{BESIII Collaboration}

\author{
M.~Ablikim$^{1}$, M.~N.~Achasov$^{4,c}$, P.~Adlarson$^{77}$, X.~C.~Ai$^{82}$, R.~Aliberti$^{36}$, A.~Amoroso$^{76A,76C}$, Q.~An$^{73,59,a}$, Y.~Bai$^{58}$, O.~Bakina$^{37}$, Y.~Ban$^{47,h}$
% ...
\\
\vspace{0.2cm}
(BESIII Collaboration)\\
\vspace{0.2cm} {\it
$^{1}$ Institute of High Energy Physics, Beijing 100049, People's Republic of China\\
% ...
$^{4}$ Budker Institute of Nuclear Physics SB RAS (BINP), Novosibirsk 630090, Russia\\
% ...
$^{36}$ Johannes Gutenberg University of Mainz, Johann-Joachim-Becher-Weg 45, D-55099 Mainz, Germany\\
$^{37}$ Joint Institute for Nuclear Research, 141980 Dubna, Moscow region, Russia\\
% ...
$^{47}$ Peking University, Beijing 100871, People's Republic of China\\
% ...
$^{58}$ Southeast University, Nanjing 211100, People's Republic of China\\
$^{59}$ State Key Laboratory of Particle Detection and Electronics, Beijing 100049, Hefei 230026, People's Republic of China\\
% ...
$^{73}$ University of Science and Technology of China, Hefei 230026, People's Republic of China\\
% ...
$^{76}$ University of Turin and INFN, (A)University of Turin, I-10125, Turin, Italy; (B)University of Eastern Piedmont, I-15121, Alessandria, Italy; (C)INFN, I-10125, Turin, Italy\\
$^{77}$ Uppsala University, Box 516, SE-75120 Uppsala, Sweden\\
% ...
$^{82}$ Zhengzhou University, Zhengzhou 450001, People's Republic of China\\

\vspace{0.2cm}
$^{a}$ Deceased\\
% ...
$^{c}$ Also at the Novosibirsk State University, Novosibirsk, 630090, Russia\\
% ...
$^{h}$ Also at State Key Laboratory of Nuclear Physics and Technology, Peking University, Beijing 100871, People's Republic of China\\
% ...

}
}

\abstract{ The strong-phase differences between \mbox{$D^0\to
K_{S(L)}^0\pi^+\pi^-$} and $\bar{D}^0\to K_{S(L)}^0\pi^+\pi^-$ decays
are one of the most important inputs in measuring the $C\!P$ violating angle $\gamma$ via $B^- \to D K^-$ decays. They also play
a key role in studies of charm mixing and indirect $C\!P$
violation. In this paper, the strong-phase differences are determined
in a model-independent way with quantum-correlated $D^0$-$\bar{D}^0$
decays from 7.93 fb$^{-1}$ of $e^+e^-$ annihilation data at
$\sqrt{s}$=3.773 GeV by the BESIII experiment. These results are the
most precise to date and are expected to significantly reduce
associated uncertainties in determining the $C\!P$ violating angle $\gamma$ and related charm mixing parameters.}

\begin{document}
\maketitle
\flushbottom

\section{Introduction}
\label{sec:intro}

In the Standard Model (SM), $C\!P$ violation in the quark sector is
only attributed to the complex phase in the Cabibbo-Kobayashi-Maskawa
(CKM) matrix~\cite{Cabibbo:1963yz,Kobayashi:1973fv}. The unitarity
nature of the CKM matrix permits its geometric representation as the
Unitary Triangle (UT) in the complex plane. Specifically, the
$C\!P$ violating angle $\gamma =
\text{arg}(-V_{us}V^{*}_{ub}/V_{cs}V^{*}_{cb})$ in the UT can be
directly measured in the tree-level decays. These decays are expected to have negligible theoretical uncertainty~\cite{Brod:2013sga}.  The
$\gamma$ angle can also be indirectly inferred from the information of
other CKM matrix elements~\cite{Charles:2004jd}, which are more
susceptible to new physics effects. Hence, the direct measurements
of $\gamma$ provide a stringent test of the CKM matrix unitarity and a
potential avenue in the search for new physics beyond the SM~\cite{Li:2021iwf}.

The leading decay channel for the direct measurement of the angle
$\gamma$ is $B^\pm \to D K^\pm$, $D \to \kspipi$~\cite{Giri:2003ty},
where $D$ represents a superposition of $\Dz$ and $\Dzb$. The
amplitude of the $B^-$ decay is written as
\begin{equation}\label{eq_BDecay} \begin{aligned} f_{B}(m_{+}^{2},
m_{-}^{2}) \propto f_{\Dz}(m_{+}^{2}, m_{-}^{2}) +
r_{B}e^{i(\delta_{B}-\gamma)}f_{\Dzb}(m_{+}^{2}, m_{-}^{2}).
\end{aligned} \end{equation} Here, $m_{\pm}^{2}$ is the squared mass
of $K_{S}^{0}\pi^{\pm}$, $f_{\Dz/\Dzb}(m_{+}^{2}, m_{-}^{2})$ denotes
the amplitudes of the $\Dz/\Dzb \to \kspipi$ decays, $\delta_{B}$ is
the strong-phase difference between the color-favoured and
color-suppressed amplitudes of $B^\pm \to D K^\pm$, and $r_{B}$ is the
modulus of the suppressed to favoured amplitudes. The amplitude of the $\Dz
\to \kspipi$ decay $f_{\Dz}(m_{+}^{2}, m_{-}^{2})$ is described by the
absolute value $|f|$ and the phase $\delta_{D}$ as $f_{\Dz}(m_{+}^{2},
m_{-}^{2}) = |f|e^{i\delta_{D}}$. The $\Dzb$ amplitude is written as
$f_{\Dzb}(m_{+}^{2}, m_{-}^{2})$ = $f_{\Dz}(m_{-}^{2}, m_{+}^{2})$
ignoring the small second-order charm mixing~\cite{Bondar:2010qs} and
direct $C\!P$ violation effects~\cite{Giri:2003ty}. The amplitude of
the $B^- \to \Dz K^-$ decay is rewritten as
\begin{equation}\label{eq_Bdetail} \begin{aligned} f_{B}(m_{+}^{2},
m_{-}^{2}) \propto |f_{\Dz}(m_{+}^{2}, m_{-}^{2})|
e^{i\delta_{D}(m_{+}^{2}, m_{-}^{2})} + r_{B}e^{i(\delta_{B}-\gamma)}
|f_{\Dz}(m_{-}^{2}, m_{+}^{2})| e^{i\delta_{D}(m_{-}^{2}, m_{+}^{2})})
\\ \propto |f_{\Dz}(m_{+}^{2}, m_{-}^{2})| +
r_{B}e^{i(\delta_{B}-\gamma)} |f_{\Dz}(m_{-}^{2}, m_{+}^{2})|
e^{i(\delta_{D}(m_{+}^{2}, m_{-}^{2}) - \delta_{D}(m_{-}^{2},
m_{+}^{2}))} \end{aligned} \end{equation} Therefore, determination of
the $\gamma$ angle requires input of the strong-phase difference
between $\Dz$ and $\Dzb \to \kspipi$ decays, $i.e.$,
\mbox{$\Delta\delta_{D} = \delta_{D}(m_{+}^{2},
m_{-}^{2})-\delta_{D}(m_{-}^{2}, m_{+}^{2})$}. In addition, due to the
abundant intermediate processes in the $D \to \kspipi$ decay,
$\Delta\delta_{D}$ varies in phase space, making this channel the
most sensitive to the $\gamma$ angle.

The strong-phase difference $\Delta\delta_{D}$ between $\Dz$ and $\Dzb
\to \kspipi$ decays can be modeled using an amplitude analysis based
on the experimental data of $\Dz\to \kspipi$. However, this approach
introduces unavoidable model dependence, complicating the estimation
of the systematic uncertainty associated with different model choices
in the $\gamma$ measurement~\cite{Battaglieri:2014gca}. Alternatively,
quantum-correlated (QC) $D\bar{D}$ pairs produced at the $\psi(3770)$
resonance provides an ideal environment to determine
$\Delta\delta_{D}$~\cite{Giri:2003ty}. This method allows for a
model-independent measurement of the $\gamma$
angle~\cite{LHCb:2020yot}, where the uncertainty from the strong-phase
difference can be reliably estimated.

Furthermore, the strong-phase difference $\Delta\delta_{D}$ between
$\Dz$ and $\Dzb \to \kspipi$ decays can provide inputs to determine
another UT angle $\beta$~\cite{Belle:2016ckr}, for the study of charm
mixing and $C\!P$ violation
phenomena~\cite{LHCb:2021ykz}. Additionally, it aids in the
measurement of the strong-phase differences in various other $D^0$
hadronic decays~\cite{BESIII:2022qkh}.

The CLEO experiment determined the strong-phase difference between
$\Dz$ and $\Dzb\to K_{S}^0\pi^+\pi^-$ decays for the first time, using
a dataset collected at the $\psi(3770)$ resonance with an integrated
luminosity of 0.818~$\rm{fb}^{-1}$~\cite{CLEO:2010iul}. The BESIII
experiment subsequently measured these parameters using a larger
$\psi(3770)$ dataset of
2.93~$\rm{fb}^{-1}$~\cite{BESIII:2020khq}. Recently, BESIII has
further expanded its data collection, and the total integrated
luminosity of the $\psi(3770)$ data sample is
7.93~$\rm{fb}^{-1}$~\cite{besiiicollaboration2024measurementintegratedluminositydata}. This
enables a more precise determination of the strong-phase differences,
thereby reducing the associated uncertainty in determining the angle
$\gamma$.

The strong-phase difference parameters in the $D^0\to
K_{L}^0\pi^+\pi^-$ decay were also measured in the previous analysis
to improve the determination of those in the $D^0\to
K_{S}^0\pi^+\pi^-$ decay~\cite{CLEO:2010iul,BESIII:2020khq}. A
model-dependent constraint, derived from the differences between the
strong-phase parameters in $D^0\to\kspipi$ and $D^0\to\klpipi$ decays,
was employed to improve the precision. Recently, the BESIII experiment
has implemented the amplitude analysis of the $D^0\to\klpipi$
decay~\cite{BESIII:2022qvy}, and the resultant model provides an
improved estimation of the model-constraint in this
analysis. Moreover, with the larger $\psi(3770)$ data sample at
BESIII, it is now feasible to remove the model-dependent constraint
entirely. This development allows for a detailed study of its impact
on the measurement of the strong-phase parameters and the angle
$\gamma$, and for improved knowledge of the strong-phase differences
in both $D^0\to\kspipi$ and $D^0\to\klpipi$ decays.

In this paper, we present an improved measurement of the strong-phase
parameters in $D\to K_{S,L}^0\pi^+\pi^-$ decays using the BESIII
$\psi(3770)$ dataset with an integrated luminosity of
7.93~$\rm{fb}^{-1}$. The paper is organized as follows: in
Section~\ref{sec:formula}, the definition and theoretical formalism of
strong-phase difference parameters are discussed.
Section~\ref{sec:detector} provides an introduction to the BESIII
detector, the data samples and the simulated Monte Carlo (MC)
samples. The most recent amplitude models are implemented in these MC
samples, resulting in an improved simulation of the QC effects, thereby reducing the associated systematic
uncertainties in this study. The event criteria, background estimation
and fitted yields are detailed in Section~\ref{sec:evt}. The
model-independent results of the strong-phase difference parameters
and the related systematic uncertainties are presented in
Section~\ref{sec:cisi}. Finally, the impact of the strong-phase
parameters on the $\gamma$ measurement is assessed in
Section~\ref{sec:gamma}.

\section{Formalism}
\label{sec:formula}

\begin{figure}[!htb]
     \centering     \includegraphics[width=.32\textwidth]{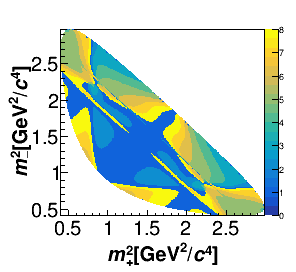}
    \includegraphics[width=.32\textwidth]{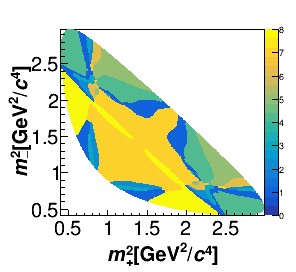}
    \includegraphics[width=.32\textwidth]{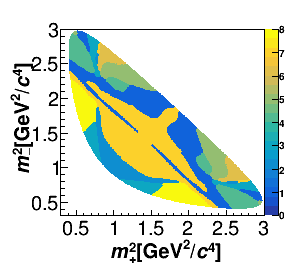}\label{fig:binningscheme3}
     \caption{Binned regions marked as different colors indexed from 1 to 8 in the equal binning scheme (left), optimal binning scheme (middle) and modified-optimal binning scheme (right).}
     \label{fig:binningscheme}
 \end{figure}

This analysis utilizes $D \to \kspipi$ phase space, which is
divided into bins according to three schemes identical to those used
in Ref.~\cite{CLEO:2010iul}. These bins in the Dalitz plot are
symmetric with respect to the $m_{+}^{2}= m_{-}^{2}$ axis and are
indexed by $i$ from $-8$ to $8$, excluding zero. Positive (negative)
bins are located in the $m_{+}^{2}>m_{-}^{2}$ ($m_{+}^{2}<m_{-}^{2}$)
region. The binning schemes are illustrated in
Figure~\ref{fig:binningscheme}, denoted as the equal binning scheme,
the optimal binning scheme and the modified optimal binning
scheme. The detailed information on the choice of these regions is
given in Ref.~\cite{CLEO:2010iul}.

The strong-phase difference $\Delta\delta_{D}$, which quantifies the
interference between the amplitudes of $\Dz$ and $\Dzb$ decays, is
parameterized using the amplitude-weighted averages of
$\cos\Delta\delta_{D}$ and $\sin\Delta\delta_{D}$ in each bin. These
parameters are defined as
\begin{equation}
\label{eq_cidefine}
    \begin{aligned}
        c_{i} &= \frac{1}{\sqrt{F_{i}F_{-i}}}\int_{i} |f_{\Dz}(m_{+}^{2}, m_{-}^{2})| |f_{\Dz}(m_{-}^{2}, m_{+}^{2})| \cos[\Delta\delta_{D}(m_{+}^{2}, m_{-}^{2})]\rm{d} \mathit{m}_{+}^{2}\rm{d} 
        \mathit{m}_{-}^{2},  \\
         s_{i} &= \frac{1}{\sqrt{F_{i}F_{-i}}}\int_{i} |f_{\Dz}(m_{+}^{2}, m_{-}^{2})| |f_{\Dz}(m_{-}^{2}, m_{+}^{2})| \sin[\Delta\delta_{D}(m_{+}^{2}, m_{-}^{2})]\rm{d} \mathit{m}_{+}^{2}\rm{d} \mathit{m}_{-}^{2},
    \end{aligned}
\end{equation}
where $F_{i}$ represents the fraction of events found in the $i^{\rm th}$ bin of the flavour-specific decay $\Dz \to \kspipi$.

Using the large dataset collected at the $\psi(3770)$ resonance,
where neutral $D\bar{D}$ decays exhibit $C$-odd correlations, the
strong-phase parameters are accessed by tagging both $D$
mesons~\cite{Giri:2003ty}. This approach is denoted as the double-tag
(DT) method. The tag modes reconstructed against the signal
$K_{S/L}^0\pi^+\pi^-$ decays fall into the categories of flavour,
$C\!P$ eigenstate or self-conjugate, as detailed in
Table~\ref{tab:tag}.

\begin{table}[tp!]
\centering
\begin{tabular}{lc}
\hline
Flavour & $\kenu$,  $\kthreepi$, $\kpipio$, $\kpi$ \\
$C\!P$-even & $\kk$, $\pipi$, $\pipipio$, $\kspiopio$, $\klpio$ \\
$C\!P$-odd  & $\kspio$, $ K_{S}^{0} \eta_{\gamma\gamma}$, $ K_{S}^{0} \eta_{\pipipio}$, $\ksomega$, $ K_{S}^{0} \eta^{\prime}_{\gamma\rho}$, $ K_{S}^{0}\eta^{\prime}_{\pi\pi\eta}$, $\klpiopio$ \\
Self-conjugate & $\kspipi$, $K_{S}^{0} (\pi)_{\rm miss}\pi$, $(\pi^{0}(\pi^{0})_{\rm miss})_{K_S^0}\pi\pi$  \\
\hline
\end{tabular}
\caption{Tag modes reconstructed against the signal decays. The partially-reconstructed tag modes of $\kenu$, $\klpio$, $\klpiopio$, $K_{S}^{0} (\pi)_{\rm miss}\pi$ and $(\pi^{0}(\pi^{0})_{\rm miss})_{K_S^0}\pi\pi$ tags are not used for the $\klpipi$ decay.\label{tab:tag}}
\end{table}

For DT events in which one $D$ meson decays into the signal mode
$K_{S}^0\pi^+\pi^-$ and the other into a $C\!P$ eigenstate, the decay
amplitude is expressed as
\begin{equation}
        \label{ampl_ci}
        \begin{aligned}
            f_{CP\pm}(m^{2}_{+},m^{2}_{-}) = \frac{1}{\sqrt{2}}(f_{D}(m^{2}_{+},m^{2}_{-})\pm f_{D}(m^{2}_{-},m^{2}_{+})),
        \end{aligned}
\end{equation} 
where $\pm$ denotes the $C\!P$ eigenvalue of different tag modes. The
expected number of signal events in the
    $i^{\rm th}$ bin, denoted as $M_{i}$ for the signal mode
    $K_{S}^0\pi^+\pi^-$, is derived by integrating the square of the
    amplitude over the bin \begin{equation}
    \label{eq_ksci}
    \begin{aligned}
            M_{i} \propto K_{i} + K_{-i} -(2F^{+}_{CP}-1)\times2c_{i}\sqrt{K_{i}K_{-i}}, \\
     \end{aligned}
     \end{equation} where $K_{i} \propto \int_{i} |f_{\Dz}(m_{+}^{2},
     m_{-}^{2})|^{2} dm_{+}^{2}dm_{-}^{2}$ represents the
     flavour-specific $\kspipi$ decay events in the $i^{\rm th}$
     bins. $F^{+}_{CP}$ denotes the $C\!P$-even fraction of the tag
     mode, taking a value of 1 (0) for $C\!P$-even (odd) eigenstate.

To further constrain the strong-phase parameters, the signal mode
$\kspipi$ is tagged with self-conjugate tags, {\it i.e.} $\kspipi$
versus $\kspipi$. In this context, the decay amplitude is written as
    \begin{equation}
        \label{ampl_si}
        \begin{aligned}
            f = \frac{1}{\sqrt{2}}\left[f_{D}(m^{2}_{+},m^{2}_{-})f_{D}(m^{2\dagger}_{-},m^{2\dagger}_{+})-f_{D}(m^{2\dagger}_{+},m^{2\dagger}_{-})f_{D}(m^{2}_{-},m^{2}_{+})\right],
        \end{aligned}
    \end{equation} where the symbol $\dagger$ differentiates the
    Dalitz plot coordinates between the two $\kspipi$ decays. The
    event rate $M_{ij}$, defined as the event yields observed in the
    $i^{\rm th}$ bin of the first and $j^{\rm th}$ bin of the second
    $D\to \kspipi$ Dalitz plot, is expressed as
 \begin{equation}
    \label{eq_kssi}
    \begin{aligned}
    M_{ij} \propto K_{i}K_{-j} + K_{-i}K_{j} - 2\sqrt{K_{i}K_{-i}K_{j}K_{-j}}(c_{i}c_{j}+s_{i}s_{j}). \\
     \end{aligned}
     \end{equation} In this analysis, the $\kspipi$ tag is
     reconstructed using three independent selection methods to
     improve precision. These methods are denoted as the fully
     reconstructed $\kspipi$ tag, the missing $\pi$ tag $K_{S}^{0}
     (\pi)_{\rm miss}\pi$ and the missing $\pi^{0}$ tag
     $(\pi^{0}(\pi^{0})_{\rm miss})_{K_S^0}\pi\pi$.

Taking advantage of the $C\!P$ relationship between $D \to \klpipi $
and $D \to \kspipi$ decays, it is helpful to include $D \to \klpipi $
as a self-conjugate tag mode to further enhance the precision of the
measurement. The event rate of $\kspipi$ versus $\klpipi$ DT sample,
denoted as $M^{\prime}_{ij}$, is given by
  \begin{equation}
    \label{eq_klsi}
    \begin{aligned}
 M_{ij}^{\prime} \propto K_{i}K_{-j}^{\prime} + K_{-i}K_{j}^{\prime} + 2\sqrt{K_{i}K_{-i}K_{j}^{\prime}K_{-j}^{\prime}}(c_{i}c_{j}^{\prime}+s_{i}s_{j}^{\prime}).
 \end{aligned}
     \end{equation}
Here, $c_{i}^{\prime}$ and $s_{i}^{\prime}$ represent the strong-phase
parameters associated with the $D\to \klpipi$ decay, and
$K^{\prime}_{i}$ denotes the yield of flavour-specific $\klpipi$ DT
decays. They are defined analogously to those for the $D \to \kspipi$
decay, following the binning schemes shown in
Figure~\ref{fig:binningscheme}.

To better constrain the strong-phase difference parameters in the
$D\to \klpipi$ decay, the DT events of $\klpipi$ versus $C\!P$
eigenstates are selected. The decay rate $M_{i}^{\prime}$ is expressed
as \begin{equation}
    \label{eq_klci}
    \begin{aligned}
  M_{i}^{\prime} \propto K_{i}^{\prime} + K_{-i}^{\prime} +(2F^{+}_{CP}-1)\times2c_{i}^{\prime}\sqrt{K_{i}^{\prime}K_{-i}^{\prime}}.
 \end{aligned}
     \end{equation}

The expected DT yields in Eqs.~\eqref{eq_ksci} and
~\eqref{eq_kssi},~\eqref{eq_klsi} and \eqref{eq_klci} are normalized
using the single-tag (ST) yields. ST events are defined for the events
that have only one of the two $D$-meson decays detected.  By
normalizing DT signal yields with the ST yields, the associated
systematic uncertainties are significantly reduced. For the tag modes
that are partially reconstructed in Table~\ref{tab:tag}, the ST yields
are calculated based on the knowledge of the branching fractions of
the tag modes~\cite{ParticleDataGroup:2024cfk}.

\section{BESIII detector}
\label{sec:detector}

The BESIII detector~\cite{BESIII:2009fln} records symmetric $e^+e^-$
collisions provided by the BEPCII storage ring~\cite{Yu:2016cof} in
the center-of-mass energy range from 1.84 to 4.95~GeV~ with a peak
luminosity of $1.1 \times 10^{33}\;\text{cm}^{-2}\text{s}^{-1}$
achieved at $\sqrt{s} = 3.773\;\text{GeV}$. BESIII has collected large
data samples in this energy
region~\cite{BESIII:2020nme,Lu:2020,Zhang:2022bdc}. The cylindrical
core of the BESIII detector covers 93\% of the full solid angle and
consists of a helium-based
 multilayer drift chamber~(MDC), a plastic scintillator time-of-flight
 system~(TOF), and a CsI(Tl) electromagnetic calorimeter~(EMC), which
 are all enclosed in a superconducting solenoidal magnet providing a
 1.0~T magnetic field, which was 0.9~T in 2012. The solenoid is
 supported by an octagonal flux-return yoke with resistive plate
 counter muon identification modules interleaved with steel.

The charged-particle momentum resolution at $1~{\rm GeV}/c$ is
$0.5\%$, and the ${\rm d}E/{\rm d}x$ resolution is $6\%$ for electrons
from Bhabha scattering. The EMC measures photon energies with a
resolution of $2.5\%$ ($5\%$) at $1$~GeV in the barrel (end cap)
region. The time resolution in the TOF barrel region is 68~ps, while
that in the end cap region was 110~ps.  The end cap TOF system was
upgraded in 2015 using multigap resistive plate chamber technology,
providing a time resolution of 60~ps~\cite{BESIII:2017TOF}. This
update benefits 63\% of the data used in this analysis. More details
can be checked in Ref.~\cite{BESIII:2009fln}.

Monte Carlo (MC) simulated data samples produced with {\sc
geant4}-based~\cite{AGOSTINELLI2003250} software, which
includes the geometric description of the BESIII detector and the
detector response, are used to determine detection efficiencies
and to estimate backgrounds. The simulation models the beam
energy spread and initial state radiation (ISR) in the $e^+e^-$
annihilations with the generator {\sc
kkmc}~\cite{Jadach:2000ir,Jadach:1999vf}. 

Inclusive MC samples are produced including the $D\bar{D}$ pairs
corrected for the QC effects, the non-$D\bar{D}$ decays of the
$\psi(3770)$, the ISR production of the $J/\psi$ and $\psi(3686)$
states, and the continuum processes incorporated in {\sc
kkmc}~\cite{Jadach:2000ir,Jadach:1999vf}. All particle decays in the
inclusive MC sample are modeled with {\sc
evtgen}~\cite{Lange:2001uf,Ping:2008zz} using branching fractions
either taken from the Particle Data
Group~\cite{ParticleDataGroup:2024cfk}, when available, or otherwise
estimated with {\sc
lundcharm}~\cite{Chen:2000tv,Yang:2014vra}. In particular, the
$D^0\to\pipi\pipi$ and $D^0\to\pipi\pi^{0}\pi^{0}$ decays, which are
the major peaking backgrounds in the $\kspipi$ signal decay, are
modified in the inclusive MC samples according to their amplitude
models~\cite{BESIII:2023exz}. Final state radiation from charged final
state particles is incorporated using {\sc photos}~\cite{Barberio:1990ms}.

The signal MC samples are simulated for the DT $\kslpipi$ versus tag
modes, where the QC effects are implemented. The $\kspipi$ decay is
simulated with the amplitude model measured by the Belle
experiment~\cite{BaBar:2018cka}. The $\klpipi$ decay is simulated with
the $\kspipi$ model implemented with the U-spin breaking
parameters~\cite{BESIII:2022qvy}. Multibody tag-side decays are
simulated with {\sc evtgen}~\cite{Lange:2001uf,Ping:2008zz} according
to the most recent models from experimental studies.

\section{Event selection and signal yield }
\label{sec:evt}
\subsection{Event selection}

Final-state particles in ST and DT decays are reconstructed from
candidates of charged tracks and photons. The selection criteria for
charged tracks and photons and intermediate resonances such as
$\pi^0$, $K_S^0$, $\eta$, $\eta'$ and $\omega$ are identical to those
described in Ref.~\cite{BESIII:2020khq}. Specifically, the invariant
mass $M_{\pi^+\pi^-}$ for the $K_{S}^{0}$ candidates is optimized to
be within (0.487, 0.511) $\gevcc$ and the invariant mass
$M_{\gamma\pi^+\pi^-}$ for the $\eta^{\prime}$ candidates must be
within (0.940, 0.970) $\gevcc$.

Decays of $D$ mesons are reconstructed into either tag modes as
detailed in Table~\ref{tab:tag} or the $\kslpipi$ signal mode. When all
final state particles are reconstructed, the tags are
denoted as fully reconstructed tags. To suppress combinatorial
background in the fully reconstructed tags, a mode-specific
requirement on the energy difference between the beam and the final
states $\Delta E =E_{D^0}-\sqrt{s}/2$ is
applied~\cite{BESIII:2020khq}, where $\sqrt{s}/2$ represents the
energy of the electron beam, and $E_{D^0}$ is the reconstructed energy
of the $D$ candidate. If multiple candidates are selected
in an event, the one with the smallest $|\Delta E|$ is chosen as the
best candidate. The beam-constrained mass, \mbox{$M_{\rm BC}
=\sqrt{({\sqrt{s}/2})^2-|\boldsymbol{p}_{D^0}|^2}$}, is utilized as
the variable to determine both ST and DT yields, where
$\boldsymbol{p}_{D^0}$ represents the reconstructed momentum of the
$D$ candidate.

When a missing particle, such as the $K^{0}_{L}$ meson, is present in
the final states, the tags are partially reconstructed. In such cases,
the ``missing mass" method described in Ref.~\cite{BESIII:2020khq} is
utilized. The squared missing mass $M^2_{{\rm miss}}$ is defined as
$M^2_{{\rm miss}}= E^2_{{\rm miss}}-|\boldsymbol{p}_{{\rm miss}}|^2$,
where $E_{{\rm miss}}=\sqrt{s}/2-E_{{\rm other}}$,
$\boldsymbol{p}_{{\rm miss}}=-\boldsymbol{p}_{\rm
tag}-\boldsymbol{p}_{{\rm other}}$. Here, $\boldsymbol{p}_{\rm tag}$
represents the momentum of the fully reconstructed signal $D\to
\kspipi$ decay, while $\boldsymbol{p}_{{\rm other}}$ and $E_{{\rm
other}}$ are the momentum and energy of the other reconstructed
particles that form the partially-reconstructed tags. Correctly
reconstructed events are expected to peak at the squared mass of the
missing particle in the $M^2_{{\rm miss}}$ distribution. Therefore the
$M^2_{{\rm miss}}$ variable is used to determine the DT yield. In the
selection of the $\kenu$ mode, the variable $U_{{\rm miss}}= E_{{\rm
miss}}-|\boldsymbol{p}_{{\rm miss}}|$ is employed as the fit
variable. Correctly reconstructed events with a missing neutrino are
expected to peak around $U_{{\rm miss}}= 0$.

For DT candidates where $\kspipi$ signal events are tagged by fully
reconstructed modes, the selection is based on the $\Delta E$ and
$M_{\rm BC}$ variables for both signal and tag sides. The selection
criteria of $\Delta E$ are identical to those applied to ST
candidates.

For partially-reconstructed DT candidates whose final states contain a
$K_L^0$ meson, it is required that there are no extra charged tracks
or $\gamma\gamma$ candidates for $\pi^{0}$ or $\eta$ particles. For
the $\kspipi$ signal tagged by the partially-reconstructed $\kspipi$
mode, DT events containing extra charged tracks are rejected. For the
signal $\kspipi$ events tagged with the $\klpiopio$ decay, the recoil
mass of either $\pi^{0}$ is required to be larger than 0.7 $\gevcc$ to
suppress the background of $ D^{0} \to \pi^{0}\pi^{0}\pi^{0}$
decay. Furthermore, the mass of any $K^{0}_{S}\pi^{\pm}\pi^{0}$
combination is required to be out of the $D^+$ mass region (1.85,
1.89)$\gevcc$ to suppress the $D^{+}\to K^{0}_{S/L}\pi^{+}\pi^{0}$
versus $D^{-}\to K^{0}_{L/S}\pi^{-}\pi^{0}$ backgrounds. The DT
$\klpipi$ versus $\kspiopio$ events are also subjected to the above
requirement.

To mitigate resolution effects in DT samples, kinematic fits are
performed to constrain the $D$ mass, $K^{0}_{S}$ mass and $K^{0}_{L}$
mass. If there are multiple candidates in a DT event, the best
candidate is selected following the methodology used in
Ref.~\cite{BESIII:2020khq}.


\subsection{Background study}
\label{sec:bkg}

The inclusive MC
samples are utilized to investigate the possible sources of peaking
background. For the ST flavour mode $\kthreepi$, the dominant peaking
background originates from the $K^{0}_{S}K^{\pm}\pi^{\mp}$ decay, with
the background fractions of approximately 2.4$\%$. In the
$C\!P$-eigenstate channels, $\pipi\pi^{0}\pi^{0}$, $\pipipio$ and
$\kspio$ are the major peaking backgrounds in the $\kspiopio$,
$\kspio$ and $\pipipio$ tags, with the background fractions of around
4.2$\%$, 0.5$\%$ and 6.7$\%$, respectively. In the $\kseta_{\pipipio}$
and $\ksomega$ tag modes, non-resonant $K_{S}\pipipio$ decay is
identified as the dominant peaking background. The corresponding
exclusive MC samples are simulated to obtain the contamination rates,
which are found to be about 8$\%$ and 12$\%$ in the
$\kseta_{\pipipio}$ and $\ksomega$ tags, respectively.

In DT $\kspipi$ decays, the major peaking backgrounds on the signal
side arise from the $\pipi\pipi$ and $K^{0}_{S}K^{0}_{S}$ decays, with
the background fractions of around 0.25$\%$ and 0.4$\%$,
respectively. The tag side of fully reconstructed modes exhibits the
same sources of peaking background as observed in ST events. In
partially reconstructed tag modes where the tag side is $K^{0}_{L}X
(X=\pi^{0}, \pi^{0}\pi^{0})$ decays, the major peaking background
comes from $K^{0}_{S}X$, primarily through the decay $K^{0}_{S} \to
\pi^{0}\pi^{0}$, with the background fractions of approximately 8$\%$
and 16$\%$, respectively. Furthermore, the dominant background becomes
$\pipi\pipi$ for $K_{S}^{0} (\pi)_{\rm miss}\pi$ and
$\pipi\pi^{0}\pi^{0}$ for $(\pi^{0}(\pi^{0})_{\rm
miss})_{K_S^0}\pi\pi$ tag modes, with the background fractions of
approximately 3$\%$ and 13$\%$, respectively.

In DT $\klpipi$ decays, the $\kspipi$ decay is identified as the principal
source of peaking background on the signal side, with a contamination
rate of approximately 10$\%$. To minimize the systematic
uncertainties, the background contribution is assessed using the
signal yields determined in this analysis. Additionally,
$\pipi\eta$ constitutes a notable peaking background, with a fraction
of 1$\%$. Since the $D \to \klpipi$ decays are tagged by fully
reconstructed modes, the peaking backgrounds on the tag side are the
same as those in the corresponding selected ST events.

\subsection{Signal yields and efficiencies}
\label{sec:Signal yield}

\begin{figure}
\centering
\parbox{1.0\textwidth}{\centering 
 \includegraphics[width=1.0\textwidth]{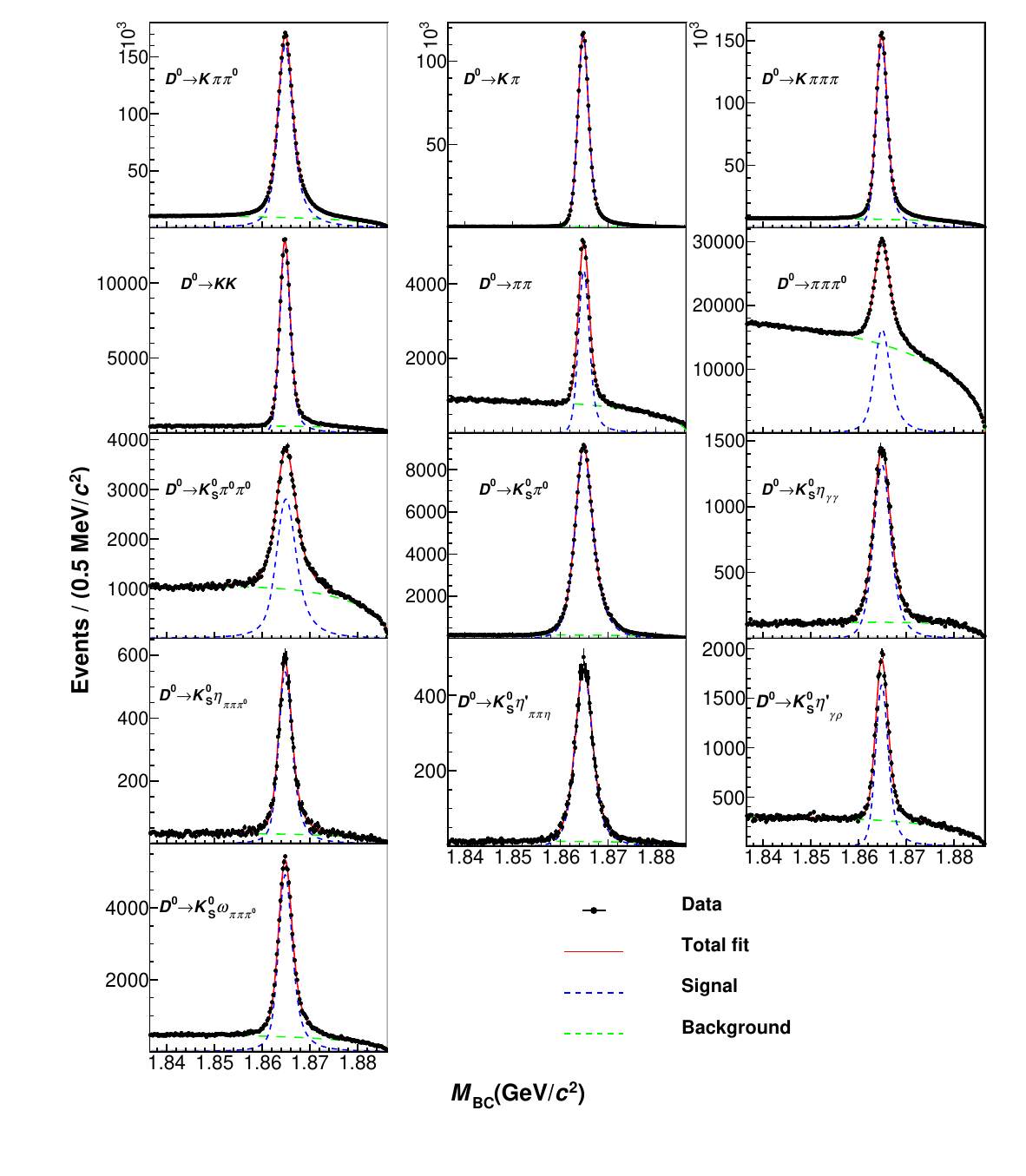}}
\caption{Fits to the $M_{\rm BC}$ distributions of ST signal events. The black points with error bars represent data where peaking background has been subtracted. The red solid lines represent the fit to data. The signal and background components are shown as blue and green dashed lines. } 
\label{fig:ST}
\end{figure}

The ST yields of the fully reconstructed tag modes are obtained
through unbinned maximum likelihood fits on the $M_{\rm BC}$
distributions. The signal shape is obtained from the signal MC
samples, but is convolved with Gaussian functions to accommodate the
resolution differences between data and MC simulations. The mean
values and widths of the Gaussian functions are free parameters of the
fits. Peaking backgrounds are subtracted according to the inclusive MC
samples. An ARGUS function~\cite{ARGUS:1990hfq}, with its endpoint
fixed at the beam energy and other parameters determined from the fit,
models the combinatorial backgrounds. The outcomes of these fits are
depicted in Figure~\ref{fig:ST}, and the
 resulting signal yields are tabulated in Table~\ref{tab:yields}.

\begin{figure}
\centering
\parbox{1.0\textwidth}{\centering 
 \includegraphics[width=1.0\textwidth]{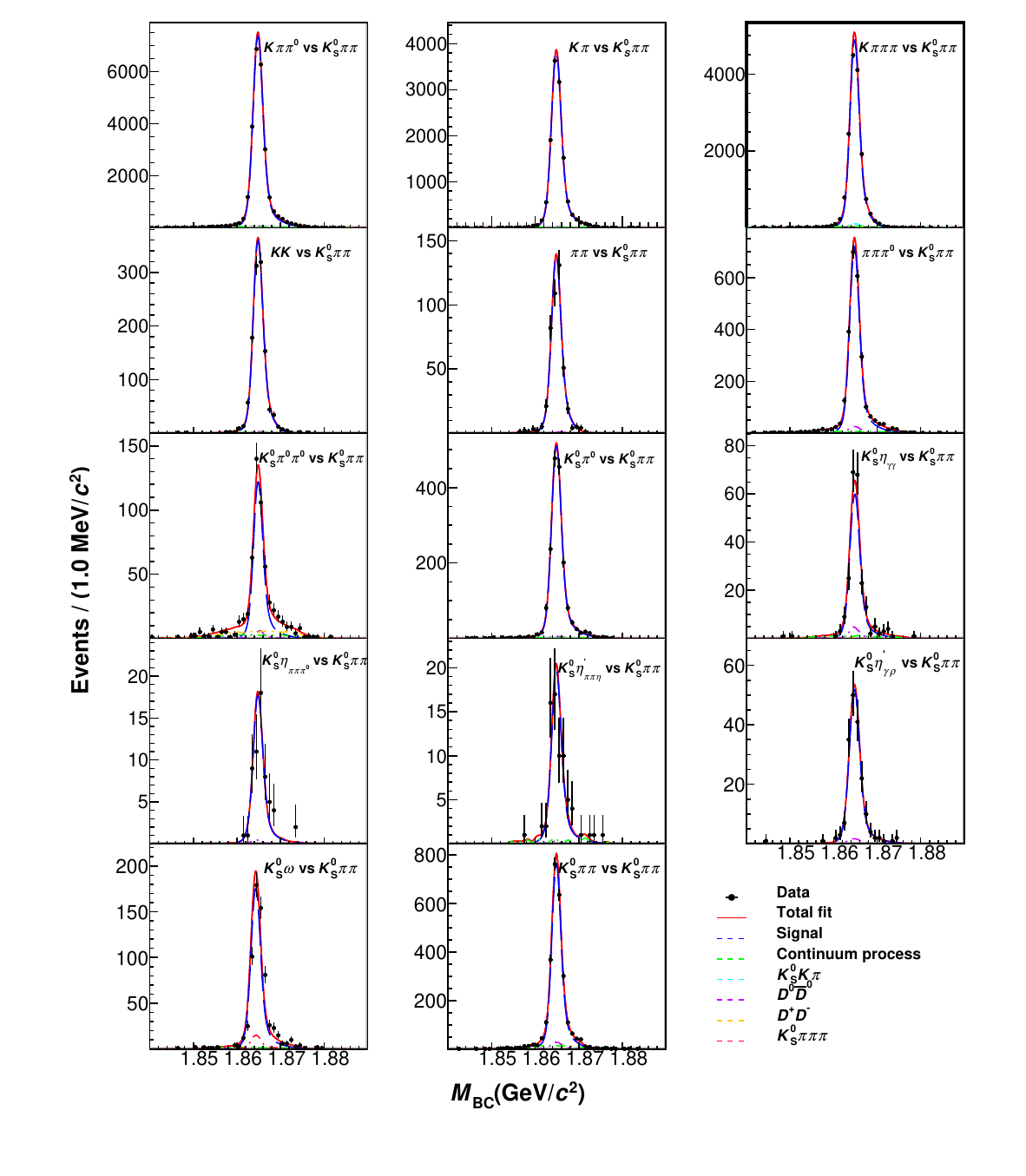}}
\caption{The 1D unbinned maximum likelihood fits for $\kspipi$ versus full reconstructed modes on $M_{\rm BC}$. The black points with error bars represent data. The dashed lines indicate the signal and background shapes. The red lines represent the sum of signal and background shapes. The fitting plots in logarithm scale are shown in Figure~\ref{fig:ksfitted2_log}.} 
\label{fig:ksfitted2}
\end{figure}

The DT yields are determined both within every bin of the phase space
and across the entire phase space. For each fully reconstructed mode,
signal yields are obtained through one-dimensional (1D) unbinned
maximum likelihood fits of the signal-side $M_{\rm BC}$
distributions. The fits across entire phase space are shown in
Figure~\ref{fig:ksfitted2} for all fully reconstructed DT events and the logarithmic-scale fits are shown in Figure~\ref{fig:ksfitted2_log}. The
signal shapes are derived from signal MC samples and convolved with
Gaussian functions. Peaking background and the background from
$D^{+}D^{-}$ decays are fixed according to the inclusive MC
simulations. Other continuum backgrounds, dominated by the $e^{+}e^{-}
\to q\bar{q}$ process, are modeled using the MC-simulated shape with
their contribution left free in the fits.
 
\begin{figure}
\centering
\parbox{1.0\textwidth}{\centering 
 \includegraphics[width=1.0\textwidth]{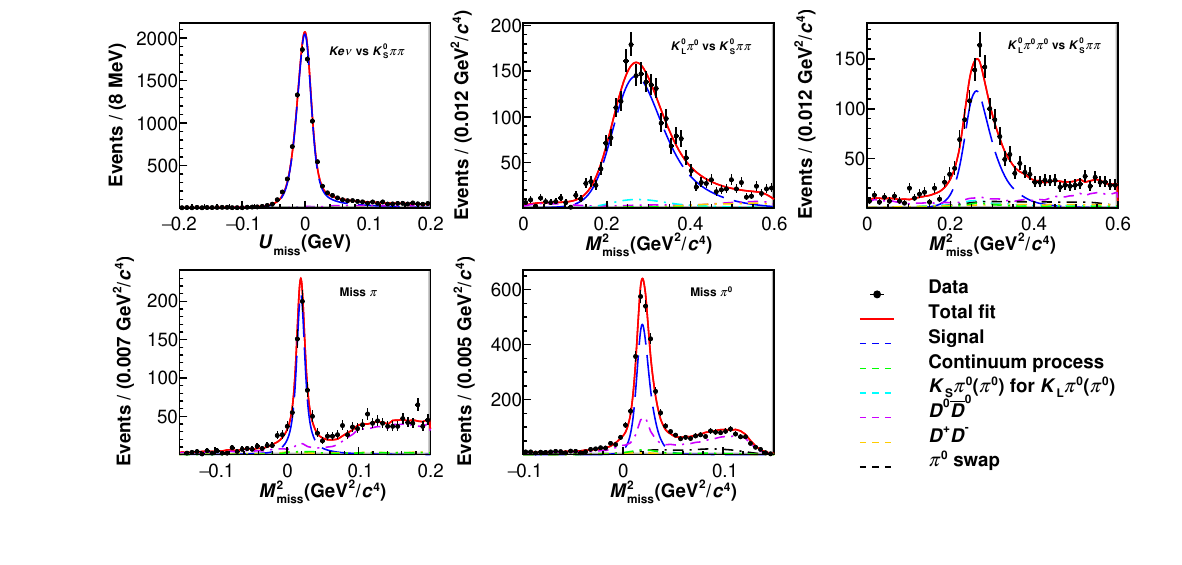}}
\caption{The 1D unbinned maximum likelihood fits for $\kspipi$ versus partially reconstructed tags on $M^{2}_{\rm miss}/U_{\rm miss}$. The black points with error bars represent data. The dashed lines indicate the signal and background shapes. The red lines represent the sum of signal and background shapes. The fitting plots in logarithm scale are shown in Figure~\ref{fig:ksfitted_log}.} 
\label{fig:ksfitted}
\end{figure}

In the partially reconstructed modes, such as $K_L^0X$, missing $\pi$
and missing $\pi^0$, 1D unbinned maximum likelihood fits are performed
to the $M^2_{{\rm miss}}$ distributions. For the $\kenu$ tag mode, the
fit is applied to the $U_{\rm miss}$ distribution. These fit results
are shown in Figure~\ref{fig:ksfitted} and the logarithmic-scale fits are shown in Figure~\ref{fig:ksfitted_log}. The determination of the
signal component, peaking background,  $D^{+}D^{-}$ background and continuum processes follow the same methodology as used for fully reconstructed tags. Notably, in signal events tagged by the $K_L^0\pi^0\pi^0$ mode or the
$(\pi^{0}(\pi^{0})_{\rm miss})_{K_S^0}\pi\pi$ mode, photons might swap
between the two $\pi^{0}$'s in the tag side, which introduces a flat
combinatorial background. The ratio of this background to the signal
yield is fixed according to signal MC simulations.
 
\begin{figure}[!htbp]
\centering
\parbox{1.0\textwidth}{\centering 
 \includegraphics[width=1.0\textwidth]{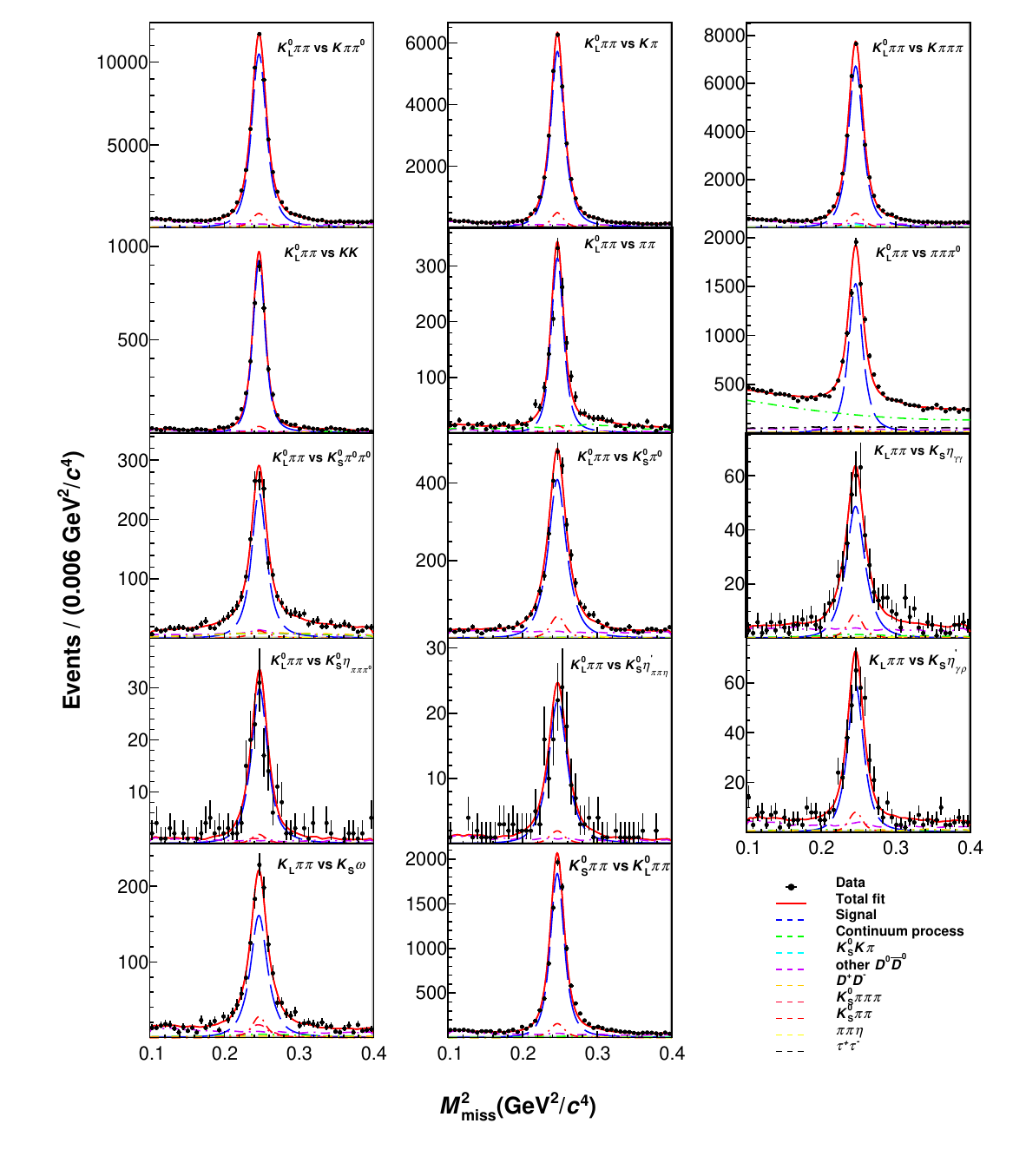}}
\caption{The 1D unbinned maximum likelihood fits for $\klpipi$ versus various tag modes on $M^{2}_{\rm miss}$. The black points with error bars represent data. The dashed lines indicate the signal and background shapes. The red lines represent the sum of signal and background shapes. The fitting plots in logarithm scale are shown in Figure~\ref{fig:klfitted_log}.} 
\label{fig:klfitted}
\end{figure}

The fits to the $M^2_{{\rm miss}}$ distributions for $\klpipi$ decays
tagged with fully reconstructed tag modes are shown in
Figure~\ref{fig:klfitted}. The fit results in logarithm scale are plotted in Figure~\ref{fig:klfitted_log}. Peaking background is fixed according to the inclusive MC
simulations. Combinatorial backgrounds in these fits arise from $D^{+}D^{-}$ decays, $\Dz\Dzb$ decays and continuum
processes. The shapes of both signal and combinatorial background
components are derived from the MC samples. For the DT $\klpipi$
versus $\pipipio$ case, the background contribution from the process
$e^+e^-\to\tau^{+}\tau^{-}$, which is non-negligible, is fixed
according to the inclusive MC samples.

The ST and DT efficiencies are determined through simulated MC
samples. For the tracking and PID efficiencies of pions and
$K^{0}_{S}$ reconstruction efficiency, their differences between data
and MC simulations are studied with a control sample of hadronic
$D\bar{D}$ events. These differences are corrected for in the
detection efficiency of the decay modes of $\kspipi$, $K_{S}^{0}
(\pi)_{\rm miss}\pi$, $(\pi^{0}(\pi^{0})_{\rm miss})_{K_S^0}\pi\pi$,
and $\klpipi$. However, the efficiency corrections are not applied to
the fully reconstructed tag modes, as their detection efficiencies
largely cancel out in the ratio of the ST and DT yields. Furthermore,
the difference of reconstruction efficiency is found to be negligible
for the $\kenu$ tag. The resultant efficiencies are listed in
Table~\ref{tab:yields}.

\begin{table}[!ht]
\begin{center}
\scriptsize
\begin{tabular}{lr@{\,$\pm$\,}lr@{\,$\pm$\,}lr@{\,$\pm$\,}lr@{\,$\pm$\,}lr@{\,$\pm$\,}lr@{\,$\pm$\,}l}
\hline
Mode & \multicolumn{2}{c}{$N_{\rm ST}$} &  \multicolumn{2}{c}{$\epsilon_{\rm ST}$} & \multicolumn{2}{c}{$N_{\rm DT}^{\kspipi}$} & \multicolumn{2}{c}{$\epsilon_{\rm DT}^{\kspipi}$}  & \multicolumn{2}{c}{$N_{\rm DT}^{\klpipi}$} & \multicolumn{2}{c}{$\epsilon_{\rm DT}^{\klpipi}$}  \\
\hline
$\kpi$ & 1449322& 1262 & 65.3&0.0 & 12159 &122 & 26.1 &0.1 & 27492 & 180 & 40.5 & 0.1 \\
$\kpipio$ & 2913155& 2011 &35.6&0.0 &23337 &172 & 13.6 &0.1 & 53536 & 257 &  21.6 & 0.1 \\
$\kthreepi$ & 1944170 & 1571 &40.8&0.0 &15341 &143 & 16.0 &0.1 & 33178 & 205 & 21.8 & 0.1 \\
$\kenu$& \multicolumn{2}{c}{-} &  \multicolumn{2}{c}{-} & 8835 &107 & 21.1 & 0.1 &  \multicolumn{2}{c}{-} &  \multicolumn{2}{c}{-} \\
\hline
$\pipipio$ & 300905 & 1131 & 37.1 & 0.1 & 2280 &53 & 15.3 & 0.0 & 7540 & 127 & 25.2 & 0.0  \\
$\pipi$ & 55461 & 296 & 64.5\,&\,0.1  & 431 &21 & 25.1 & 0.1 & 1322 & 43 & 41.3 & 0.1  \\
$\kk$ &  150181 & 421 & 61.0\,&\,0.1 & 1124 &37 & 24.3 & 0.1 & 3783 & 66 & 38.7 & 0.1 \\
$\kspiopio$ & 65631 & 409 & 16.6\,&\,0.2 & 392 &24 & 6.2 & 0.0 & 1325 & 47 & 9.2 & 0.0 \\
$\klpio$ & \multicolumn{2}{c}{-} & \multicolumn{2}{c}{-} & 1969 & 48 & 17.9 & 0.1 & \multicolumn{2}{c}{-} & \multicolumn{2}{c}{-} \\
\hline
$\kspio$ & 190361 & 471 & 40.3\,&\,0.3  & 1640 &43 & 14.8 & 0.0 & 2547 & 66 & 22.6 & 0.1  \\
$ K_{S}^{0} \eta_{\gamma\gamma}$ & 26016 & 195 & 33.4 & 0.2  &  221 &15 & 12.0 & 0.1 & 313 & 25 & 19.2 & 0.1 \\
$ K_{S}^{0} \eta^{\prime}_{\pi\pi\eta}$ & 9262 & 105 & 14.5 & 0.0  &  67 &9 & 5.2 & 0.1 & 126 & 13 & 8.2 & 0.1 \\
$ K_{S}^{0} \eta_{\pi\pi\pi^{0}}$  & 9006 & 119 & 19.6 & 0.1 & 57 &8 & 6.5 & 0.1 & 121 & 15 & 10.5 & 0.1 \\
$ K_{S}^{0} \eta^{\prime}_{\pi\pi\gamma}$  & 23015 & 195 & 19.9 & 0.1 & 175 &14 & 7.2 & 0.1 & 320 & 22 & 11.4 & 0.1 \\
$\ksomega$ & 72787 & 375 & 17.3 & 0.1 & 564 &27 & 6.0 & 0.0 & 967 & 44 & 9.8 & 0.1 \\
$\klpiopio$ & \multicolumn{2}{c}{-} & \multicolumn{2}{c}{-} & 908 &33 & 6.6 & 0.0 & \multicolumn{2}{c}{-} & \multicolumn{2}{c}{-}  \\
\hline
$\kspipi$ & \multicolumn{2}{c}{-} & \multicolumn{2}{c}{-}  & 2302 &57 &  18.5 & 0.0 & 8506 & 103 & 22.0 & 0.0 \\
$K_{S}^{0} (\pi)_{\rm miss}\pi$ & \multicolumn{2}{c}{-} & \multicolumn{2}{c}{-}  & 572 &28 &  4.4 & 0.0 &  \multicolumn{2}{c}{-} &  \multicolumn{2}{c}{-}  \\
$(\pi^{0}(\pi^{0})_{\rm miss})_{K_S^0}\pi\pi$ & \multicolumn{2}{c}{-} & \multicolumn{2}{c}{-}   &  1715 &49 & 16.6 & 0.0 & \multicolumn{2}{c}{-} & \multicolumn{2}{c}{-} \\
\hline
\end{tabular}
\end{center}
\caption{Results of the ST yields ($N_{\rm ST}$), ST efficiency ($\epsilon_{\rm ST}$), DT yields ($N_{\rm DT}^{\kspipi}$) for $\kspipi$ signals and the corresponding DT efficiency ($\epsilon_{\rm DT}^{\kspipi}$), as well as DT yields ($N_{\rm DT}^{\klpipi}$) for $\klpipi$ signals and the corresponding DT efficiency ($\epsilon_{\rm DT}^{\klpipi}$), where the uncertainties are statistical. The efficiencies are in units of percentage.}
\label{tab:yields}
\end{table}

\section{\boldmath Measurement of $c^{(\prime)}_{i}$ and $s^{(\prime)}_{i}$ parameters}
\label{sec:cisi}

\subsection{\boldmath Determination of the $K_{i}^{(\prime)}$ parameters}

 The efficiency-corrected binned yields of flavour-specific
 $D^0\to\kslpipi$ decays, denoted as $K_{i}^{(\prime)}$, are necessary
 to determine the expected yields in the decays sensitive to the
 strong-phase difference parameters. $K_{i}^{(\prime)}$ is calculated
 from the observed yields $N^{\text{obs}({\prime})}_{i}$ in the
 $i^{\rm th}$ bin of flavour tags via $K_{i}^{(\prime)}=\Sigma^{8}_{j=-8}[(\epsilon^{(\prime)})^{-1}]_{ij}N_{j}^{\text{obs}(\prime)}$, where the efficiency matrix $\epsilon_{ij}^{(\prime)}$ denotes the efficiency correction matrix determined with MC simulations. This
matrix describes the experimental migration effect when a
signal event is produced in the $j^{\rm th}$ bin but reconstructed in
the $i^{\rm th}$ bin.
 
\begin{figure}
\centering
\parbox{1.0\textwidth}{\centering 
 \includegraphics[width=1.0\textwidth]{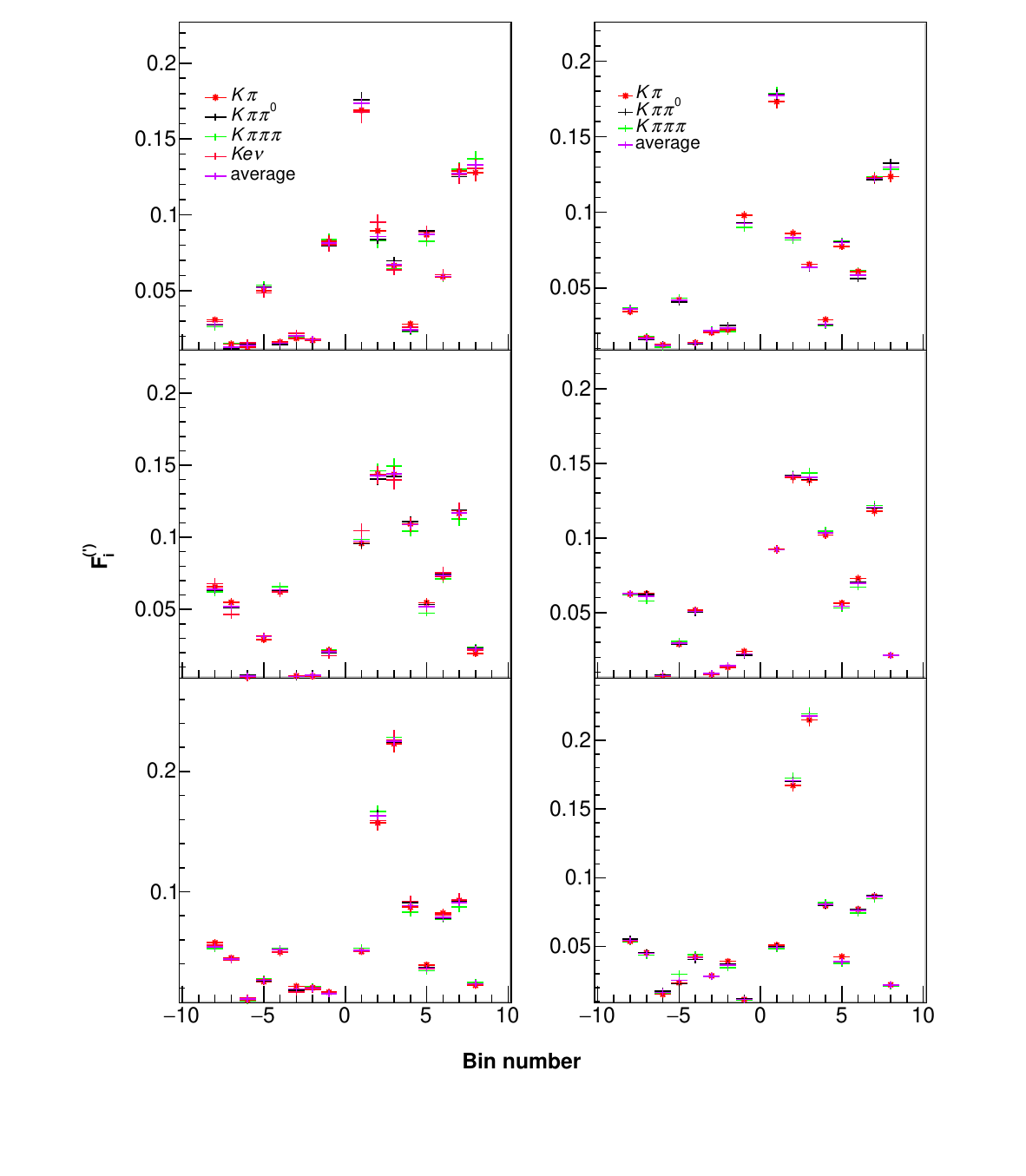}}
\caption{The $F^{(\prime)}_{i}$ distribution for the $\kspipi$ decay (left) and the $\klpipi$ decay (right) under the equal binning scheme (top), the optimal binning scheme (middle) and the modified optimal binning scheme (bottom). The different colors represent the $F^{(\prime)}_{i}$ results of different tags and the averaged values.} 
\label{fig:Fi}
\end{figure}

Doubly Cabibbo-suppressed (DCS) processes include the hadronic
flavoured tags $\kpi$, $\kpipio$ and $\kthreepi$. The DCS contribution
must be subtracted because they lead to incorrect assignments of the
charm quantum number. To account for this effect, the $\kslpipi$
signal events are scaled by correction factors. These factors are
calculated according to Eq.~(19) from Ref.~\cite{BESIII:2020khq},
incorporating the updated strong-phase parameters
~\cite{BESIII:2021eud,schwartz2022effectd0bard0mixingcabibbofavored}
and amplitude models
~\cite{BaBar:2018cka,BESIII:2022qvy}. Furthermore, the binned
fractional signal yields $ F_{i}^{(\prime)} =
\frac{K_{i}^{(\prime)}}{\sum_{j}{K_{j}^{(\prime)}}}$ for each flavour
tag are determined. As demonstrated in Figure~\ref{fig:Fi}, these
yields are in good agreement across bins. The averaged $
F_{i}^{(\prime)}$ values are summarized in Table~\ref{tab:kslfi}.

\begin{table}[!htpb]
    \centering
    \small

    \resizebox{\linewidth}{!}{
    \begin{tabular}{ccccccc}
    \hline
         & \multicolumn{2}{c}{Equal binning scheme} & \multicolumn{2}{c}{Optimal binning scheme} & \multicolumn{2}{c}{Modified optimal binning scheme} \\
         \hline
        Bin &  $F_{i}$ & $F^{\prime}_{i}$ & $F_{i}$ & $F^{\prime}_{i}$ & $F_{i}$ & $F^{\prime}_{i}$ \\
        \hline
        $-$8 & 0.028\,$\pm$\,0.001 & 0.036\,$\pm$\,0.001 & 0.064\,$\pm$\,0.002 & 0.063\,$\pm$\,0.001 & 0.054\,$\pm$\,0.001 & 0.054\,$\pm$\,0.001 \\
        $-$7 & 0.013\,$\pm$\,0.001 & 0.017\,$\pm$\,0.001 & 0.052\,$\pm$\,0.002 & 0.061\,$\pm$\,0.001 & 0.044\,$\pm$\,0.001 & 0.046\,$\pm$\,0.001  \\
        $-$6 & 0.014\,$\pm$\,0.001 & 0.012\,$\pm$\,0.001 & 0.004\,$\pm$\,0.001 & 0.008\,$\pm$\,0.001 & 0.011\,$\pm$\,0.001 & 0.017\,$\pm$\,0.001 \\
        $-$5 & 0.052\,$\pm$\,0.002 & 0.042\,$\pm$\,0.001 & 0.031\,$\pm$\,0.001 & 0.030\,$\pm$\,0.001 & 0.026\,$\pm$\,0.001 & 0.025\,$\pm$\,0.001 \\
        $-$4 & 0.016\,$\pm$\,0.001 & 0.014\,$\pm$\,0.001 & 0.064\,$\pm$\,0.002 & 0.051\,$\pm$\,0.001 & 0.052\,$\pm$\,0.001 & 0.043\,$\pm$\,0.001 \\
        $-$3 & 0.020\,$\pm$\,0.001 & 0.022\,$\pm$\,0.001 & 0.004\,$\pm$\,0.001 & 0.009\,$\pm$\,0.001 & 0.019\,$\pm$\,0.001 & 0.028\,$\pm$\,0.001 \\
        $-$2 & 0.018\,$\pm$\,0.001 & 0.024\,$\pm$\,0.001 & 0.004\,$\pm$\,0.001 & 0.014\,$\pm$\,0.001 & 0.020\,$\pm$\,0.001 & 0.037\,$\pm$\,0.001 \\
        $-$1 & 0.081\,$\pm$\,0.002 & 0.093\,$\pm$\,0.002 & 0.020\,$\pm$\,0.001 & 0.022\,$\pm$\,0.001 & 0.015\,$\pm$\,0.001 & 0.011\,$\pm$\,0.001\\
        1 & 0.174\,$\pm$\,0.003 & 0.177\,$\pm$\,0.003 & 0.097\,$\pm$\,0.002  & 0.092\,$\pm$\,0.002 & 0.051\,$\pm$\,0.002 & 0.049\,$\pm$\,0.001 \\
        2 & 0.086\,$\pm$\,0.002 & 0.083\,$\pm$\,0.002 & 0.143\,$\pm$\,0.003 & 0.142\,$\pm$\,0.002 & 0.163\,$\pm$\,0.003 & 0.171\,$\pm$\,0.002 \\
        3 & 0.067\,$\pm$\,0.002 & 0.064\,$\pm$\,0.001 & 0.144\,$\pm$\,0.003 & 0.141\,$\pm$\,0.002 & 0.225\,$\pm$\,0.004 & 0.218\,$\pm$\,0.003 \\
        4 & 0.025\,$\pm$\,0.001 & 0.026\,$\pm$\,0.001 & 0.109\,$\pm$\,0.002 & 0.103\,$\pm$\,0.002 & 0.088\,$\pm$\,0.002 & 0.081\,$\pm$\,0.001 \\
        5 & 0.087\,$\pm$\,0.002 & 0.080\,$\pm$\,0.001 & 0.052\,$\pm$\,0.001 & 0.054\,$\pm$\,0.001 & 0.037\,$\pm$\,0.001 & 0.039\,$\pm$\,0.001 \\
        6 & 0.060\,$\pm$\,0.002 & 0.059\,$\pm$\,0.001 & 0.074\,$\pm$\,0.002 & 0.070\,$\pm$\,0.001 & 0.079\,$\pm$\,0.002 & 0.076\,$\pm$\,0.001 \\
        7 & 0.127\,$\pm$\,0.003 & 0.122\,$\pm$\,0.002 & 0.117\,$\pm$\,0.002 & 0.120\,$\pm$\,0.002 & 0.091\,$\pm$\,0.002 & 0.086\,$\pm$\,0.002 \\
        8 & 0.133\,$\pm$\,0.003 & 0.130\,$\pm$\,0.002 & 0.023\,$\pm$\,0.001 & 0.021\,$\pm$\,0.001 & 0.024\,$\pm$\,0.001 & 0.022\,$\pm$\,0.001 \\
        \hline
    \end{tabular}
    }
    \caption{The averaged $ F_{i}^{(\prime)}$ values for the three binning schemes, where the uncertainties are only statistical.}
    \label{tab:kslfi}
\end{table}

\subsection{Expected number of binned signal events}

Following Eqs.~\eqref{eq_ksci} and \eqref{eq_klci}, the binned signal
yields of the $\kslpipi$ decays tagged by $C\!P$ eigenstates are
expressed in terms of the ST yields and reconstruction efficiencies as
\begin{equation}
    \label{eq_ksci_eff}
    \begin{aligned}
            M_{i}=h_{CP}\Sigma^{8}_{j=1} \epsilon_{ij} (K_{j} + K_{-j} -(2F^{+}_{CP}-1)\times2c_{j}\sqrt{K_{j}K_{-j}}), \\
            M_{i}^{\prime}=h_{CP}^{\prime}\Sigma^{8}_{j=1} \epsilon_{ij}^{\prime} (K_{j}^{\prime} + K_{-j}^{\prime} +(2F^{+}_{CP}-1)\times2c_{j}^{\prime}\sqrt{K_{j}^{\prime}K_{-j}^{\prime}}).
     \end{aligned}
     \end{equation}
Here, the normalization factor $h_{C\!P}^{(\prime)}$ is determined by
the ratio of the efficiency-corrected ST yields of the $C\!P$
eigenstates to those of the flavour modes, given by
$S_{C\!P}/S^{(\prime)}_{F}$. To accurately account for the DCS
contributions and charm mixing effects in the ST yields, $S_{C\!P}$ is
corrected with the factor of ($1 - (2F^{+}_{C\!P}-1) y + y^2 $) and
$S^{(\prime)}_{F}$ is corrected with ($1 + (r_D^{F})^2 - 2y
R_{F}r^{F}_D \cos\delta^{F}_D + y^2$). Here, $r_D^{F}$, $R_F$ and
$\delta_D^F$ are the decay parameters in the three hadronic flavour
tags as determined in
Refs.~\cite{BESIII:2021eud,schwartz2022effectd0bard0mixingcabibbofavored}. The
charm mixing parameter $y$ is equal to $(0.647\pm0.024)\%$ according to
Ref.~\cite{schwartz2022effectd0bard0mixingcabibbofavored}.

For the DT $\kspipi$ versus $\kspipi$, there are 72 observables due to
the symmetry of the Dalitz plot. In contrast, the DT $\kspipi$ versus
$\klpipi$ involves 128 observables. The expected signal yields are
expressed as
 \begin{equation}
    \label{eq_kssi_eff}
    \begin{aligned}
            M_{ij} = h_{\rm corr}\Sigma^{m=72}_{m\to(i,j)=0}\epsilon_{mn}(K_{i}K_{-j} + K_{-i}K_{j} - 2\sqrt{K_{i}K_{-i}K_{j}K_{-j}}(c_{i}c_{j}+s_{i}s_{j})), \\
            M_{ij}^{\prime} = h_{\rm corr}^{\prime}\Sigma^{m=128}_{m\to(i,j)=0}\epsilon_{mn}^{\prime}(K_{i}K_{-j}^{\prime} + K_{-i}K_{j}^{\prime} + 2\sqrt{K_{i}K_{-i}K_{j}^{\prime}K_{-j}^{\prime}}(c_{i}c_{j}^{\prime}+s_{i}s_{j}^{\prime})).
     \end{aligned} \end{equation} 
Here, the normalization factors are
     given by
     $h_{\rm corr}= \frac{N_{\Dz\Dzb}}{S^{2}_{\rm flav}} \times \alpha
     \beta $ and $h_{\rm corr}^{\prime} = \frac{ 2N_{\Dz\Dzb}}{S_{\rm
     flav}S^{\prime}_{\rm flav}}$. The parameter $\alpha$ is
     determined from the symmetry of Dalitz plots and is 1 for the bin
     pair $(i,j)$ where $|i| = |j|$ and 2 otherwise. $\beta =
     \frac{2Br({K_{S}^{0}\to \pi^{0}\pi^{0}})}{Br({K_{S}^{0}\to
     \pipi)}}$ when one $K_{S}^{0}$ is reconstructed with the decay
     $\pi^{0}\pi^{0}$. Here, $Br({K_{S}^{0}\to \pi^{0}\pi^{0}})$ and
     $Br({K_{S}^{0}\to \pipi)}$ are the branching fractions of
     $K_{S}^{0}\to \pi^{0}\pi^{0}$ and $K_{S}^{0}\to \pipi$. The total
     number of $\Dz\Dzb$ pairs $N_{\Dz\Dzb}$ is estimated to be (28660
     $\pm$ 247) $\times 10^{3}$ for the utilized data
     set~\cite{Ablikim__2024}.

\subsection{\boldmath Fit to the $c^{(\prime)}_{i}$ and
$s^{(\prime)}_{i}$ parameters} 

The strong-phase difference parameters $c_{i}^{(\prime)}$ and
$s_{i}^{(\prime)}$ are determined via a maximum likelihood fit to the
observed and expected numbers of signal events. Under the assumption
that the sum of binned signal yields and peaking background follows a
Poisson distribution, the likelihood function is formulated as
\begin{equation} \label{eq_llfit}
\centering
\begin{aligned}
     -2\rm{ln}\mathscr{L} = & -2\Sigma_{i=1}^{8}{\rm ln}(\text{Poisson}(N_{i}^{\rm obs},N_{i}^{\rm exp})_{\kslpipi ~ \text{versus} ~ C\!P }\\
    & -2\Sigma_{i=1}^{72}{\rm ln}(\text{Poisson}(N_{i}^{\rm obs},N_{i}^{\rm exp})_{\kspipi ~ \text{versus} ~ \kspipi } \\  
   & -2\Sigma_{i=1}^{128}{\rm ln}(\text{Poisson}(N_{i}^{\rm obs},N_{i}^{\rm exp})_{\kspipi ~ \text{versus} ~ \klpipi}.\\
\end{aligned}
\end{equation}

Here, $N^{\rm exp}_{i}$ represents the sum of the expected signal
yields $M_{i}^{(\prime)}$ and the peaking background contributions. However, since the non-peaking background in the $\klpipi$ versus $\pipipio$ mode is significantly larger than that in other modes,  the number of non-peaking background events is added to $N^{\rm exp}_{i}$. To
avoid large statistical fluctuations, self-conjugated tags $\kspipi$,
$K_{S}^{0} (\pi)_{\rm miss}\pi$, $(\pi^{0}(\pi^{0})_{\rm
miss})_{K_S^0}\pi\pi$ are combined. For $C\!P$ tags, $K_{S}^{0}
\eta_{\gamma\gamma}$ and $K_{S}^{0} \eta_{\pipipio}$ tags are combined
similarly, and $K_{S}^{0} \eta^{\prime}_{\gamma\rho}$ and $
K_{S}^{0}\eta^{\prime}_{\pi\pi\eta}$ tags are also combined. The
results of the model-independent fits to $c^{(\prime)}_{i}$ and
$s^{(\prime)}_{i}$ parameters under the three binning schemes are
listed in Table~\ref{tab:cifit}. Discussions on systematic
uncertainties are presented in the next section. The correlation matrices
of statistical uncertainties are shown in Appendix~\ref{app:corr}.

\begin{table}[]
    \centering
    \small
    \resizebox{\linewidth}{!}{
    \begin{tabular}{cr@{\,$\pm$\,}lr@{\,$\pm$\,}lr@{\,$\pm$\,}lr@{\,$\pm$\,}l}
    \hline
        Bin & \multicolumn{2}{c}{$c_{i}$} & \multicolumn{2}{c}{$s_{i}$} & \multicolumn{2}{c}{$c^{\prime}_{i}$} & \multicolumn{2}{c}{$s^{\prime}_{i}$}  \\
        \hline
        \multicolumn{9}{c}{Equal binning scheme} \\
     \hline
         1 & 0.682&0.017\,$\pm$\,0.008 & $-$0.007&0.052\,$\pm$\,0.006 & 0.755&0.018\,$\pm$\,0.009 & 0.025&0.087\,$\pm$\,0.006 \\
        2 & 0.602&0.035\,$\pm$\,0.013 & 0.270&0.120\,$\pm$\,0.022 & 0.668&0.034\,$\pm$\,0.012 & 0.557&0.143\,$\pm$\,0.018 \\
        3 & 0.060&0.034\,$\pm$\,0.009 & 0.666&0.088\,$\pm$\,0.015 & 0.265&0.034\,$\pm$\,0.007 & 1.031&0.150\,$\pm$\,0.018 \\
        4 &$-$0.553&0.043\,$\pm$\,0.012 & 0.681&0.123\,$\pm$\,0.023 & $-$0.575&0.045\,$\pm$\,0.008 & 0.722&0.196\,$\pm$\,0.030 \\
        5 & $-$0.955&0.017\,$\pm$\,0.007 & $-$0.117&0.056\,$\pm$\,0.005 & $-$0.923&0.020\,$\pm$\,0.010 & $-$0.148&0.118\,$\pm$\,0.010 \\
        6 & $-$0.574&0.043\,$\pm$\,0.014 & $-$0.431&0.110\,$\pm$\,0.018 & $-$0.343&0.047\,$\pm$\,0.010 & $-$0.604&0.214\,$\pm$\,0.033 \\
        7 &  0.072&0.045\,$\pm$\,0.014 & $-$0.810&0.102\,$\pm$\,0.018 & 0.313&0.043\,$\pm$\,0.011 & $-$0.717&0.183\,$\pm$\,0.036 \\
        8 &  0.522&0.030\,$\pm$\,0.011 & $-$0.363&0.080\,$\pm$\,0.010 & 0.609&0.032\,$\pm$\,0.011 & $-$0.352&0.134\,$\pm$\,0.020 \\
\hline

      \multicolumn{9}{c}{Optimal binning scheme} \\
       \hline
       1 & 0.034&0.036\,$\pm$\,0.011 & $-$0.741&0.090\,$\pm$\,0.015 & 0.179&0.036\,$\pm$\,0.007 & $-$0.686&0.155\,$\pm$\,0.029 \\
        2 & 0.946&0.066\,$\pm$\,0.034 & $-$0.116&0.179\,$\pm$\,0.033 & 0.823&0.044\,$\pm$\,0.021 & 0.129&0.167\,$\pm$\,0.024 \\
        3 & 0.244&0.078\,$\pm$\,0.025 & $-$0.622&0.168\,$\pm$\,0.030 & 0.613&0.053\,$\pm$\,0.024 & $-$0.390&0.200\,$\pm$\,0.035 \\
        4 & $-$0.909&0.016\,$\pm$\,0.008 & $-$0.195&0.053\,$\pm$\,0.006 & $-$0.869&0.018\,$\pm$\,0.010 & $-$0.258&0.097\,$\pm$\,0.015 \\
        5 & $-$0.215&0.031\,$\pm$\,0.010 & 0.883&0.083\,$\pm$\,0.016 & $-$0.150&0.031\,$\pm$\,0.005 & 0.906&0.127\,$\pm$\,0.020 \\
        6 & 0.342&0.071\,$\pm$\,0.021 & 0.566&0.185\,$\pm$\,0.039 & 0.646&0.053\,$\pm$\,0.022 & 0.911&0.204\,$\pm$\,0.033 \\
        7 &  0.871&0.017\,$\pm$\,0.008 & 0.128&0.071\,$\pm$\,0.011 & 0.875&0.019\,$\pm$\,0.010 & 0.186&0.088\,$\pm$\,0.007 \\
        8 &  0.845&0.025\,$\pm$\,0.011 & $-$0.197&0.096\,$\pm$\,0.012 & 0.801&0.029\,$\pm$\,0.013 & $-$0.083&0.139\,$\pm$\,0.021 \\
        \hline
        \multicolumn{9}{c}{Modified optimal binning scheme} \\
     \hline
        1 & $-$0.283&0.044\,$\pm$\,0.012 & $-$0.276&0.128\,$\pm$\,0.020 & $-$0.180&0.049\,$\pm$\,0.006 & $-$0.239&0.188\,$\pm$\,0.035 \\
        2 & 0.828&0.029\,$\pm$\,0.016 & $-$0.042&0.094\,$\pm$\,0.017 & 0.860&0.024\,$\pm$\,0.012 & 0.036&0.098\,$\pm$\,0.015 \\
        3 & 0.098&0.034\,$\pm$\,0.016 & $-$0.687&0.082\,$\pm$\,0.015 & 0.389&0.028\,$\pm$\,0.010 & $-$0.507&0.117\,$\pm$\,0.022 \\
        4 & $-$0.965&0.016\,$\pm$\,0.006 & $-$0.195&0.060\,$\pm$\,0.006 & $-$0.894&0.020\,$\pm$\,0.010 & $-$0.134&0.113\,$\pm$\,0.012 \\
        5 & $-$0.427&0.034\,$\pm$\,0.010 & 0.845&0.102\,$\pm$\,0.019 & $-$0.473&0.034\,$\pm$\,0.005 & 0.800&0.150\,$\pm$\,0.020 \\
        6 & 0.241&0.043\,$\pm$\,0.013 & 0.730&0.114\,$\pm$\,0.019 & 0.516&0.037\,$\pm$\,0.015 & 0.937&0.154\,$\pm$\,0.026\\
        7 &  0.727&0.022\,$\pm$\,0.010 & 0.109&0.090\,$\pm$\,0.011 & 0.713&0.025\,$\pm$\,0.010 & 0.272&0.115\,$\pm$\,0.011 \\
        8 &  0.786&0.027\,$\pm$\,0.010 & $-$0.201&0.103\,$\pm$\,0.012 & 0.729&0.030\,$\pm$\,0.013 & $-$0.148&0.145\,$\pm$\,0.021 \\
   \hline
    \end{tabular}
    }
    \caption{The $c_{i}^{(\prime)}$ and $s_{i}^{(\prime)}$ parameters
      determined from the model-independent measurement under three binning schemes. The first uncertainties are statistical and the second are systematic.}
    \label{tab:cifit}
    \end{table}

\begin{figure}
\centering
\parbox{0.8\textwidth}{\centering 
 \includegraphics[width=0.8\textwidth]{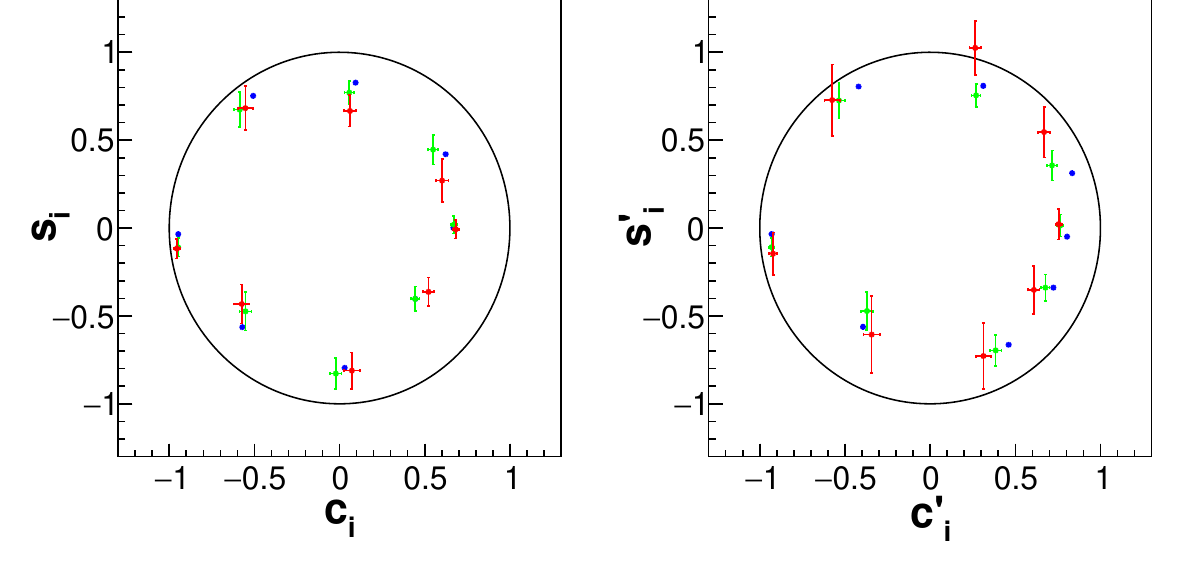}}
\parbox{0.8\textwidth}{\centering 
 \includegraphics[width=0.8\textwidth]{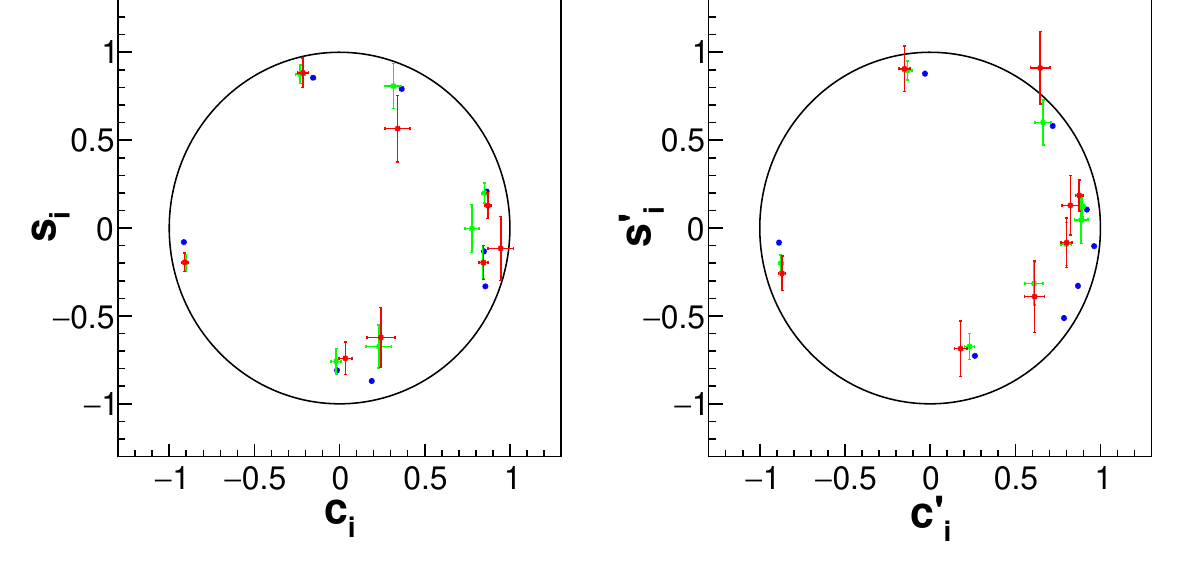}}
\parbox{0.8\textwidth}{\centering 
 \includegraphics[width=0.8\textwidth]{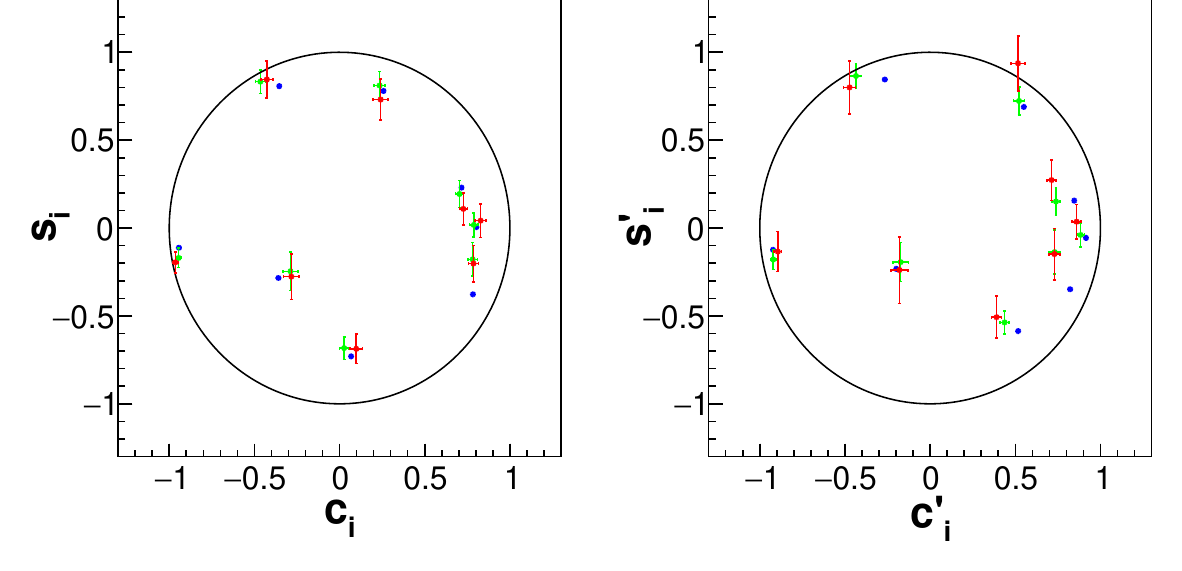}}
\caption{The $c^{(\prime)}_{i}/s_i^{(\prime)}$ parameters determined without constraints (red), with constraints (green) and predicted by the amplitude models~\cite{BaBar:2018cka} (blue) under the equal binning scheme (top), the optimal binning scheme (middle) and the modified optimal binning scheme (bottom).} 
\label{fig:2Dci_nocs}
\end{figure}

Model-predicted differences in the strong-phase parameters between
$\kspipi$ and $\klpipi$ decays, denoted as $\Delta c_{i}$ and $\Delta
s_{i}$, have been used to improve the precision of the
$c^{(\prime)}_{i}$ and $s^{(\prime)}_{i}$
parameters~\cite{BESIII:2020khq}. Using the amplitude model of the
$\kspipi$ decay~\cite{BaBar:2018cka} and the U-spin breaking
parameters for the $\klpipi$ decay~\cite{BESIII:2022qvy}, the
model-predicted $\Delta c_{i}$ and $\Delta s_{i}$ parameters have been
updated, as detailed in Table~\ref{tab:dcs} in Appendix A. The differences between the 
$\kspipi$ amplitude models measured by the Belle and BESIII experiments~\cite{BaBar:2018cka,BESIII:2022qvy}
and the uncertainties of the U-spin breaking parameters are assigned as 
their uncertainties. The $\Delta
c_{i}$ and $\Delta s_{i}$ parameters determined by the unconstrained
$c_{i}$ and $s_{i}$ results, which are consistent with the
model-predicted parameters within 2$\sigma$, are shown in
Table~\ref{tab:dcs_data} in Appendix A. To incorporate these constraints into the likelihood fit, 
a $\chi^{2}$ term as follows is added to the likelihood function in
Eq.~\eqref{eq_llfit} \begin{equation} \centering \begin{aligned}
    \chi^2 = & \frac{(c_i^{\prime}-c_i-\Delta c_i)^2}{\sigma_{\Delta
        c_i}^2} + \frac{(s_i^{\prime}-s_i-\Delta
      s_i)^2}{\sigma_{\Delta s_i}^2}.
\end{aligned}
\end{equation}
The fit results incorporating the model-predicted $\Delta c_{i}$ and
$\Delta s_{i}$ constraints are summarized in
Table~\ref{tab:dcsfit}.  Compared to the unconstrained $c_{i}$ and
$s_{i}$ results, the constrained $c_{i}$ and $s_{i}$ results are
closer to the model-prediction with a significantly smaller
uncertainty. Both sets of results are consistent within
2$\sigma$ as illustrated in Figure ~\ref{fig:2Dci_nocs}.
The constrained results are consistent with the previous BESIII
measurement~\cite{BESIII:2020khq} within 2$\sigma$.

\begin{table}[]
    \centering
    \small
    \resizebox{\linewidth}{!}{
    \setlength\tabcolsep{6pt}
    \begin{tabular}{cr@{\,$\pm$\,}lr@{\,$\pm$\,}lr@{\,$\pm$\,}lr@{\,$\pm$\,}lcc}
    \hline
        Bin & \multicolumn{2}{c}{$c_{i}$} & \multicolumn{2}{c}{$s_{i}$} & \multicolumn{2}{c}{$c^{\prime}_{i}$} & \multicolumn{2}{c}{$s^{\prime}_{i}$}  \\
        \hline
        \multicolumn{9}{c}{Equal binning scheme} \\
        \hline
        1 & 0.671&0.015\,$\pm$\,0.008 & 0.019&0.050\,$\pm$\,0.004 & 0.763&0.015\,$\pm$\,0.007 & 0.014&0.061\,$\pm$\,0.004 \\
        2 & 0.549&0.028\,$\pm$\,0.010 & 0.446&0.082\,$\pm$\,0.009 & 0.716&0.027\,$\pm$\,0.010 & 0.356&0.083\,$\pm$\,0.010 \\
        3 & 0.056&0.026\,$\pm$\,0.007 & 0.770&0.064\,$\pm$\,0.010 & 0.270&0.026\,$\pm$\,0.006 & 0.754&0.065\,$\pm$\,0.010 \\
        4 & $-$0.586&0.034\,$\pm$\,0.009 & 0.674&0.097\,$\pm$\,0.011 & $-$0.535&0.036\,$\pm$\,0.007 & 0.726&0.101\,$\pm$\,0.015 \\
        5 & $-$0.948&0.013\,$\pm$\,0.007 & $-$0.110&0.052\,$\pm$\,0.004 & $-$0.930&0.014\,$\pm$\,0.007 & $-$0.109&0.053\,$\pm$\,0.004 \\
        6 & $-$0.552&0.034\,$\pm$\,0.010 & $-$0.474&0.107\,$\pm$\,0.016 & $-$0.371&0.035\,$\pm$\,0.009 & $-$0.472&0.108\,$\pm$\,0.016 \\
        7 & $-$0.023&0.032\,$\pm$\,0.011 & $-$0.827&0.087\,$\pm$\,0.013 & 0.385&0.032\,$\pm$\,0.009 & $-$0.696&0.088\,$\pm$\,0.012 \\
        8 &  0.444&0.024\,$\pm$\,0.009 & $-$0.402&0.070\,$\pm$\,0.007 & 0.677&0.024\,$\pm$\,0.009 & $-$0.339&0.075\,$\pm$\,0.007 \\
\hline

      \multicolumn{9}{c}{Optimal binning scheme} \\
        \hline
         1 & $-$0.020&0.028\,$\pm$\,0.009 & $-$0.758&0.071\,$\pm$\,0.014 & 0.231&0.028\,$\pm$\,0.007 & $-$0.674&0.072\,$\pm$\,0.014 \\
        2 & 0.777&0.037\,$\pm$\,0.021 & $-$0.003&0.136\,$\pm$\,0.014 & 0.887&0.036\,$\pm$\,0.019 & 0.046&0.132\,$\pm$\,0.014 \\
        3 & 0.231&0.071\,$\pm$\,0.024 & $-$0.674&0.120\,$\pm$\,0.019 & 0.610&0.050\,$\pm$\,0.021 & $-$0.316&0.120\,$\pm$\,0.019 \\
        4 & $-$0.903&0.012\,$\pm$\,0.007 & $-$0.197&0.047\,$\pm$\,0.005 & $-$0.876&0.013\,$\pm$\,0.007 & $-$0.200&0.047\,$\pm$\,0.005 \\
        5 & $-$0.231&0.026\,$\pm$\,0.007 & 0.876&0.052\,$\pm$\,0.009 & $-$0.133&0.026\,$\pm$\,0.005 & 0.897&0.054\,$\pm$\,0.009 \\
        6 & 0.316&0.045\,$\pm$\,0.017 & 0.808&0.123\,$\pm$\,0.031 & 0.663&0.043\,$\pm$\,0.016 & 0.600&0.124\,$\pm$\,0.032 \\
        7 &  0.851&0.014\,$\pm$\,0.008 & 0.199&0.057\,$\pm$\,0.006 & 0.895&0.013\,$\pm$\,0.008 & 0.124&0.061\,$\pm$\,0.005 \\
        8 &  0.843&0.025\,$\pm$\,0.011 & $-$0.195&0.095\,$\pm$\,0.012 & 0.800&0.028\,$\pm$\,0.013 & $-$0.091&0.120\,$\pm$\,0.018 \\
        \hline
        \multicolumn{9}{c}{Modified optimal binning scheme} \\
        \hline
        1 & $-$0.288&0.041\,$\pm$\,0.012 & $-$0.246&0.109\,$\pm$\,0.013 & $-$0.174&0.045\,$\pm$\,0.006 & $-$0.193&0.110\,$\pm$\,0.014 \\
        2 & 0.788&0.021\,$\pm$\,0.012 & 0.018&0.067\,$\pm$\,0.007 & 0.882&0.020\,$\pm$\,0.010 & $-$0.039&0.067\,$\pm$\,0.007 \\
        3 & 0.026&0.025\,$\pm$\,0.014 & $-$0.682&0.063\,$\pm$\,0.012 & 0.437&0.024\,$\pm$\,0.009 & $-$0.537&0.064\,$\pm$\,0.012 \\
        4 & $-$0.945&0.013\,$\pm$\,0.007 & $-$0.170&0.055\,$\pm$\,0.008 & $-$0.921&0.013\,$\pm$\,0.007 & $-$0.179&0.056\,$\pm$\,0.008 \\
        5 & $-$0.465&0.028\,$\pm$\,0.009 & 0.833&0.066\,$\pm$\,0.010 & $-$0.435&0.029\,$\pm$\,0.005 & 0.865&0.069\,$\pm$\,0.010 \\
        6 & 0.235&0.031\,$\pm$\,0.012 & 0.811&0.077\,$\pm$\,0.017 & 0.522&0.030\,$\pm$\,0.010 & 0.722&0.077\,$\pm$\,0.018 \\
        7 &  0.704&0.021\,$\pm$\,0.008 & 0.194&0.076\,$\pm$\,0.012 & 0.739&0.022\,$\pm$\,0.009 & 0.151&0.080\,$\pm$\,0.009 \\
        8 &  0.781&0.026\,$\pm$\,0.010 & $-$0.177&0.095\,$\pm$\,0.013 & 0.731&0.030\,$\pm$\,0.012 & $-$0.137&0.123\,$\pm$\,0.018 \\
        \hline
    \end{tabular}
    }
    \caption{The $c_{i}^{(\prime)}$ and $s_{i}^{(\prime)}$ parameters constrained by $\Delta c_{i}$ and $\Delta s_{i}$ parameters under three binning schemes. The first uncertainties are statistical and the second are systematic.}
    \label{tab:dcsfit}
    \end{table}

\subsection{Systematic uncertainties} \label{sec:syst} 

The dominant sources of systematic uncertainties are identified as
follows: statistical fluctuations in the $K_{i}^{(\prime)}$
parameters, inputs of the hadronic parameters for flavour tags and
charm mixing parameters, statistical fluctuations in the ST yields,
limited statistics of the signal MC samples, and background
estimation.

To determine these uncertainties, pseudo-experiments are
performed. Specifically, 1000 sets of values for the
$K_{i}^{(\prime)}$ parameters, input parameters, ST yields, signal
efficiencies, and background contributions are generated individually
from Gaussian distributions. These distributions have means equal to
the nominal parameter values and standard deviations reflecting their
respective uncertainties.

For each set of generated values, the likelihood fit to obtain the
$c^{(\prime)}_{i}$ and $s^{(\prime)}_{i}$ parameters is repeated for
each uncertainty contribution. No significant bias is observed in the
re-fitted $c^{(\prime)}_{i}$ and $s^{(\prime)}_{i}$ parameters. The
standard deviations of the resultant distributions are taken as the
corresponding systematic uncertainties.

Most of the systematic uncertainties associated with the PID and
tracking efficiencies are found to be negligible. This is due to the
corrections from control samples having been applied to the $K^{0}_{S}$
selection efficiency and the tracking and PID efficiencies of $\pi$'s
not originating from $K^{0}_{S}$. However, the tracking efficiencies
of $\pi$'s coming from $K^{0}_{S}$ introduce a non-negligible
uncertainty. The signal efficiencies are varied by the tracking
efficiency corrections estimated with $\pi^{\pm}$ control samples. The
differences between the nominal and refitted $c^{(\prime)}_{i}$ and
$s^{(\prime)}_{i}$ parameters are then taken as the systematic
uncertainties.

The above systematic uncertainties are determined separately for both
the constrained and unconstrained $c^{(\prime)}_{i}$ and
$s^{(\prime)}_{i}$ results under three different binning schemes,
details of which can be found in Appendix~\ref{app:syst}. The
correlation matrices of the systematic uncertainties, derived from the
pseudo-experiments, are detailed in Appendix~\ref{app:corr}.

\section{\boldmath Impact of the results on the $\gamma$ measurement}
\label{sec:gamma}

The decay mode $B^- \to DK^-$, with $D \to \kspipi$ gives the most
precise model-independent angle $\gamma$ measurement channel, where
the $\gamma$ is determined through comparing the expected and observed
binned yields of $B$ decay. The expected binned yield of $B^- \to
DK^-$, $D\to \kspipi$~\cite{Giri:2003ty} is calculated using the input
parameters of $c_{i}$, $s_{i}$ and $F_{i}$ from $D$ decays and the
parameters $\gamma$, $\delta_{B}$, $r_{B}$ from $B$ decays, which are
formulated as
\begin{equation}\small
  \label{eq:Bexpyields}
   \begin{aligned}
       N^{\rm exp}_{i}(B^{-} \to K^{-}D) &= h^{-}_{B}[F_{i}+r^{2}_{B}F_{-i}+2r_{B}\sqrt{F_{i}F_{-i}}\times c_{i}\cos(\delta_{B}-\gamma) - s_{i}\sin(\delta_{B}-\gamma)], \\
       N^{\rm exp}_{-i}(B^{-} \to K^{-}D) &= h^{-}_{B}[F_{-i}+r^{2}_{B}F_{i}+2r_{B}\sqrt{F_{i}F_{-i}}\times c_{i}\cos(\delta_{B}-\gamma) + s_{i}\sin(\delta_{B}-\gamma)], \\
       N^{\rm exp}_{i}(B^{+} \to K^{+}D) &= h^{+}_{B}[F_{-i}+r^{2}_{B}F_{i}+2r_{B}\sqrt{F_{i}F_{-i}}\times c_{i}\cos(\delta_{B}+\gamma) + s_{i}\sin(\delta_{B}+\gamma)], \\
        N^{\rm exp}_{-i}(B^{+} \to K^{+}D) &= h^{+}_{B}[F_{i}+r^{2}_{B}F_{-i}+2r_{B}\sqrt{F_{i}F_{-i}}\times c_{i}\cos(\delta_{B}+\gamma) - s_{i}\sin(\delta_{B}+\gamma)]. \\
  \end{aligned}
  \end{equation}

To evaluate the impact of the strong-phase parameters ($c_{i}$,
$s_{i}$) on the measurement of $\gamma$, a large simulated dataset of
$B$ decays is generated by setting the input parameters to $\gamma =
68.7^{\circ}$, $r_{B}=0.0904$,
$\delta_{B}=118.3^{\circ}$~\cite{LHCb:2020yot}, and the measured
$c_{i}$, $s_{i}$, $F_{i}$. The normalization factor $h_{B}$ is set to
be a large number to suppress the influence of $B$ decay
statistics. Then, 1000 toy samples of $c_{i}$ and $s_{i}$ parameters are
generated by sampling according to their uncertainties and
correlations. For each toy sample of generated $c_{i}$ and $s_{i}$
parameters, the $\gamma$ values are fitted using the pseudo-data
binned yields of $B$ decays, the measured $F_{i}$ parameters as well
as the toy $c_{i}$ and $s_{i}$ parameters. The standard deviation of
the resulting $\gamma$ value is taken as the uncertainty contribution
of the strong-phase parameters.

\begin{figure}[!tp]
\centering
\parbox{0.45\textwidth}{\centering 
 \includegraphics[width=0.4\textwidth]{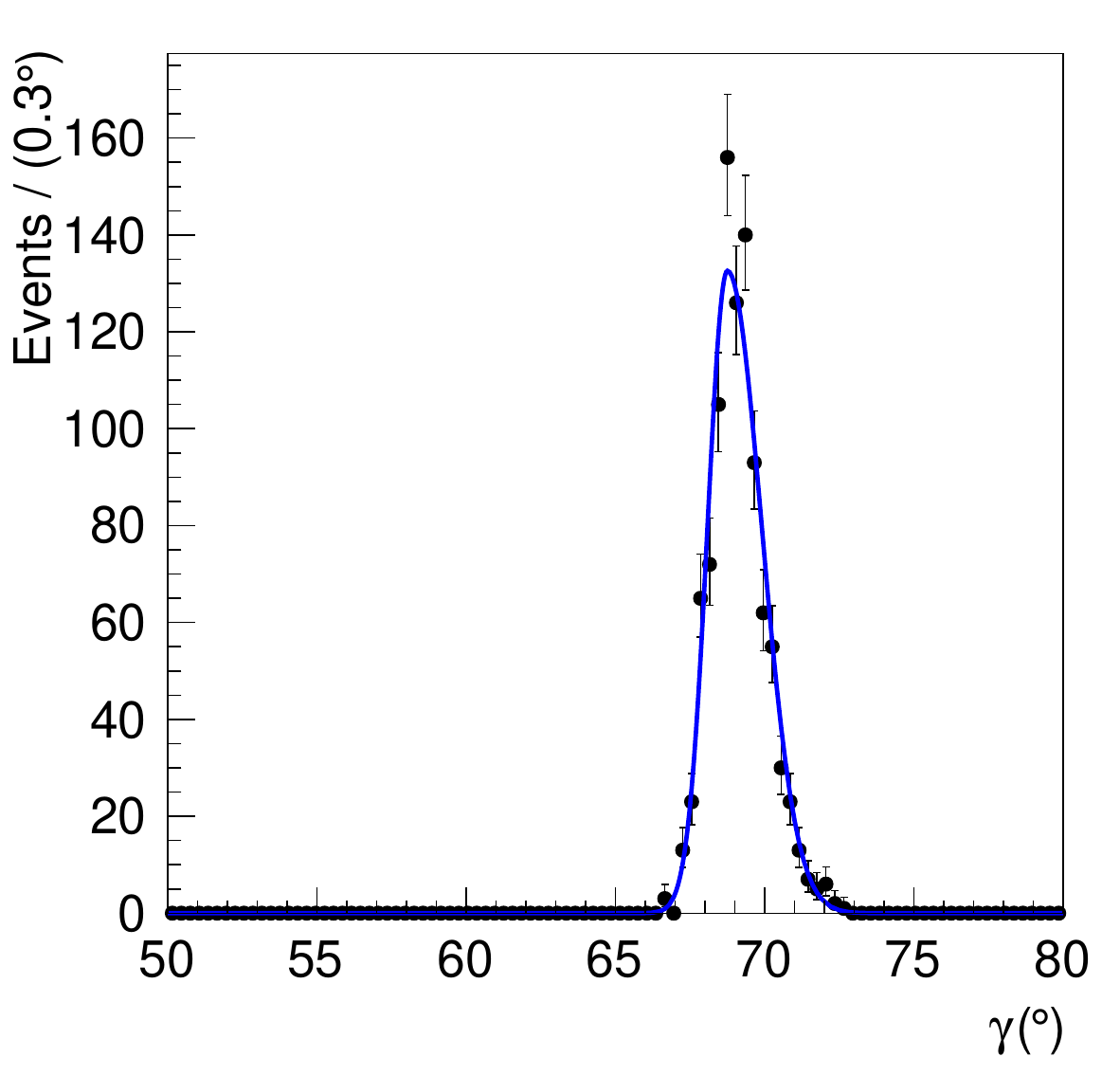}}
\parbox{0.45\textwidth}{\centering 
 \includegraphics[width=0.4\textwidth]{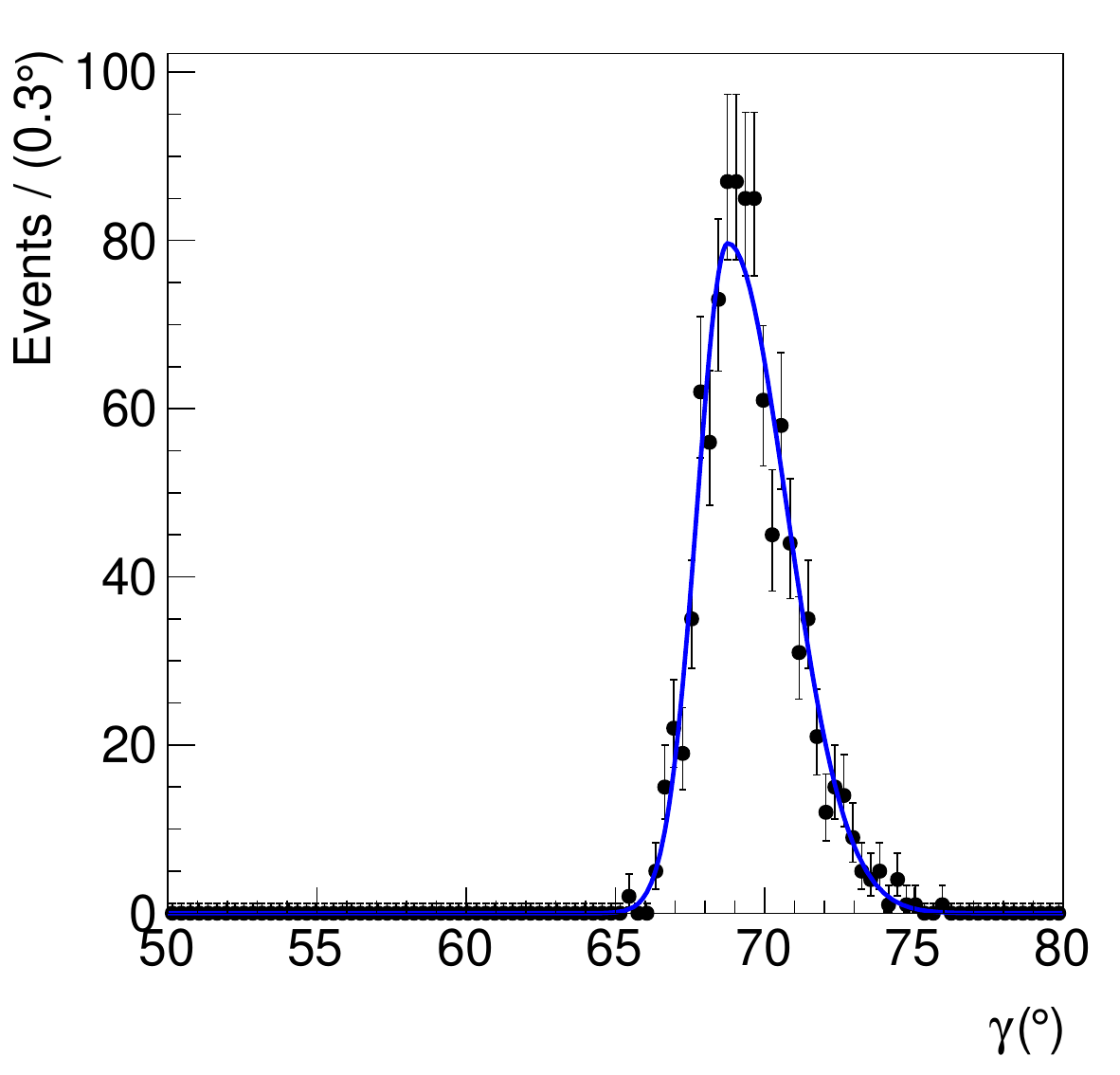}}
 \caption{The impact on the $\gamma$ measurement for the optimal binning scheme with (left) and without (right) constraints.} 
\label{fig:gamma}
\end{figure}

The uncertainties contributed by the constrained and unconstrained
strong-phase parameters to the $\gamma$ measurement in the optimal
binning scheme are found to be 0.9$^{\circ}$ and 1.5$^{\circ}$,
respectively, as shown in Figure~\ref{fig:gamma}. Furthermore, the
difference of the $\gamma$ mean value by taking the constrained and
unconstrained strong-phase parameters as input is found to be around
0.4$^{\circ}$ for the optimal binning scheme. The uncertainty
contributions under the equal binning scheme and modified optimal
binning scheme are also studied, which are
0.7$^{\circ}$(0.9$^{\circ}$) and 0.8$^{\circ}$(1.1$^{\circ}$) by
taking the (un)constrained strong-phase parameters as
inputs. The difference of the $\gamma$ angles by taking the
constrained and unconstrained strong-phase parameters as inputs are
found to be 0.3$^{\circ}$ for the equal binning scheme and
0.4$^{\circ}$ for the modified binning scheme. Furthermore, the
updated result can provide better input for the $\gamma$ measurement
in the upcoming $B$ factory upgrades
~\cite{lhcbcollaboration2019physicscaselhcbupgrade, Kou_2019} and
will not become the dominant source of uncertainty.

\section{Summary}

An improved measurement of the strong-phase parameters between $D^{0}$
and $\bar{D}^{0} \to K^{0}_{S/L}\pi^{+}\pi^{-}$ decays has been
preformed using BESIII $\psi(3770)$ dataset of 7.93
$\rm{fb}^{-1}$. The $c_{i}^{(\prime)}$ and $s_{i}^{(\prime)}$
parameters have been measured with or without the constraints of
model-predicted differences between the strong-phase parameters in the
$\Dz \to \kslpipi$ decays. These results provide essential inputs for
the angle $\gamma$ measurement at LHCb and Belle. The propagated
uncertainty from the constrained strong-phase inputs contributing to
the $\gamma$ measurement is found to be $0.9^{\circ}$ based on optimal
binning scheme, where it is $1.5^{\circ}$ from the unconstrained
result. In this case, constraints in the strong-phase parameters are
found to have a negligible effect on the current $\gamma$ measurement
at LHCb and Belle, which has the best statistical uncertainty of about
$5^{\circ}$~\cite{LHCb:2020yot}.

\section{Acknowledgement}

The BESIII Collaboration thanks the staff of BEPCII (https://cstr.cn/31109.02.BEPC) and the IHEP computing center for their strong support. This work is supported in part by National Key R\&D Program of China under Contracts Nos. 2023YFA1606000, 2023YFA1606704; National Natural Science Foundation of China (NSFC) under Contracts Nos. 11635010, 11935015, 11935016, 11935018, 12025502, 12035009, 12035013, 12061131003, 12192260, 12192261, 12192262, 12192263, 12192264, 12192265, 12221005, 12225509, 12235017, 12361141819, 12405112; the Chinese Academy of Sciences (CAS) Large-Scale Scientific Facility Program; CAS under Contract No. YSBR-101; 100 Talents Program of CAS; The Institute of Nuclear and Particle Physics (INPAC) and Shanghai Key Laboratory for Particle Physics and Cosmology; Agencia Nacional de Investigación y Desarrollo de Chile (ANID), Chile under Contract No. ANID PIA/APOYO AFB230003; German Research Foundation DFG under Contract No. FOR5327; Istituto Nazionale di Fisica Nucleare, Italy; Knut and Alice Wallenberg Foundation under Contracts Nos. 2021.0174, 2021.0299; Ministry of Development of Turkey under Contract No. DPT2006K-120470; National Research Foundation of Korea under Contract No. NRF-2022R1A2C1092335; National Science and Technology fund of Mongolia; National Science Research and Innovation Fund (NSRF) via the Program Management Unit for Human Resources \& Institutional Development, Research and Innovation of Thailand under Contract No. B50G670107; Polish National Science Centre under Contract No. 2024/53/B/ST2/00975; Swedish Research Council under Contract No. 2019.04595; U. S. Department of Energy under Contract No. DE-FG02-05ER41374

\bibliographystyle{JHEP}
\bibliography{body/Ref}

\appendix
\newpage
\section{\boldmath Fitting plots in logarithmic scale}
\label{app:logfit}

 \begin{figure}[!htbp]
\centering
\parbox{1.0\textwidth}{\centering 
 \includegraphics[width=1.0\textwidth]{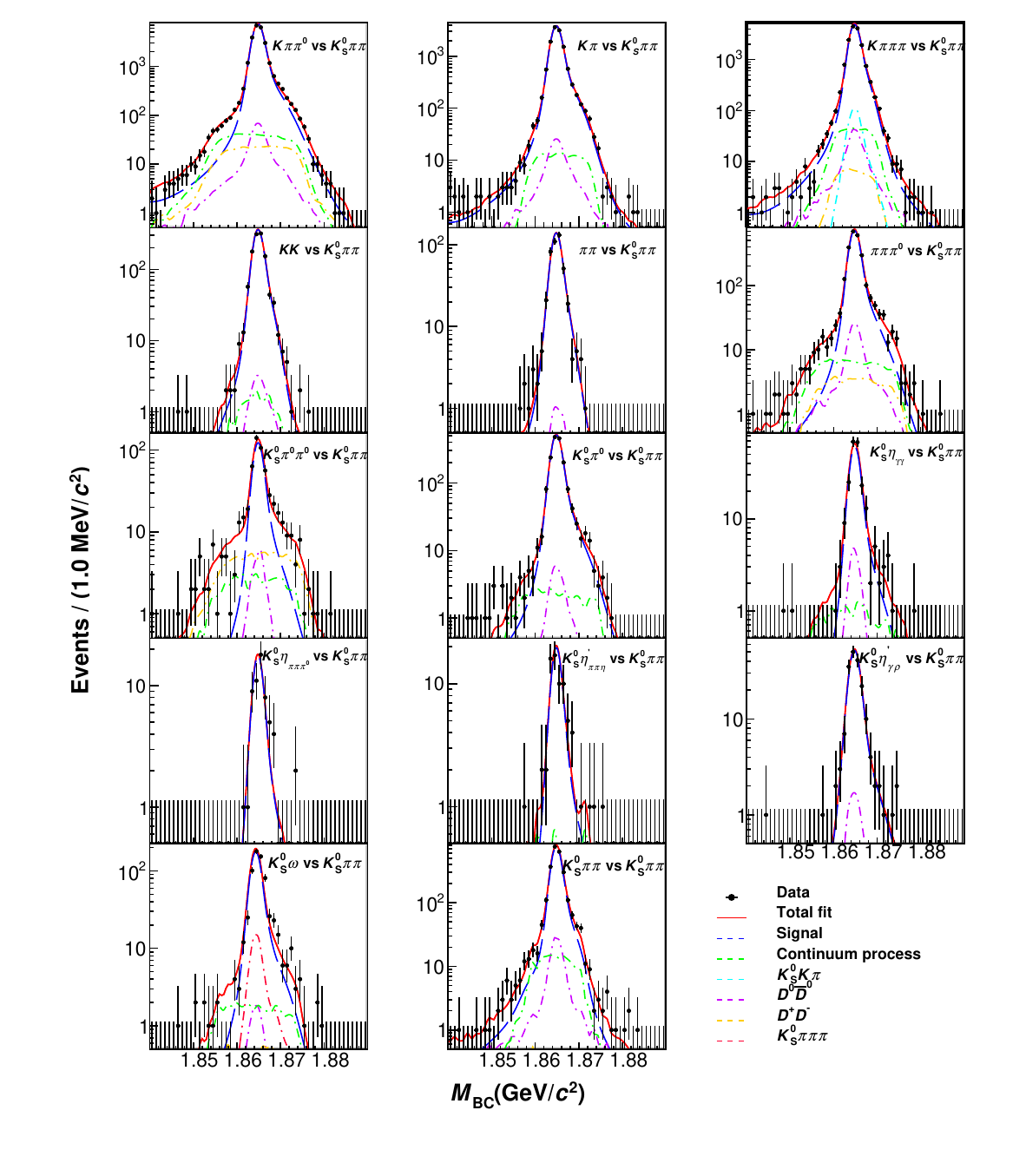}}
\caption{The logarithmic scale 1D unbinned maximum likelihood fits for $\kspipi$ versus full reconstructed modes on $M_{\rm BC}$. The black points with error bars represent data. The dashed lines indicate the signal and background shapes. The red lines represent the sum of signal and background shapes.} 
\label{fig:ksfitted2_log}
\end{figure}

\begin{figure}
\centering
\parbox{1.0\textwidth}{\centering 
 \includegraphics[width=1.0\textwidth]{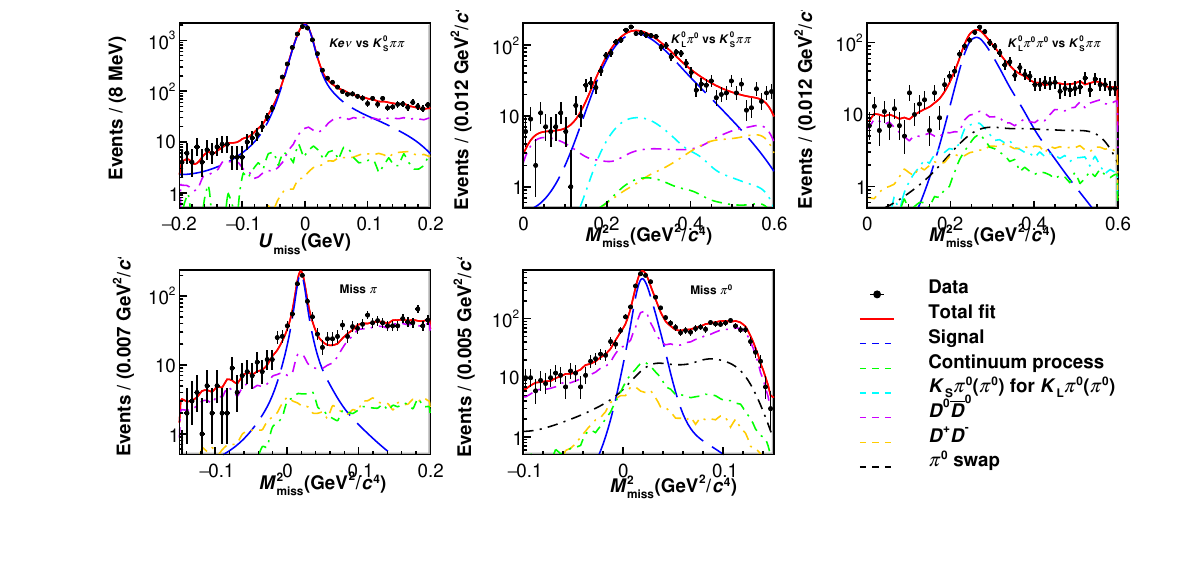}}
\caption{The logarithmic scale 1D unbinned maximum likelihood fits for $\kspipi$ versus partially reconstructed tags on $M^{2}_{\rm miss}/U_{\rm miss}$. The black points with error bars represent data. The dashed lines indicate the signal and background shapes. The red lines represent the sum of signal and background shapes.} 
\label{fig:ksfitted_log}
\end{figure}

\begin{figure}
\centering
\parbox{1.0\textwidth}{\centering 
 \includegraphics[width=1.0\textwidth]{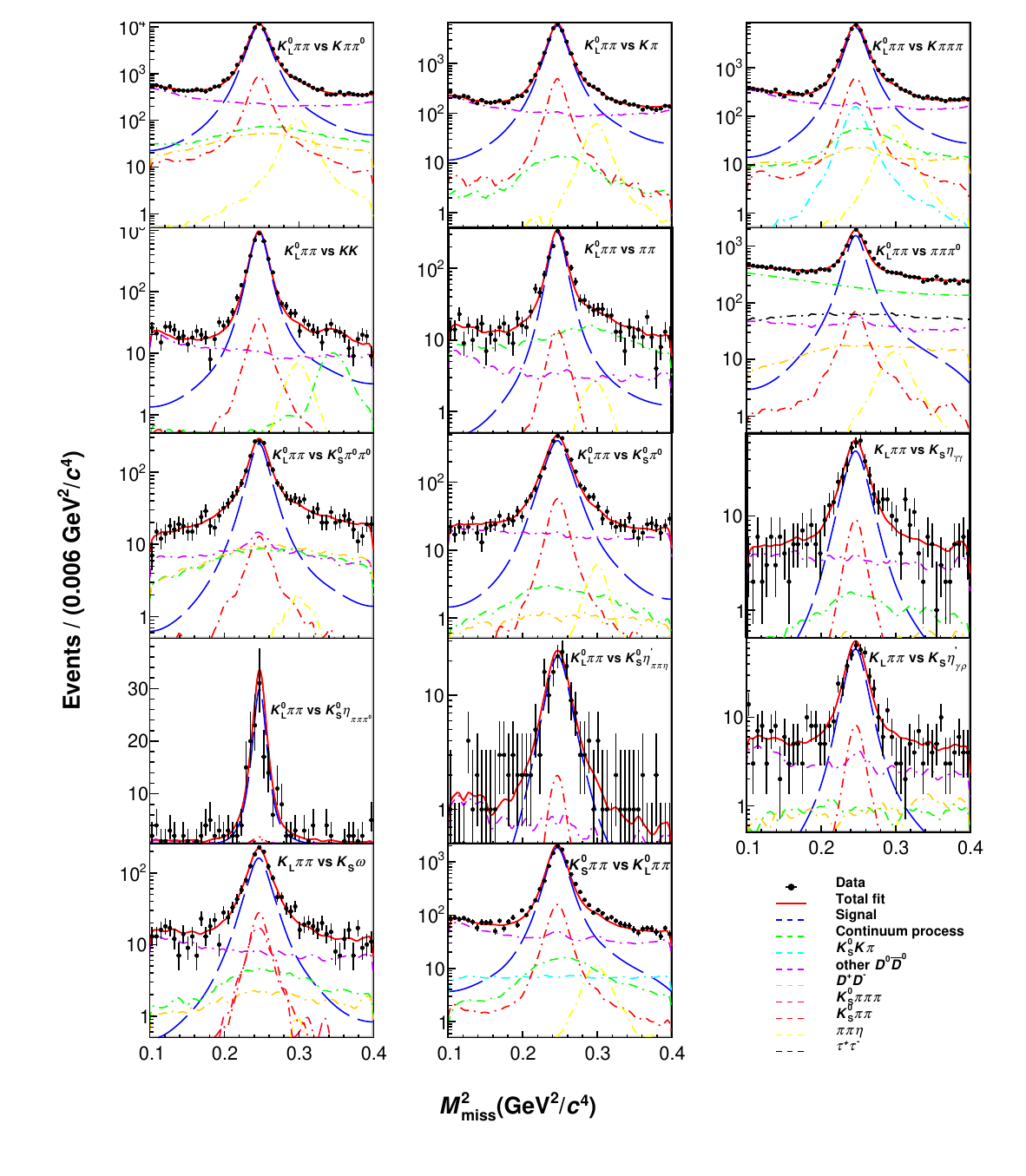}}
\caption{The logarithmic scale 1D unbinned maximum likelihood fits for $\klpipi$ versus various tag modes on $M^{2}_{\rm miss}$. The black points with error bars represent data. The dashed lines indicate the signal and background shapes. The red lines represent the sum of signal and background shapes.} 
\label{fig:klfitted_log}
\end{figure}

\clearpage
\section{\boldmath Values of $\Delta c_{i}$ and $\Delta s_{i}$ constraints}
\label{app:dcs}
\begin{table}[!htbp]
    \centering
    \scriptsize
    \begin{tabular}{cr@{\,$\pm$\,}cr@{\,$\pm$\,}cr@{\,$\pm$\,}cr@{\,$\pm$\,}cr@{\,$\pm$\,}cr@{\,$\pm$\,}c}
    \hline
        Bin & \multicolumn{2}{c}{$\Delta c_{i}$}&\multicolumn{2}{c}{$\Delta s_{i}$} & \multicolumn{2}{c}{$\Delta c_{i}$} & \multicolumn{2}{c}{$\Delta s_{i}$} & \multicolumn{2}{c}{$\Delta c_{i}$} & \multicolumn{2}{c}{$\Delta s_{i}$}\\
        \hline
         & \multicolumn{4}{c}{Equal} & \multicolumn{4}{c}{Optimal} & \multicolumn{4}{c}{Modified optimal} \\
         \hline
         1 & 0.136&0.024 & $-$0.051&0.085 & 0.278&0.025& \;0.083&0.014& 0.160&0.102& \;0.052&0.032 \\
         2 & 0.211&0.031 & $-$0.108&0.043 & 0.115&0.011& \;0.030&0.074& 0.112&0.019& $-$0.061&0.023 \\
         3 & 0.219&0.023 & $-$0.018&0.015 & 0.597&0.162& \;0.358&0.029& 0.449&0.025& \;0.144&0.018 \\
         4 & 0.087&0.041 & \;0.053&0.049 & 0.027&0.007& $-$0.003&0.012& 0.021&0.006& $-$0.012&0.018 \\
         5 & 0.017&0.006 & \;0.001&0.013 & 0.125&0.036& \;0.023&0.029& 0.088&0.041& \;0.038&0.041 \\
         6 & 0.177&0.027 & \;0.002&0.008 & 0.335&0.028& $-$0.211&0.022& 0.292&0.027& $-$0.091&0.016 \\
         7 & 0.431&0.021 & \;0.131&0.016 & 0.056&0.013& $-$0.102&0.055& 0.130&0.046& $-$0.074&0.048 \\
         8 & 0.279&0.022 & \;0.063&0.050 & 0.012&0.132& \;0.003&0.244& 0.039&0.129 & \;0.029&0.243 \\
         \hline
    \end{tabular}
    \caption{The values of $\Delta c_{i}$ and $\Delta s_{i}$ parameters under three binning schemes. The differences between the $\kspipi$ amplitude models measured by the Belle and BESIII experiments and the uncertainties of the U-spin breaking parameters are assigned as the uncertainties.}
    \label{tab:dcs}
\end{table}

\begin{table}[!htbp]
    \centering
    \scriptsize
    \begin{tabular}{cr@{\,$\pm$\,}cr@{\,$\pm$\,}cr@{\,$\pm$\,}cr@{\,$\pm$\,}cr@{\,$\pm$\,}cr@{\,$\pm$\,}c}
    \hline
        Bin & \multicolumn{2}{c}{$\Delta c_{i}$}&\multicolumn{2}{c}{$\Delta s_{i}$} & \multicolumn{2}{c}{$\Delta c_{i}$} & \multicolumn{2}{c}{$\Delta s_{i}$} & \multicolumn{2}{c}{$\Delta c_{i}$} & \multicolumn{2}{c}{$\Delta s_{i}$}\\
        \hline
         & \multicolumn{4}{c}{Equal} & \multicolumn{4}{c}{Optimal} & \multicolumn{4}{c}{Modified optimal} \\
         \hline
         1 & 0.105&0.026 & \;0.028&0.098 & \;0.202&0.053& \;0.057&0.194 & \;0.156&0.067& \;0.031&0.228 \\
         2 & 0.114&0.051 & \;0.270&0.202 & $-$0.043&0.088 & \;0.239&0.234 & \;0.079&0.043& 0.069&0.142 \\
         3 & 0.251&0.049 & \;0.358&0.197 & \;0.474&0.100 & \;0.224&0.280 & \;0.365&0.048& \;0.180&0.155 \\
         4 & 0.006&0.064 & \;0.052&0.256 & \;0.057&0.025 & $-$0.069&0.112& \;0.086&0.027& \;0.054&0.128 \\
         5 & 0.046&0.028 & $-$0.030&0.130 & \;0.109&0.044 & \;0.020&0.187 & $-$0.011&0.049& $-$0.042&0.213 \\
         6 & 0.281&0.066 & $-$0.179&0.244 & \;0.377&0.094 & \;0.328&0.294 & \;0.331&0.060& \;0.191&0.211 \\
         7 & 0.313&0.064 & \;0.085&0.228 & \;0.030 &0.027 & \;0.049&0.114 & \;0.023&0.035& \;0.156&0.131 \\
         8 & 0.136&0.046 & \;0.015&0.162 & $-$0.012&0.041 & \;0.120&0.159 & $-$0.023&0.042 & \;0.060&0.174 \\
         \hline
    \end{tabular}
    \caption{The $\Delta c_{i}$ and $\Delta s_{i}$ parameters calculated with the unconstrained strong-phase parameters as listed in Table~\ref{tab:cifit} under three binning schemes.}
    \label{tab:dcs_data}
\end{table}

\section{Systematic uncertainties}
\label{app:syst}
\begin{table}[!ht]
    \centering
    \tiny
    \begin{tabular}{ccccccccccccccccc}
    \hline
       Uncertainty($10^{-4}$) & $c_{1}$ & $c_{2}$ & $c_{3}$ & $c_{4}$ & $c_{5}$ & $c_{6}$ & $c_{7}$ & $c_{8}$ & $s_{1}$ & $s_{2}$ & $s_{3}$ & $s_{4}$ & $s_{5}$ & $s_{6}$ & $s_{7}$ & $s_{8}$  \\
       \hline
        $K_{i}^{(\prime)}$ inputs &  26 & 79 & 27 & 55 & 37 & 97 & 45 & 56 & 37 & 134 & 88 & 98 & 27 & 32 & 102 & 49 \\
       Input parameters & 5 & 18 & 17 & 13 & 3 & 25 & 27 & 19 & 37 & 122 & 18 & 12 & 22 & 32 & 41 & 49  \\
       ST yields & 60 & 79 & 75 & 55 & 29 & 63 & 104 & 78 & 16 & 73 & 53 & 73 & 7 & 64 & 61 & 41\\
       MC statistics & 37 & 43 & 14 & 63 & 34 & 46 & 18 & 37  & 11 & 73 & 97 & 172 & 11 & 107 & 112 & 41 \\
       Background  & 17 & 14 & 20 & 50 & 45 & 50 & 27 & 22 & 21 & 61 & 44 & 86 & 27 & 12 & 51 & 32 \\
       Reconstruction & 21 & 31 & 38 & 27 & 9 & 45 & 66 & 39 & 0 & 10 & 6 & 4 & 6 & 2 & 6 & 19 \\
       \hline
       Total  & 80 & 127 & 93 & 116 & 73 & 144 & 138 & 114 & 60 & 218 & 149 & 229 & 46 & 177 & 176 & 97 \\
       \hline
        Uncertainty($10^{-4}$)   & $c^{\prime}_{1}$ & $c^{\prime}_{2}$ & $c^{\prime}_{3}$ & $c^{\prime}_{4}$ & $c^{\prime}_{5}$ & $c^{\prime}_{6}$ & $c^{\prime}_{7}$ & $c^{\prime}_{8}$  & $s^{\prime}_{1}$ & $s^{\prime}_{2}$ & $s^{\prime}_{3}$ & $s^{\prime}_{4}$ & $s^{\prime}_{5}$ & $s^{\prime}_{6}$ & $s^{\prime}_{7}$ & $s^{\prime}_{8}$ \\
       \hline
       $K_{i}^{(\prime)}$ inputs & 31 & 76 & 40 & 32 & 33 & 38 & 71 & 62 &  52 & 72 & 61 & 60 & 85 & 195 & 256 & 146 \\
       Input parameters & 9 & 13 & 10 & 14 & 9 & 14 & 34 & 25 & 9 & 43 & 30 & 0 & 12 & 22 & 73 & 67\\
        ST yields & 36 & 56 & 37 & 28 & 21 & 38 & 59 & 53 & 17 & 29 & 15 & 40 & 12 & 7 & 74 & 32 \\
       MC statistics  & 22 & 30 & 13 & 55 & 46 & 72 & 17 & 22 & 17 & 57 & 91 & 121 & 36 & 109 & 73 & 40  \\
       Background & 70 & 59 & 40 & 37 & 79 & 34 & 38 & 62 & 26 & 143 & 137 & 261 & 36 & 239 & 220 & 106  \\
       Reconstruction & 3 & 8 & 9 & 4 & 0 & 7 & 10 & 7 & 13 & 43 & 25 & 2 & 0 & 5 & 0 & 3\\
       \hline
       Total & 88 & 116 & 70 & 81 & 100 & 97 & 107 & 108  & 64 & 183 & 180 & 297 & 101 & 327 & 361 & 200 \\
       \hline
    \end{tabular}
    \caption{Systematic uncertainties for the equal binning scheme without $\Delta c_{i}$ and $\Delta s_{i}$ constraints.}
    \label{tab:syst_eq}
\end{table}

\begin{table}[!ht]
    \centering
    \tiny
    \begin{tabular}{ccccccccccccccccc}
    \hline
       Uncertainty($10^{-4}$) & $c_{1}$ & $c_{2}$ & $c_{3}$ & $c_{4}$ & $c_{5}$ & $c_{6}$ & $c_{7}$ & $c_{8}$ & $s_{1}$ & $s_{2}$ & $s_{3}$ & $s_{4}$ & $s_{5}$ & $s_{6}$ & $s_{7}$ & $s_{8}$  \\
       \hline
        $K_{i}^{(\prime)}$ inputs &  32 & 284 & 109 & 33 & 33 & 128 & 43 & 78 & 71 & 201 & 252 & 26 & 61 & 299 & 36 & 49 \\
       Input parameters & 22 & 33 & 55 & 3 & 15 & 64 & 5 & 10 & 36 & 238 & 118 & 21 & 9 & 94 & 36 & 49  \\
       ST yields & 86 & 152 & 172 & 29 & 66 & 121 & 54 & 62 & 90 & 37 & 34 & 27 & 83 & 56 & 65 & 39\\
       MC statistics & 18 & 40 & 31 & 45 & 24 & 21 & 37 & 34  & 80 & 73 & 67 & 16 & 104 & 187 & 14 & 29 \\
       Background  & 18 & 26 & 47 & 39 & 42 & 50 & 24 & 29 & 53 & 73 & 84 & 31 & 52 & 131 & 71 & 87 \\
       Reconstruction & 49 & 91 & 121 & 13 & 35 & 76 & 16 & 19 & 10 & 17 & 8 & 11 & 6 & 9 & 11 & 7 \\
       \hline
       Total  & 109 & 340 & 250 & 75 & 96 & 209 & 83 & 111 & 154 & 331 & 300 & 57 & 156 & 392 & 110 & 122 \\
       \hline
        Uncertainty($10^{-4}$)   & $c^{\prime}_{1}$ & $c^{\prime}_{2}$ & $c^{\prime}_{3}$ & $c^{\prime}_{4}$ & $c^{\prime}_{5}$ & $c^{\prime}_{6}$ & $c^{\prime}_{7}$ & $c^{\prime}_{8}$  & $s^{\prime}_{1}$ & $s^{\prime}_{2}$ & $s^{\prime}_{3}$ & $s^{\prime}_{4}$ & $s^{\prime}_{5}$ & $s^{\prime}_{6}$ & $s^{\prime}_{7}$ & $s^{\prime}_{8}$ \\
       \hline
       $K_{i}^{(\prime)}$ inputs & 43 & 172 & 198 & 29 & 19 & 189 & 37 & 73 &  140 & 184 & 260 & 50 & 41 & 145 & 52 & 70 \\
       Input parameters & 18 & 59 & 104 & 6 & 9 & 41 & 9 & 14 & 31 & 117 & 12 & 40 & 14 & 83 & 26 & 56 \\
        ST yields & 43 & 73 & 83 & 15 & 31 & 68 & 34 & 31 & 15 & 66 & 79 & 98 & 51 & 40 & 17 & 55 \\
       MC statistics  & 18 & 25 & 10 & 57 & 25 & 20 & 36 & 30 & 140 & 67 & 80 & 30 & 149 & 207 & 26 & 42  \\
       Background & 25 & 76 & 47 & 68 & 16 & 77 & 80 & 103 & 203 & 50 & 160 & 80 & 122 & 186 & 26 & 181  \\
       Reconstruction   & 8 & 12 & 16 & 2 & 6 & 14 & 0 & 2 & 9 & 28 & 14 & 9 & 4 & 48 & 19 & 1\\
       \hline
       Total & 71 & 212 & 244 & 95 & 48 & 220 & 101 & 133  & 286 & 244 & 347 & 145 & 203 & 331 & 74 & 212 \\
       \hline
    \end{tabular}
    \caption{Systematic uncertainties for the optimal binning scheme without $\Delta c_{i}$ and $\Delta s_{i}$ constraints.}
    \label{tab:syst_op}
\end{table}

\begin{table}[]
    \centering
    \tiny
    \begin{tabular}{ccccccccccccccccc}
    \hline
       Uncertainty($10^{-4}$) & $c_{1}$ & $c_{2}$ & $c_{3}$ & $c_{4}$ & $c_{5}$ & $c_{6}$ & $c_{7}$ & $c_{8}$ & $s_{1}$ & $s_{2}$ & $s_{3}$ & $s_{4}$ & $s_{5}$ & $s_{6}$ & $s_{7}$ & $s_{8}$  \\
       \hline
        $K_{i}^{(\prime)}$ inputs &  60 & 111 & 41 & 33 & 36 & 53 & 35 & 56 & 51 & 92 & 97 & 23 & 62 & 116 & 36 & 61 \\
       Input parameters & 17 & 9 & 31 & 3 & 13 & 31 & 5 & 8 & 51 & 110 & 81 & 17 & 10 & 46 & 36 & 41  \\
       ST yields & 77 & 90 & 112 & 22 & 59 & 92 & 69 & 65 & 51 & 47 & 57 & 30 & 91 & 80 & 36 & 42 \\
       MC statistics & 26 & 39 & 10 & 32 & 26 & 22 & 39 & 34  & 165 & 37 & 41 & 23 & 124 & 70 & 27 & 31 \\
       Background  & 56 & 18 & 31 & 35 & 50 & 31 & 32 & 31 & 76 & 74 & 49 & 34 & 93 & 93 & 82 & 71 \\
       Reconstruction & 42 & 68 & 101 & 2 & 36 & 51 & 26 & 28 & 0 & 7 & 9 & 6 & 0 & 13 & 8 & 13 \\
       \hline
       Total  & 124 & 164 & 162 & 61 & 97 & 127 & 96 & 102 & 202 & 172 & 152 & 59 & 190 & 189 & 107 & 117 \\
       \hline
        Uncertainty($10^{-4}$)   & $c^{\prime}_{1}$ & $c^{\prime}_{2}$ & $c^{\prime}_{3}$ & $c^{\prime}_{4}$ & $c^{\prime}_{5}$ & $c^{\prime}_{6}$ & $c^{\prime}_{7}$ & $c^{\prime}_{8}$  & $s^{\prime}_{1}$ & $s^{\prime}_{2}$ & $s^{\prime}_{3}$ & $s^{\prime}_{4}$ & $s^{\prime}_{5}$ & $s^{\prime}_{6}$ & $s^{\prime}_{7}$ & $s^{\prime}_{8}$ \\
       \hline
       $K_{i}^{(\prime)}$ inputs & 20 & 76 & 64 & 32 & 18 & 119 & 43 & 64 &  97 & 109 & 167 & 48 & 31 & 79 & 46 & 58 \\
       Input parameters & 10 & 28 & 42 & 8 & 11 & 25 & 7 & 12 & 77 & 69 & 71 & 36 & 16 & 47 & 35 & 44 \\
        ST yields & 39 & 48 & 56 & 19 & 24 & 47 & 41 & 32 & 56 & 20 & 46 & 11 & 60 & 46 & 11 & 72 \\
       MC statistics  & 20 & 16 & 8 & 40 & 32 & 54 & 24 & 26 & 193 & 40 & 24 & 48 & 62 & 79 & 23 & 44  \\
       Background & 29 & 76 & 31 & 80 & 14 & 58 & 77 & 99 & 251 & 50 & 107 & 95 & 172 & 221 & 93 & 175  \\
       Reconstruction   & 7 & 6 & 11 & 0 & 4 & 11 & 4 & 4 & 4 & 24 & 5 & 7 & 9 & 44 & 21 & 6\\
       \hline
       Total & 58 & 122 & 101 & 97 & 47 & 153 & 101 & 125  & 345 & 147 & 217 & 122 & 195 & 260 & 114 & 208 \\
       \hline
    \end{tabular}
    \caption{Systematic uncertainties for the modified optimal binning scheme without $\Delta c_{i}$ and $\Delta s_{i}$ constraints.}
    \label{tab:syst_modi}
\end{table}

\begin{table}[]
    \centering
    \tiny
    \begin{tabular}{ccccccccccccccccc}
    \hline
       Uncertainty($10^{-4}$) & $c_{1}$ & $c_{2}$ & $c_{3}$ & $c_{4}$ & $c_{5}$ & $c_{6}$ & $c_{7}$ & $c_{8}$ & $s_{1}$ & $s_{2}$ & $s_{3}$ & $s_{4}$ & $s_{5}$ & $s_{6}$ & $s_{7}$ & $s_{8}$  \\
       \hline
        $K_{i}^{(\prime)}$ inputs &  21 & 50 & 23 & 31 & 21 & 54 & 48 & 43 & 25 & 49 & 25 & 38 & 15 & 31 & 71 & 27 \\
       Input parameters & 5 & 11 & 8 & 7 & 5 & 7 & 29 & 19 & 30 & 41 & 0 & 10 & 15 & 20 & 27 & 21  \\
       ST yields &  56 & 67 & 55 & 44 & 17 & 44 & 70 & 60 & 4 & 17 & 39 & 39 & 10 & 54 & 61 & 42 \\
       MC statistics & 32 & 34 & 5 & 51 & 39 & 44 & 10 & 26 & 10 & 42 & 62 & 94 & 25 & 107 & 41 & 18 \\
       Background  & 27 & 25 & 18 & 44 & 52 & 40 & 16 & 24 & 15 & 49 & 64 & 97 & 15 & 104 & 72 & 35 \\
       Reconstruction & 22 & 30 & 37 & 25 & 10 & 44 & 62 & 38 & 0 & 13 & 5 & 0 & 6 & 4 & 6 & 21 \\
       \hline
       Total  & 76 & 99 & 70 & 89 & 71 & 102 & 111 & 92 & 44 & 94 & 101 & 108 & 37 & 163 & 128 & 68 \\
       \hline
        Uncertainty($10^{-4}$)   & $c^{\prime}_{1}$ & $c^{\prime}_{2}$ & $c^{\prime}_{3}$ & $c^{\prime}_{4}$ & $c^{\prime}_{5}$ & $c^{\prime}_{6}$ & $c^{\prime}_{7}$ & $c^{\prime}_{8}$  & $s^{\prime}_{1}$ & $s^{\prime}_{2}$ & $s^{\prime}_{3}$ & $s^{\prime}_{4}$ & $s^{\prime}_{5}$ & $s^{\prime}_{6}$ & $s^{\prime}_{7}$ & $s^{\prime}_{8}$ \\
       \hline
       $K_{i}^{(\prime)}$ inputs & 26 & 57 & 29 & 25 & 21 & 49 & 54 & 48 & 25 & 42 & 25 & 30 & 15 & 31 & 71 & 37 \\
       input parameters & 8 & 10 & 5 & 7 & 6 & 3 & 31 & 18 & 6 & 34 & 0 & 10 & 15 & 21 & 27 & 14\\
        ST yields & 36 & 57 & 47 & 37 & 15 & 39 & 64 & 50 & 18 & 17 & 39 & 30 & 11 & 54 & 52 & 38 \\
       MC statistics  & 21 & 30 & 5 & 47 & 38 & 49 & 10 & 22 & 16 & 44 & 64 & 94 & 24 & 102 & 40 & 12  \\
       Background & 55 & 42 & 18 & 35 & 56 & 37 & 22 & 41 & 12 & 59 & 64 & 101 & 15 & 104 & 72 & 35  \\
       Reconstruction   & 3 & 7 & 8 & 3 & 0 & 6 & 11 & 6 & 13 & 32 & 19 & 0 & 3 & 2 & 3 & 3\\
       \hline
       Total & 74 & 97 & 60 & 74 & 73 & 88 & 93 & 85  & 39 & 98 & 104 & 145 & 38 & 160 & 123 & 66\\
       \hline
    \end{tabular}
    \caption{Systematic uncertainties for the equal binning scheme with $\Delta c_{i}$ and $\Delta s_{i}$ constraints.}
    \label{tab:syst_eq_dcs}
\end{table}

\begin{table}[]
    \centering
    \tiny
    \begin{tabular}{ccccccccccccccccc}
    \hline
       Uncertainty($10^{-4}$) & $c_{1}$ & $c_{2}$ & $c_{3}$ & $c_{4}$ & $c_{5}$ & $c_{6}$ & $c_{7}$ & $c_{8}$ & $s_{1}$ & $s_{2}$ & $s_{3}$ & $s_{4}$ & $s_{5}$ & $s_{6}$ & $s_{7}$ & $s_{8}$  \\
       \hline
        $K_{i}^{(\prime)}$ inputs &  25 & 148 & 94 & 21 & 21 & 121 & 30 & 72 & 65 & 106 & 132 & 17 & 43 & 274 & 24 & 31 \\
       Input parameters & 14 & 39 & 49 & 5 & 8 & 13 & 4 & 8 & 22 & 54 & 12 & 19 & 5 & 36 & 23 & 19  \\
       ST yields & 64 & 83 & 170 & 17 & 55 & 81 & 48 & 65 & 56 & 14 & 48 & 19 & 42 & 50 & 23 & 28\\
       MC statistics & 14 & 18 & 21 & 35 & 21 & 14 & 26 & 33 & 67 & 58 & 77 & 20 & 55 & 108 & 15 & 30 \\
       Background  & 11 & 60 & 42 & 50 & 31 & 35 & 45 & 31 & 86 & 54 & 94 & 33 & 45 & 83 & 41 & 103 \\
       Reconstruction & 48 & 93 & 118 & 15 & 34 & 73 & 15 & 17 & 3 & 2 & 14 & 9 & 3 & 19 & 15 & 12 \\
       \hline
       Total  & 87 & 207 & 236 & 69 & 72 & 167 & 78 & 109 & 140 & 144 & 186 & 51 & 92 & 313 & 59 & 116 \\
       \hline
        Uncertainty($10^{-4}$)   & $c^{\prime}_{1}$ & $c^{\prime}_{2}$ & $c^{\prime}_{3}$ & $c^{\prime}_{4}$ & $c^{\prime}_{5}$ & $c^{\prime}_{6}$ & $c^{\prime}_{7}$ & $c^{\prime}_{8}$  & $s^{\prime}_{1}$ & $s^{\prime}_{2}$ & $s^{\prime}_{3}$ & $s^{\prime}_{4}$ & $s^{\prime}_{5}$ & $s^{\prime}_{6}$ & $s^{\prime}_{7}$ & $s^{\prime}_{8}$ \\
       \hline
       $K_{i}^{(\prime)}$ inputs & 30 & 148 & 158 & 22 & 16 & 127 & 31 & 72 & 65 & 107 & 133 & 18 & 43 & 278 & 24 & 61 \\
       input parameters & 14 & 41 & 96 & 5 & 3 & 16 & 7 & 13 & 22 & 26 & 12 & 15 & 5 & 36 & 12 & 36\\
        ST yields & 53 & 83 & 75 & 16 & 34 & 77 & 38 & 30 & 57 & 26 & 60 & 19 & 39 & 50 & 12 & 60 \\
       MC statistics  & 14 & 17 & 14 & 37 & 21 & 15 & 21 & 20 & 66 & 62 & 77 & 20 & 54 & 107 & 18 & 36  \\
       Background & 14 & 61 & 43 & 55 & 18 & 45 & 56 & 96 & 86 & 52 & 95 & 34 & 47 & 83 & 30 & 155  \\
        Reconstruction  & 9 & 12 & 16 & 0 & 6 & 14 & 2 & 3 & 9 & 26 & 17 & 7 & 5 & 29 & 18 & 0\\
       \hline
       Total & 66 & 186 & 205 & 72 & 47 & 158 & 78 & 126  & 139 & 142 & 192 & 50 & 92 & 316 & 49 & 184 \\
       \hline
    \end{tabular}
    \caption{Systematic uncertainties for the optimal binning scheme with $\Delta c_{i}$ and $\Delta s_{i}$ constraints.}
    \label{tab:syst_op_dcs}
\end{table}

\begin{table}[]
    \centering
    \tiny
    \begin{tabular}{ccccccccccccccccc}
    \hline
       Uncertainty($10^{-4}$) & $c_{1}$ & $c_{2}$ & $c_{3}$ & $c_{4}$ & $c_{5}$ & $c_{6}$ & $c_{7}$ & $c_{8}$ & $s_{1}$ & $s_{2}$ & $s_{3}$ & $s_{4}$ & $s_{5}$ & $s_{6}$ & $s_{7}$ & $s_{8}$  \\
       \hline
        $K_{i}^{(\prime)}$ inputs &  57 & 67 & 34 & 23 & 27 & 60 & 29 & 53 & 30 & 51 & 77 & 31 & 41 & 91 & 60 & 36 \\
       Input parameters & 12 & 10 & 25 & 5 & 8 & 9 & 4 & 7 & 20 & 20 & 32 & 29 & 13 & 8 & 70 & 19 \\
       ST yields & 78 & 63 & 80 & 14 & 57 & 81 & 46 & 65 & 33 & 13 & 50 & 27 & 40 & 54 & 15 & 47 \\
       MC statistics & 33 & 24 & 5 & 33 & 30 & 12 & 35 & 30  & 84 & 27 & 43 & 41 & 62 & 71 & 20 & 25 \\
       Background  & 53 & 42 & 13 & 52 & 39 & 19 & 34 & 29 & 90 & 27 & 52 & 46 & 54 & 108 & 70 & 106 \\
       Reconstruction & 42 & 67 & 103 & 5 & 36 & 51 & 26 & 27 & 0 & 2 & 4 & 11 & 3 & 13 & 6 & 8 \\
       \hline
       Total  & 123 & 124 & 138 & 68 & 88 & 116 & 78 & 98 & 132 & 68 & 118 & 80 & 101 & 168 & 119 & 126  \\
       \hline
        Uncertainty($10^{-4}$)   & $c^{\prime}_{1}$ & $c^{\prime}_{2}$ & $c^{\prime}_{3}$ & $c^{\prime}_{4}$ & $c^{\prime}_{5}$ & $c^{\prime}_{6}$ & $c^{\prime}_{7}$ & $c^{\prime}_{8}$  & $s^{\prime}_{1}$ & $s^{\prime}_{2}$ & $s^{\prime}_{3}$ & $s^{\prime}_{4}$ & $s^{\prime}_{5}$ & $s^{\prime}_{6}$ & $s^{\prime}_{7}$ & $s^{\prime}_{8}$ \\
       \hline
       $K_{i}^{(\prime)}$ inputs & 19 & 63 & 46 & 23 & 17 & 75 & 37 & 63 & 31 & 52 & 78 & 31 & 40 & 90 & 60 & 53 \\
       Input parameters & 4 & 18 & 32 & 5 & 6 & 12 & 8 & 11 & 20 & 13 & 33 & 30 & 7 & 8 & 45 & 38 \\
        ST yields & 41 & 53 & 62 & 17 & 26 & 57 & 44 & 31 & 22 & 13 & 57 & 22 & 35 & 78 & 14 & 74 \\
       MC statistics & 21 & 20 & 4 & 33 & 24 & 13 & 22 & 23 & 86 & 28 & 42 & 47 & 60 & 70 & 21 & 29  \\
       Background & 22 & 55 & 16 & 53 & 20 & 32 & 67 & 89 & 103 & 27 & 53 & 42 & 56 & 108 & 45 & 152  \\
       Reconstruction   & 7 & 5 & 11 & 0 & 4 & 10 & 4 & 3 & 4 & 20 & 4 & 11 & 8 & 26 & 19 & 1 \\
       \hline
       Total &  55 & 103 & 86 & 69 & 45 & 102 & 91 & 116  & 141 & 70 & 123 & 80 & 98 & 177 & 93 & 184 \\
       \hline
    \end{tabular}
    \caption{Systematic uncertainties for the modified optimal binning scheme with $\Delta c_{i}$ and $\Delta s_{i}$ constraints. }
    \label{tab:syst_modi_dcs}
\end{table}
\section{Correlation matrices of statistical and systematic uncertainties}
\label{app:corr}
\begin{table}[!htbp]
\begin{sideways}
    \scriptsize
    \setlength\tabcolsep{2.3pt}
    \centering
    \begin{tabular}{cccccccccccccccccccccccccccccccc}
         \hline       
         &$c_{2}$&$c_{3}$&$c_{4}$&$c_{5}$&$c_{6}$&$c_{7}$&$c_{8}$&$s_{1}$&$s_{2}$&$s_{3}$&$s_{4}$&$s_{5}$&$s_{6}$&$s_{7}$&$s_{8}$&$c'_{1}$& $c'_{2}$&$c'_{3}$&$c'_{4}$&$c'_{5}$&$c'_{6}$&$c'_{7}$&$c'_{8}$&$s'_{1}$&$s'_{2}$&$s'_{3}$&$s'_{4}$&$s'_{5}$&$s'_{6}$&$s'_{7}$&$s'_{8}$\\
         \hline
        $c_{1}$&$-$11&$-$1&0&0&$-$1&0&$-$15&0&0&0&0&0&0&0&0&$-$1&0&0&3&3&0&0&0&0&0&0&0&0&0&0&0\\
$c_{2}$& &$-$7&0&0&0&$-$2&$-$1&0&$-$1&0&0&0&1&0&$-$1&1&0&0&0&2&2&$-$1&$-$1&$-$1&0&0&0&3&1&0&0\\
$c_{3}$& & &$-$6&$-$1&$-$1&$-$1&$-$2&0&0&2&1&1&0&0&$-$2&0&$-$1&0&1&0&0&$-$1&0&$-$1&$-$1&0&$-$1&3&2&$-$1&$-$2\\
$c_{4}$& & & &$-$6&$-$2&$-$1&1&$-$1&0&$-$1&2&0&1&1&0&1&2&1&0&$-$1&0&0&2&$-$1&1&1&$-$1&0&0&$-$1&$-$3\\
$c_{5}$& & & & &$-$10&0&0&0&0&0&0&$-$5&1&0&0&5&3&0&0&$-$1&0&0&3&0&0&0&0&0&0&0&0\\
$c_{6}$& & & & & &$-$9&0&0&1&0&0&$-$1&$-$5&1&1&$-$1&2&1&0&$-$1&0&0&1&2&2&0&0&0&$-$1&0&1\\
$c_{7}$& & & & & & &$-$13&2&0&0&0&0&0&$-$1&4&0&0&1&0&0&$-$1&$-$1&0&4&6&3&$-$3&$-$2&$-$1&1&2\\
$c_{8}$& & & & & & & &1&0&0&0&$-$1&0&1&1&0&0&0&0&1&0&0&$-$1&1&1&1&$-$1&$-$3&$-$1&1&1\\
$s_{1}$& & & & & & & & &8&11&$-$6&0&0&$-$4&$-$10&0&$-$1&$-$3&$-$2&1&2&3&0&8&$-$6&$-$1&$-$2&21&6&12&$-$5\\
$s_{2}$& & & & & & & & & &2&3&6&$-$11&$-$1&0&0&0&$-$1&$-$2&1&4&1&0&20&$-$12&$-$18&0&30&36&$-$3&$-$9\\
$s_{3}$& & & & & & & & & & &$-$2&$-$1&$-$3&$-$8&$-$16&0&0&0&0&0&0&1&0&1&$-$9&$-$25&$-$16&7&16&20&10\\
$s_{4}$& & & & & & & & & & & &$-$3&$-$7&$-$8&$-$8&0&0&$-$1&$-$1&0&$-$1&$-$1&$-$1&$-$2&$-$18&$-$23&$-$16&$-$1&9&17&25\\
$s_{5}$& & & & & & & & & & & & &$-$13&5&$-$3&0&2&2&$-$1&0&$-$1&0&$-$1&8&7&$-$1&$-$6&6&4&$-$8&$-$5\\
$s_{6}$& & & & & & & & & & & & & &$-$4&1&0&1&4&1&0&$-$1&$-$1&0&1&9&25&5&$-$3&$-$1&$-$8&$-$4\\
$s_{7}$& & & & & & & & & & & & & & &$-$10&0&$-$1&0&0&0&0&0&0&7&24&28&22&$-$4&$-$7&$-$16&$-$6\\
$s_{8}$& & & & & & & & & & & & & & & &0&$-$1&$-$4&$-$1&0&1&1&0&2&5&24&15&$-$4&$-$8&$-$8&$-$6\\
$c'_{2}$& & & & & & & & & & & & & & & & &$-$14&1&$-$1&$-$1&$-$1&3&$-$25&$-$1&0&0&0&0&0&0&0\\
$c'_{3}$& & & & & & & & & & & & & & & & & &$-$8&0&$-$1&$-$1&0&2&0&$-$2&0&0&0&0&0&0\\
$c'_{4}$& & & & & & & & & & & & & & & & & & &$-$7&0&$-$1&0&0&0&1&0&$-$1&$-$1&0&$-$1&0\\
$c'_{5}$& & & & & & & & & & & & & & & & & & & &$-$5&0&0&0&$-$1&0&0&2&$-$1&$-$1&0&0\\
$c'_{6}$& & & & & & & & & & & & & & & & & & & & &$-$10&0&$-$1&0&0&0&0&0&1&0&0\\
$c'_{7}$& & & & & & & & & & & & & & & & & & & & & &$-$16&2&1&0&0&0&2&$-$1&0&$-$1\\
$c'_{8}$& & & & & & & & & & & & & & & & & & & & & & &$-$24&0&0&0&0&1&1&2&$-$1\\
$c'_{9}$& & & & & & & & & & & & & & & & & & & & & & & &0&0&0&0&0&0&0&1\\
$s'_{1}$& & & & & & & & & & & & & & & & & & & & & & & & &$-$10&0&1&7&7&$-$1&$-$19\\
$s'_{2}$& & & & & & & & & & & & & & & & & & & & & & & & & &11&9&$-$5&$-$8&$-$10&$-$5\\
$s'_{3}$& & & & & & & & & & & & & & & & & & & & & & & & & & &12&$-$8&$-$14&$-$14&$-$9\\
$s'_{4}$& & & & & & & & & & & & & & & & & & & & & & & & & & & &$-$5&$-$6&$-$9&$-$7\\
$s'_{5}$& & & & & & & & & & & & & & & & & & & & & & & & & & & & &6&3&$-$3\\
$s'_{6}$& & & & & & & & & & & & & & & & & & & & & & & & & & & & & &$-$7&1\\
$s'_{7}$& & & & & & & & & & & & & & & & & & & & & & & & & & & & & & &$-$10\\
 \hline
    \end{tabular}
\end{sideways}
    \caption{Correlation matrix ($\%$) of statistical uncertainties for the equal binning scheme without $\Delta c_{i}$ and $\Delta s_{i}$ constraints.}
    \label{tab:eqstatnocs}
\end{table}

\begin{table}[!htbp]
\begin{sideways}
    \scriptsize

    \setlength\tabcolsep{3pt}
    \centering
    \begin{tabular}{cccccccccccccccccccccccccccccccc}
     \hline        
         &$c_{2}$&$c_{3}$&$c_{4}$&$c_{5}$&$c_{6}$&$c_{7}$&$c_{8}$&$s_{1}$&$s_{2}$&$s_{3}$&$s_{4}$&$s_{5}$&$s_{6}$&$s_{7}$&$s_{8}$&$c'_{1}$& $c'_{2}$&$c'_{3}$&$c'_{4}$&$c'_{5}$&$c'_{6}$&$c'_{7}$&$c'_{8}$&$s'_{1}$&$s'_{2}$&$s'_{3}$&$s'_{4}$&$s'_{5}$&$s'_{6}$&$s'_{7}$&$s'_{8}$\\
       \hline
        $c_{1}$&$-$2&$-$7&$-$3&$-$1&$-$1&$-$2&$-$2&$-$3&3&1&$-$2&$-$1&0&2&$-$1&$-$1&$-$1&0&1&$-$1&1&0&1&1&3&3&$-$2&$-$1&6&4&$-$3\\
$c_{2}$& &$-$13&0&0&$-$8&$-$4&0&0&$-$3&1&0&0&0&0&0&1&0&$-$2&2&1&$-$1&1&1&0&0&0&$-$1&0&0&0&0\\
$c_{3}$& & &0&$-$1&$-$3&$-$1&$-$1&1&6&$-$4&$-$2&$-$1&1&0&$-$2&0&$-$2&$-$1&$-$1&$-$1&0&0&1&1&3&2&0&0&3&1&$-$2\\
$c_{4}$& & & &$-$3&0&0&$-$1&1&0&0&$-$5&0&0&0&0&1&3&1&$-$1&0&1&7&3&0&0&1&0&0&0&1&$-$1\\
$c_{5}$& & & & &$-$5&$-$1&0&0&$-$1&0&$-$1&2&0&1&2&0&0&0&$-$1&1&$-$1&1&0&$-$1&$-$1&$-$1&4&$-$3&0&1&2\\
$c_{6}$& & & & & &0&0&0&$-$3&0&1&1&$-$2&0&1&0&1&$-$1&0&2&$-$2&0&1&$-$1&$-$3&$-$2&2&0&0&$-$1&1\\
$c_{7}$& & & & & & &0&1&0&0&0&0&0&$-$5&0&0&$-$1&$-$1&7&$-$1&0&$-$1&0&$-$1&0&$-$1&1&1&0&0&0\\
$c_{8}$& & & & & & & &0&0&0&0&0&0&0&3&$-$1&0&0&3&$-$1&$-$1&$-$1&0&$-$1&0&0&1&0&0&0&1\\
$s_{1}$& & & & & & & & &$-$8&$-$7&$-$4&$-$6&$-$5&$-$17&4&0&0&0&$-$1&$-$1&0&0&0&$-$18&$-$1&1&$-$7&19&36&15&$-$10\\
$s_{2}$& & & & & & & & & &$-$16&$-$5&$-$5&$-$4&6&$-$7&2&$-$1&1&1&$-$3&$-$2&0&0&12&14&$-$1&10&6&1&$-$4&$-$2\\
$s_{3}$& & & & & & & & & & &$-$1&$-$5&$-$9&2&5&0&0&0&0&$-$1&$-$1&0&0&$-$8&$-$2&$-$12&$-$6&8&8&2&$-$2\\
$s_{4}$& & & & & & & & & & & &0&1&$-$3&5&$-$2&0&0&0&3&1&1&0&1&6&$-$10&0&10&$-$7&4&12\\
$s_{5}$& & & & & & & & & & & & &$-$6&12&3&$-$1&0&0&0&0&0&0&0&23&$-$6&10&16&$-$49&$-$12&$-$10&2\\
$s_{6}$& & & & & & & & & & & & & &8&1&1&0&0&0&0&0&0&0&14&0&7&15&$-$1&$-$12&5&5\\
$s_{7}$& & & & & & & & & & & & & & &$-$1&2&0&1&3&$-$1&$-$2&0&0&17&$-$13&7&26&$-$14&$-$15&$-$1&3\\
$s_{8}$& & & & & & & & & & & & & & & &$-$2&0&0&$-$1&0&0&0&0&$-$12&3&2&$-$4&3&9&0&13\\
$c'_{2}$& & & & & & & & & & & & & & & & &$-$2&$-$15&$-$6&$-$1&$-$1&$-$2&$-$3&$-$1&0&0&1&0&0&0&0\\
$c'_{3}$& & & & & & & & & & & & & & & & & &$-$21&$-$1&0&$-$7&$-$11&0&0&$-$2&0&0&0&0&0&0\\
$c'_{4}$& & & & & & & & & & & & & & & & & & &$-$1&0&0&2&0&0&0&$-$1&0&0&0&0&0\\
$c'_{5}$& & & & & & & & & & & & & & & & & & & &$-$2&$-$1&$-$1&$-$1&1&0&0&$-$2&0&0&0&0\\
$c'_{6}$& & & & & & & & & & & & & & & & & & & & &$-$5&$-$1&0&0&0&0&0&$-$4&0&0&1\\
$c'_{7}$& & & & & & & & & & & & & & & & & & & & & &1&0&0&0&0&$-$1&0&1&0&0\\
$c'_{8}$& & & & & & & & & & & & & & & & & & & & & & &$-$3&0&0&0&0&0&0&$-$3&0\\
$c'_{9}$& & & & & & & & & & & & & & & & & & & & & & & &0&0&0&0&0&0&0&1\\
$s'_{1}$& & & & & & & & & & & & & & & & & & & & & & & & &$-$4&$-$7&7&$-$15&$-$13&$-$5&$-$1\\
$s'_{2}$& & & & & & & & & & & & & & & & & & & & & & & & & &$-$18&$-$3&5&$-$4&$-$8&0\\
$s'_{3}$& & & & & & & & & & & & & & & & & & & & & & & & & & &4&$-$7&$-$2&0&$-$1\\
$s'_{4}$& & & & & & & & & & & & & & & & & & & & & & & & & & & &$-$12&$-$8&$-$2&1\\
$s'_{5}$& & & & & & & & & & & & & & & & & & & & & & & & & & & & &11&7&$-$1\\
$s'_{6}$& & & & & & & & & & & & & & & & & & & & & & & & & & & & & &6&$-$4\\
$s'_{7}$& & & & & & & & & & & & & & & & & & & & & & & & & & & & & & &$-$4\\
 \hline
    \end{tabular}
    \end{sideways}
    \caption{Correlation matrix ($\%$) of statistical uncertainties for the optimal binning scheme without $\Delta c_{i}$ and $\Delta s_{i}$ constraints.}
    \label{tab:opstatnocs}

\end{table}

\begin{table}[!htbp]
\begin{sideways}
    \scriptsize       
    \setlength\tabcolsep{3pt}
    \centering
    \begin{tabular}{cccccccccccccccccccccccccccccccc}
 \hline    
         &$c_{2}$&$c_{3}$&$c_{4}$&$c_{5}$&$c_{6}$&$c_{7}$&$c_{8}$&$s_{1}$&$s_{2}$&$s_{3}$&$s_{4}$&$s_{5}$&$s_{6}$&$s_{7}$&$s_{8}$&$c'_{1}$& $c'_{2}$&$c'_{3}$&$c'_{4}$&$c'_{5}$&$c'_{6}$&$c'_{7}$&$c'_{8}$&$s'_{1}$&$s'_{2}$&$s'_{3}$&$s'_{4}$&$s'_{5}$&$s'_{6}$&$s'_{7}$&$s'_{8}$\\
     \hline
        $c_{1}$&$-$1&$-$3&$-$4&$-$1&0&$-$3&0&$-$1&0&0&0&0&0&0&0&0&0&0&$-$1&0&0&0&0&0&0&0&$-$1&0&1&0&0\\
$c_{2}$& &$-$7&1&$-$1&$-$2&$-$1&$-$1&0&$-$2&0&0&0&0&0&0&2&$-$1&$-$1&3&2&$-$1&0&1&0&0&0&$-$1&0&0&0&0\\
$c_{3}$& & &$-$1&$-$1&$-$3&$-$5&$-$2&$-$3&7&$-$1&$-$1&$-$2&1&1&$-$3&0&$-$1&0&$-$1&$-$1&$-$1&1&1&0&7&4&$-$2&$-$2&5&4&$-$4\\
$c_{4}$& & & &$-$3&0&0&0&0&1&0&$-$7&$-$1&0&0&0&0&3&2&$-$1&0&1&4&3&0&1&1&0&0&0&0&$-$1\\
$c_{5}$& & & & &$-$4&0&0&0&$-$1&1&$-$1&5&0&0&0&0&3&1&$-$1&0&1&2&1&1&$-$1&$-$1&$-$1&$-$4&1&2&0\\
$c_{6}$& & & & & &$-$5&0&2&$-$3&$-$1&0&1&$-$3&$-$1&1&1&0&0&0&1&$-$1&$-$1&0&0&$-$5&$-$4&5&0&0&$-$2&3\\
$c_{7}$& & & & & & &2&0&0&0&0&0&0&$-$2&0&0&$-$1&$-$1&6&0&0&0&$-$1&0&0&0&1&0&0&$-$1&0\\
$c_{8}$& & & & & & & &0&$-$1&0&0&0&0&0&4&$-$1&0&0&5&1&0&0&0&0&$-$1&$-$1&2&0&0&$-$1&0\\
$s_{1}$& & & & & & & & &$-$3&$-$19&$-$8&$-$2&$-$1&$-$6&$-$4&0&0&0&0&1&1&1&0&1&$-$12&$-$7&9&2&1&$-$15&14\\
$s_{2}$& & & & & & & & & &$-$7&$-$5&$-$4&7&11&$-$2&1&0&3&1&$-$2&$-$4&0&0&17&$-$2&10&9&$-$1&5&1&$-$8\\
$s_{3}$& & & & & & & & & & &3&$-$13&$-$3&6&8&0&0&0&0&$-$1&$-$2&$-$1&0&3&7&$-$14&$-$1&23&24&26&$-$9\\
$s_{4}$& & & & & & & & & & & &5&2&2&2&$-$1&0&$-$1&0&0&3&1&0&6&7&$-$8&1&0&$-$1&12&5\\
$s_{5}$& & & & & & & & & & & & &4&$-$2&$-$6&$-$1&0&$-$1&0&$-$1&0&1&0&12&$-$6&21&$-$2&$-$38&$-$40&$-$17&15\\
$s_{6}$& & & & & & & & & & & & & &5&$-$5&0&0&1&0&0&0&$-$1&0&13&7&19&19&$-$15&$-$19&8&$-$5\\
$s_{7}$& & & & & & & & & & & & & & &6&1&0&2&0&$-$3&$-$1&$-$2&0&12&$-$6&3&28&4&$-$7&20&3\\
$s_{8}$& & & & & & & & & & & & & & & &0&0&$-$1&0&$-$2&3&1&0&$-$9&$-$5&$-$9&$-$6&16&6&5&$-$1\\
$c'_{2}$& & & & & & & & & & & & & & & & &0&$-$2&$-$4&$-$1&0&$-$5&$-$1&$-$1&0&0&1&0&0&0&0\\
$c'_{3}$& & & & & & & & & & & & & & & & & &$-$13&$-$1&$-$1&$-$1&$-$3&$-$2&0&$-$3&0&0&0&0&0&0\\
$c'_{4}$& & & & & & & & & & & & & & & & & & &$-$3&$-$1&$-$1&$-$3&$-$2&1&0&0&1&0&0&1&0\\
$c'_{5}$& & & & & & & & & & & & & & & & & & & &$-$1&$-$1&$-$1&$-$1&0&0&0&$-$3&0&0&0&0\\
$c'_{6}$& & & & & & & & & & & & & & & & & & & & &$-$6&0&0&$-$1&0&0&$-$1&$-$2&0&$-$1&0\\
$c'_{7}$& & & & & & & & & & & & & & & & & & & & & &$-$6&1&$-$1&0&$-$1&0&0&$-$4&0&1\\
$c'_{8}$& & & & & & & & & & & & & & & & & & & & & & &$-$1&0&0&0&$-$1&0&0&$-$5&0\\
$c'_{9}$& & & & & & & & & & & & & & & & & & & & & & & &0&0&0&0&0&0&0&1\\
$s'_{1}$& & & & & & & & & & & & & & & & & & & & & & & & &0&4&4&$-$7&$-$6&$-$1&0\\
$s'_{2}$& & & & & & & & & & & & & & & & & & & & & & & & & &$-$10&$-$2&1&1&1&$-$6\\
$s'_{3}$& & & & & & & & & & & & & & & & & & & & & & & & & & &2&$-$14&$-$15&$-$6&$-$1\\
$s'_{4}$& & & & & & & & & & & & & & & & & & & & & & & & & & & &$-$3&$-$5&5&1\\
$s'_{5}$& & & & & & & & & & & & & & & & & & & & & & & & & & & & &18&10&$-$5\\
$s'_{6}$& & & & & & & & & & & & & & & & & & & & & & & & & & & & & &6&$-$7\\
$s'_{7}$& & & & & & & & & & & & & & & & & & & & & & & & & & & & & & &$-$6\\
 \hline
    \end{tabular}
    \end{sideways}
    \caption{Correlation matrix ($\%$) of statistical uncertainties for the modified optimal binning scheme without $\Delta c_{i}$ and $\Delta s_{i}$ constraints.}
    \label{tab:modistatnocs}

\end{table}

 \begin{table}[!htbp]
\begin{sideways}
    \scriptsize       
    \setlength\tabcolsep{3pt}
    \centering
    \begin{tabular}{cccccccccccccccccccccccccccccccc}
         \hline       
         &$c_{2}$&$c_{3}$&$c_{4}$&$c_{5}$&$c_{6}$&$c_{7}$&$c_{8}$&$s_{1}$&$s_{2}$&$s_{3}$&$s_{4}$&$s_{5}$&$s_{6}$&$s_{7}$&$s_{8}$&$c'_{1}$& $c'_{2}$&$c'_{3}$&$c'_{4}$&$c'_{5}$&$c'_{6}$&$c'_{7}$&$c'_{8}$&$s'_{1}$&$s'_{2}$&$s'_{3}$&$s'_{4}$&$s'_{5}$&$s'_{6}$&$s'_{7}$&$s'_{8}$\\
       \hline
        $c_{1}$&72&73&33&0&27&68&74&25&22&37&10&6&$-$19&$-$29&$-$28&49&43&38&$-$21&$-$23&$-$20&30&46&0&23&28&16&1&$-$1&$-$15&$-$12\\
$c_{2}$& &64&32&3&26&62&61&11&28&27&11&1&$-$18&$-$26&$-$28&30&29&25&$-$11&$-$10&$-$11&20&31&5&19&19&8&10&11&$-$6&$-$3\\
$c_{3}$& & &59&31&54&89&77&18&18&13&14&$-$11&$-$22&$-$13&$-$14&13&20&14&17&21&11&18&18&$-$1&6&5&$-$20&$-$3&14&$-$4&0\\
$c_{4}$& & & &55&53&57&48&11&4&$-$30&$-$42&$-$9&28&34&2&0&0&$-$12&29&33&13&0&$-$1&$-$10&$-$25&$-$21&$-$39&$-$5&29&2&3\\
$c_{5}$& & & & &42&32&20&3&$-$10&$-$39&$-$22&$-$31&27&36&10&$-$27&$-$21&$-$30&33&45&19&$-$15&$-$25&5&$-$27&$-$28&$-$42&0&33&23&27\\
$c_{6}$& & & & & &55&40&3&$-$3&$-$19&$-$4&$-$13&12&15&$-$2&$-$13&$-$8&$-$11&27&32&15&$-$3&$-$9&8&$-$3&$-$15&$-$34&0&27&20&25\\
$c_{7}$& & & & & & &75&15&16&12&12&$-$9&$-$17&$-$6&$-$3&11&18&13&16&18&11&16&15&5&13&9&$-$16&$-$4&10&$-$6&0\\
$c_{8}$& & & & & & & &25&24&20&7&$-$2&$-$11&$-$8&$-$14&24&27&19&$-$1&5&$-$5&17&25&$-$1&4&14&$-$7&$-$4&11&$-$10&$-$5\\
$s_{1}$& & & & & & & & &46&28&18&10&$-$9&1&6&2&22&17&1&22&19&1&6&15&$-$28&$-$16&$-$19&39&26&4&$-$8\\
$s_{2}$& & & & & & & & & &36&21&19&$-$18&$-$10&2&9&24&29&$-$1&11&20&13&15&18&$-$9&$-$19&$-$11&37&24&$-$16&$-$17\\
$s_{3}$& & & & & & & & & & &56&2&$-$51&$-$52&$-$33&25&33&45&$-$27&$-$19&$-$2&19&30&26&19&8&14&33&8&11&3\\
$s_{4}$& & & & & & & & & & & &$-$24&$-$72&$-$53&$-$19&$-$3&23&21&17&17&31&16&5&15&$-$4&$-$12&$-$17&17&12&28&21\\
$s_{5}$& & & & & & & & & & & & &28&13&23&19&12&26&$-$33&$-$33&$-$22&4&14&10&18&18&37&5&$-$15&$-$32&$-$36\\
$s_{6}$& & & & & & & & & & & & & &60&22&$-$10&$-$20&$-$27&$-$24&$-$12&$-$38&$-$33&$-$22&9&$-$9&0&10&0&21&8&9\\
$s_{7}$& & & & & & & & & & & & & & &32&$-$25&$-$31&$-$25&13&19&3&$-$28&$-$34&5&$-$13&$-$17&$-$17&5&16&5&11\\
$s_{8}$& & & & & & & & & & & & & & & &$-$19&$-$14&$-$11&6&3&16&$-$19&$-$21&8&5&10&10&0&$-$16&$-$20&$-$26\\
$c'_{2}$& & & & & & & & & & & & & & & & &57&61&$-$24&$-$57&$-$14&47&66&$-$21&28&40&39&$-$10&$-$28&$-$31&$-$26\\
$c'_{3}$& & & & & & & & & & & & & & & & & &52&$-$6&$-$24&6&43&55&2&$-$6&23&25&5&$-$8&$-$14&$-$18\\
$c'_{4}$& & & & & & & & & & & & & & & & & & &$-$3&$-$30&11&54&62&4&31&35&39&1&$-$31&$-$33&$-$32\\
$c'_{5}$& & & & & & & & & & & & & & & & & & & &68&72&14&$-$12&$-$21&$-$36&$-$32&$-$46&$-$11&10&8&10\\
$c'_{6}$& & & & & & & & & & & & & & & & & & & & &60&$-$9&$-$36&$-$6&$-$53&$-$47&$-$65&21&36&27&24\\
$c'_{7}$& & & & & & & & & & & & & & & & & & & & & &27&1&$-$9&$-$28&$-$22&$-$37&0&13&$-$3&$-$6\\
$c'_{8}$& & & & & & & & & & & & & & & & & & & & & & &44&$-$18&15&20&16&$-$14&$-$26&$-$51&$-$22\\
$c'_{9}$& & & & & & & & & & & & & & & & & & & & & & & &$-$7&22&33&33&$-$6&$-$25&$-$26&$-$45\\
$s'_{1}$& & & & & & & & & & & & & & & & & & & & & & & & &19&$-$6&8&47&33&33&26\\
$s'_{2}$& & & & & & & & & & & & & & & & & & & & & & & & & &56&58&$-$9&$-$45&$-$28&$-$13\\
$s'_{3}$& & & & & & & & & & & & & & & & & & & & & & & & & & &70&$-$28&$-$55&$-$35&$-$28\\
$s'_{4}$& & & & & & & & & & & & & & & & & & & & & & & & & & & &$-$12&$-$54&$-$38&$-$35\\
$s'_{5}$& & & & & & & & & & & & & & & & & & & & & & & & & & & & &47&33&23\\
$s'_{6}$& & & & & & & & & & & & & & & & & & & & & & & & & & & & & &60&52\\
$s'_{7}$& & & & & & & & & & & & & & & & & & & & & & & & & & & & & & &66\\
 \hline
    \end{tabular}
    \end{sideways}
    \caption{Correlation matrix ($\%$) of systematic uncertainties for the equal binning scheme without $\Delta c_{i}$ and $\Delta s_{i}$ constraints.}
    \label{tab:eqsystnocs}
\end{table}

 \begin{table}[!htbp]
\begin{sideways}
    \scriptsize       
    \setlength\tabcolsep{3pt}
    \centering
    \begin{tabular}{cccccccccccccccccccccccccccccccc}
          \hline       
         &$c_{2}$&$c_{3}$&$c_{4}$&$c_{5}$&$c_{6}$&$c_{7}$&$c_{8}$&$s_{1}$&$s_{2}$&$s_{3}$&$s_{4}$&$s_{5}$&$s_{6}$&$s_{7}$&$s_{8}$&$c'_{1}$& $c'_{2}$&$c'_{3}$&$c'_{4}$&$c'_{5}$&$c'_{6}$&$c'_{7}$&$c'_{8}$&$s'_{1}$&$s'_{2}$&$s'_{3}$&$s'_{4}$&$s'_{5}$&$s'_{6}$&$s'_{7}$&$s'_{8}$\\
         \hline
        $c_{1}$&50&85&38&84&70&49&32&$-$10&5&$-$3&5&22&2&30&2&27&11&14&23&29&11&2&0&8&$-$1&$-$5&16&$-$23&8&8&$-$8\\
$c_{2}$& &46&8&43&42&39&25&$-$17&10&$-$3&0&21&6&14&$-$7&13&14&9&$-$3&7&12&14&11&$-$6&4&$-$6&6&$-$4&12&10&$-$14\\
$c_{3}$& & &32&79&68&46&30&$-$12&11&$-$19&9&21&4&30&1&21&11&12&20&20&12&$-$1&$-$1&6&$-$3&$-$8&16&$-$20&7&9&$-$5\\
$c_{4}$& & & &54&23&$-$26&$-$17&30&$-$10&5&9&$-$22&$-$15&23&42&3&$-$19&$-$3&74&38&$-$14&$-$54&$-$46&57&$-$4&30&37&$-$44&$-$19&5&43\\
$c_{5}$& & & & &66&37&23&$-$2&6&2&14&2&$-$5&44&18&20&2&10&44&33&3&$-$18&$-$16&26&$-$11&1&29&$-$41&$-$7&3&20\\
$c_{6}$& & & & & &47&31&$-$22&$-$2&$-$12&3&25&38&29&3&17&12&12&12&14&9&4&$-$2&5&6&1&22&$-$13&12&14&$-$7\\
$c_{7}$& & & & & & &51&$-$44&6&$-$13&$-$2&42&12&20&$-$18&27&32&22&$-$30&7&25&47&35&$-$25&3&$-$18&9&8&23&6&$-$33\\
$c_{8}$& & & & & & & &$-$32&1&$-$10&$-$2&33&7&12&$-$16&17&19&12&$-$22&3&22&33&26&$-$15&9&$-$4&11&11&24&11&$-$23\\
$s_{1}$& & & & & & & & &5&25&23&$-$61&$-$29&$-$6&25&$-$20&$-$22&$-$15&28&$-$2&$-$18&$-$30&$-$24&42&3&30&8&$-$39&$-$34&3&39\\
$s_{2}$& & & & & & & & & &24&22&0&14&26&$-$31&$-$1&3&$-$2&$-$1&$-$1&10&6&4&$-$15&$-$28&$-$36&$-$21&$-$5&$-$20&$-$24&7\\
$s_{3}$& & & & & & & & & & &28&$-$19&$-$3&17&$-$9&$-$10&$-$5&$-$11&10&$-$5&5&$-$9&$-$10&5&$-$24&$-$10&$-$6&$-$17&$-$27&$-$22&26\\
$s_{4}$& & & & & & & & & & & &$-$8&4&37&10&$-$10&$-$9&$-$5&22&2&$-$6&$-$15&$-$23&21&$-$27&$-$7&27&$-$11&$-$31&$-$20&53\\
$s_{5}$& & & & & & & & & & & & &37&21&$-$23&18&18&12&$-$15&7&24&24&21&$-$31&$-$3&$-$15&4&36&57&$-$7&$-$34\\
$s_{6}$& & & & & & & & & & & & & &13&$-$38&3&8&2&$-$11&0&10&13&3&$-$30&$-$12&$-$31&$-$14&41&20&$-$12&$-$16\\
$s_{7}$& & & & & & & & & & & & & & &5&14&0&8&46&30&7&$-$19&$-$26&33&$-$28&$-$7&37&$-$34&$-$25&$-$13&39\\
$s_{8}$& & & & & & & & & & & & & & & &$-$20&$-$21&$-$8&37&2&$-$18&$-$38&$-$30&53&14&56&45&$-$41&$-$2&16&52\\
$c'_{2}$& & & & & & & & & & & & & & & & &32&27&6&62&29&34&33&$-$29&$-$13&$-$24&$-$16&$-$3&6&$-$16&$-$29\\
$c'_{3}$& & & & & & & & & & & & & & & & & &18&$-$23&22&20&40&37&$-$23&$-$13&$-$17&$-$16&9&11&$-$2&$-$30\\
$c'_{4}$& & & & & & & & & & & & & & & & & & &$-$3&29&20&23&19&$-$5&$-$2&$-$38&$-$5&2&8&$-$1&$-$16\\
$c'_{5}$& & & & & & & & & & & & & & & & & & & &45&$-$14&$-$69&$-$58&60&$-$18&22&40&$-$48&$-$27&$-$12&52\\
$c'_{6}$& & & & & & & & & & & & & & & & & & & & &22&3&5&16&$-$17&$-$8&2&$-$21&$-$3&$-$17&$-$3\\
$c'_{7}$& & & & & & & & & & & & & & & & & & & & & &31&31&$-$18&$-$2&$-$13&$-$13&5&33&$-$5&$-$24\\
$c'_{8}$& & & & & & & & & & & & & & & & & & & & & & &71&$-$51&6&$-$28&$-$34&31&25&$-$7&$-$50\\
$c'_{9}$& & & & & & & & & & & & & & & & & & & & & & & &$-$48&4&$-$20&$-$39&22&35&$-$4&$-$49\\
$s'_{1}$& & & & & & & & & & & & & & & & & & & & & & & & &23&63&65&$-$57&$-$33&27&53\\
$s'_{2}$& & & & & & & & & & & & & & & & & & & & & & & & & &52&24&18&28&65&$-$26\\
$s'_{3}$& & & & & & & & & & & & & & & & & & & & & & & & & & &57&$-$26&14&49&23\\
$s'_{4}$& & & & & & & & & & & & & & & & & & & & & & & & & & & &$-$23&$-$6&31&43\\
$s'_{5}$& & & & & & & & & & & & & & & & & & & & & & & & & & & & &49&19&$-$43\\
$s'_{6}$& & & & & & & & & & & & & & & & & & & & & & & & & & & & & &22&$-$42\\
$s'_{7}$& & & & & & & & & & & & & & & & & & & & & & & & & & & & & & &$-$15\\
 \hline
    \end{tabular}
    \end{sideways}
    \caption{Correlation matrix ($\%$) of systematic uncertainties for the optimal binning scheme without $\Delta c_{i}$ and $\Delta s_{i}$ constraints.}
    \label{tab:opsystnocs}
\end{table}

 \begin{table}[!htbp]
\begin{sideways}
    \scriptsize       
    \setlength\tabcolsep{3pt}
    \centering
    \begin{tabular}{cccccccccccccccccccccccccccccccc}
         \hline     
         &$c_{2}$&$c_{3}$&$c_{4}$&$c_{5}$&$c_{6}$&$c_{7}$&$c_{8}$&$s_{1}$&$s_{2}$&$s_{3}$&$s_{4}$&$s_{5}$&$s_{6}$&$s_{7}$&$s_{8}$&$c'_{1}$& $c'_{2}$&$c'_{3}$&$c'_{4}$&$c'_{5}$&$c'_{6}$&$c'_{7}$&$c'_{8}$&$s'_{1}$&$s'_{2}$&$s'_{3}$&$s'_{4}$&$s'_{5}$&$s'_{6}$&$s'_{7}$&$s'_{8}$\\
          \hline
        $c_{1}$&44&73&47&76&66&36&36&11&21&$-$11&$-$10&20&$-$3&29&7&17&$-$6&11&43&28&$-$11&$-$10&$-$20&39&$-$2&18&34&$-$51&$-$18&$-$14&16\\
$c_{2}$& &63&2&46&60&60&55&$-$11&12&$-$20&$-$17&21&21&15&$-$9&20&20&20&$-$3&$-$2&16&23&16&11&8&$-$6&7&$-$8&22&14&$-$17\\
$c_{3}$& & &26&76&82&59&57&$-$8&20&$-$12&$-$19&26&16&24&$-$2&23&10&18&22&19&7&7&$-$1&19&5&0&15&$-$29&10&7&$-$4\\
$c_{4}$& & & &53&21&$-$21&$-$12&23&13&9&5&$-$7&$-$31&16&23&2&$-$30&$-$11&58&36&$-$34&$-$43&$-$46&45&$-$6&34&41&$-$57&$-$47&$-$32&38\\
$c_{5}$& & & & &69&37&36&12&27&$-$5&1&5&$-$6&32&10&18&$-$7&11&46&29&$-$16&$-$12&$-$21&44&$-$21&7&39&$-$57&$-$29&$-$24&29\\
$c_{6}$& & & & & &64&59&$-$11&11&$-$28&$-$25&26&26&21&$-$5&18&12&18&16&8&6&10&1&26&8&6&22&$-$27&14&10&$-$12\\
$c_{7}$& & & & & & &72&$-$12&3&$-$35&$-$28&23&33&11&$-$22&32&43&36&$-$25&$-$9&29&51&39&9&5&$-$21&2&4&41&25&$-$40\\
$c_{8}$& & & & & & & &$-$22&6&$-$28&$-$28&31&30&13&$-$12&26&32&26&$-$16&$-$7&27&38&29&3&11&$-$8&5&$-$2&30&17&$-$26\\
$s_{1}$& & & & & & & & &$-$7&17&45&$-$60&$-$31&$-$15&11&$-$18&$-$16&$-$18&$-$1&$-$16&$-$19&$-$12&$-$8&40&$-$24&$-$1&31&$-$4&$-$13&$-$20&18\\
$s_{2}$& & & & & & & & & &21&7&28&17&52&$-$24&20&$-$6&6&39&17&13&$-$10&$-$17&12&$-$31&$-$10&2&$-$37&$-$28&$-$29&23\\
$s_{3}$& & & & & & & & & & &37&$-$20&$-$11&10&$-$2&$-$10&$-$19&$-$23&9&$-$4&6&$-$21&$-$14&$-$9&$-$15&$-$5&$-$7&4&$-$14&$-$12&23\\
$s_{4}$& & & & & & & & & & & &$-$40&$-$27&9&23&$-$32&$-$25&$-$26&5&$-$26&$-$16&$-$23&$-$19&27&$-$30&1&31&$-$5&$-$34&$-$27&47\\
$s_{5}$& & & & & & & & & & & & &42&37&$-$14&25&10&19&28&23&26&10&0&$-$14&5&17&$-$4&$-$37&$-$12&$-$10&6\\
$s_{6}$& & & & & & & & & & & & & &21&$-$33&30&27&21&$-$12&0&43&34&27&$-$21&8&$-$8&$-$17&13&30&26&$-$36\\
$s_{7}$& & & & & & & & & & & & & & &$-$13&17&$-$9&10&45&17&9&$-$15&$-$27&35&$-$21&8&34&$-$46&$-$34&$-$17&28\\
$s_{8}$& & & & & & & & & & & & & & & &$-$26&$-$20&$-$19&11&$-$2&$-$36&$-$19&$-$14&8&2&27&30&$-$9&$-$36&$-$15&38\\
$c'_{2}$& & & & & & & & & & & & & & & & &46&58&11&63&31&46&32&2&$-$4&$-$20&$-$19&$-$15&14&7&$-$25\\
$c'_{3}$& & & & & & & & & & & & & & & & & &41&$-$40&8&36&65&58&$-$20&$-$4&$-$26&$-$26&22&36&24&$-$39\\
$c'_{4}$& & & & & & & & & & & & & & & & & & &$-$5&34&34&45&35&$-$2&0&$-$34&$-$13&$-$3&18&14&$-$24\\
$c'_{5}$& & & & & & & & & & & & & & & & & & & &54&$-$28&$-$54&$-$60&42&$-$21&33&39&$-$71&$-$62&$-$49&54\\
$c'_{6}$& & & & & & & & & & & & & & & & & & & & &6&$-$3&$-$10&7&$-$10&3&$-$4&$-$39&$-$23&$-$13&9\\
$c'_{7}$& & & & & & & & & & & & & & & & & & & & & &41&39&$-$24&13&$-$19&$-$27&26&45&27&$-$33\\
$c'_{8}$& & & & & & & & & & & & & & & & & & & & & & &73&$-$27&4&$-$33&$-$30&33&43&24&$-$45\\
$c'_{9}$& & & & & & & & & & & & & & & & & & & & & & & &$-$39&9&$-$34&$-$36&48&49&36&$-$44\\
$s'_{1}$& & & & & & & & & & & & & & & & & & & & & & & & &$-$8&39&69&$-$57&$-$28&$-$29&23\\
$s'_{2}$& & & & & & & & & & & & & & & & & & & & & & & & & &40&1&27&45&51&$-$37\\
$s'_{3}$& & & & & & & & & & & & & & & & & & & & & & & & & & &55&$-$31&$-$21&$-$10&26\\
$s'_{4}$& & & & & & & & & & & & & & & & & & & & & & & & & & & &$-$49&$-$44&$-$27&44\\
$s'_{5}$& & & & & & & & & & & & & & & & & & & & & & & & & & & & &70&58&$-$55\\
$s'_{6}$& & & & & & & & & & & & & & & & & & & & & & & & & & & & & &69&$-$80\\
$s'_{7}$& & & & & & & & & & & & & & & & & & & & & & & & & & & & & & &$-$65\\
 \hline
    \end{tabular}
    \end{sideways}
    \caption{Correlation matrix ($\%$) of systematic uncertainties for the modified optimal binning scheme without $\Delta c_{i}$ and $\Delta s_{i}$ constraints.}
    \label{tab:modisystnocs}
\end{table}

 \begin{table}[!htbp]
\begin{sideways}
    \scriptsize       
    \setlength\tabcolsep{3pt}
    \centering
    \begin{tabular}{cccccccccccccccccccccccccccccccc}
          \hline      
         &$c_{2}$&$c_{3}$&$c_{4}$&$c_{5}$&$c_{6}$&$c_{7}$&$c_{8}$&$s_{1}$&$s_{2}$&$s_{3}$&$s_{4}$&$s_{5}$&$s_{6}$&$s_{7}$&$s_{8}$&$c'_{1}$& $c'_{2}$&$c'_{3}$&$c'_{4}$&$c'_{5}$&$c'_{6}$&$c'_{7}$&$c'_{8}$&$s'_{1}$&$s'_{2}$&$s'_{3}$&$s'_{4}$&$s'_{5}$&$s'_{6}$&$s'_{7}$&$s'_{8}$\\
         \hline
        $c_{1}$&$-$9&$-$1&1&3&$-$1&1&$-$14&$-$1&0&0&0&0&0&0&0&34&$-$7&0&2&3&$-$1&1&$-$13&$-$1&0&0&0&0&0&0&0\\
$c_{2}$& &$-$7&1&2&1&$-$1&0&$-$1&$-$2&0&0&2&1&$-$1&$-$1&$-$8&56&$-$6&0&1&2&$-$1&0&$-$1&$-$2&0&0&3&1&$-$1&$-$1\\
$c_{3}$& & &$-$4&0&$-$1&$-$1&$-$1&$-$2&$-$2&1&0&1&2&$-$1&$-$3&0&$-$6&69&$-$4&0&$-$1&$-$1&$-$1&$-$2&$-$2&1&0&1&2&$-$1&$-$3\\
$c_{4}$& & & &$-$5&$-$2&$-$1&1&$-$2&$-$1&0&0&0&1&$-$1&$-$2&1&1&$-$4&53&$-$5&$-$1&0&2&$-$2&$-$1&0&0&0&1&$-$1&$-$3\\
$c_{5}$& & & & &$-$10&0&2&0&0&0&0&$-$3&0&0&0&3&2&0&$-$5&91&$-$9&0&2&0&0&0&0&$-$3&0&0&0\\
$c_{6}$& & & & & &$-$11&1&1&2&0&0&$-$1&$-$4&1&2&$-$1&1&0&$-$1&$-$9&74&$-$11&1&2&2&0&0&$-$1&$-$4&1&2\\
$c_{7}$& & & & & & &$-$15&4&3&1&$-$1&$-$1&$-$2&0&3&2&0&0&$-$1&0&$-$12&81&$-$17&4&3&1&$-$1&$-$1&$-$2&0&3\\
$c_{8}$& & & & & & & &1&1&1&$-$1&$-$2&$-$1&1&1&$-$15&0&$-$1&0&2&1&$-$16&67&1&1&1&$-$1&$-$2&$-$1&1&1\\
$s_{1}$& & & & & & & & &2&7&$-$8&10&3&7&$-$10&$-$1&$-$1&$-$2&$-$2&0&1&4&1&40&1&7&$-$8&10&3&7&$-$10\\
$s_{2}$& & & & & & & & & &$-$14&$-$2&13&10&9&0&0&$-$2&$-$1&$-$1&0&2&3&0&5&87&$-$14&$-$2&13&10&9&$-$2\\
$s_{3}$& & & & & & & & & & &$-$19&$-$1&12&8&3&0&0&1&0&0&0&1&0&4&$-$13&97&$-$18&$-$1&12&8&3\\
$s_{4}$& & & & & & & & & & & &$-$7&0&14&11&0&0&0&0&0&0&$-$1&$-$1&$-$1&$-$2&$-$18&88&$-$7&0&14&13\\
$s_{5}$& & & & & & & & & & & & &$-$10&1&$-$5&0&2&1&$-$1&$-$3&$-$1&0&$-$1&8&12&$-$1&$-$7&97&$-$10&1&$-$5\\
$s_{6}$& & & & & & & & & & & & & &$-$19&$-$7&0&1&2&1&0&$-$4&$-$1&$-$1&3&8&12&0&$-$9&100&$-$19&$-$5\\
$s_{7}$& & & & & & & & & & & & & & &$-$23&0&0&$-$1&0&0&1&0&0&8&9&8&14&1&$-$19&98&$-$20\\
$s_{8}$& & & & & & & & & & & & & & & &0&$-$1&$-$3&$-$2&0&1&3&1&$-$9&0&3&11&$-$5&$-$7&$-$23&79\\
$c'_{2}$& & & & & & & & & & & & & & & & &$-$11&0&1&3&$-$1&2&$-$20&$-$1&0&0&0&0&0&0&0\\
$c'_{3}$& & & & & & & & & & & & & & & & & &$-$7&0&2&1&$-$1&0&$-$1&$-$2&0&0&2&1&0&$-$1\\
$c'_{4}$& & & & & & & & & & & & & & & & & & &$-$5&0&0&0&$-$1&$-$1&$-$1&1&0&1&2&$-$1&$-$3\\
$c'_{5}$& & & & & & & & & & & & & & & & & & & &$-$5&$-$1&$-$1&1&$-$1&$-$1&0&0&$-$1&1&0&$-$2\\
$c'_{6}$& & & & & & & & & & & & & & & & & & & & &$-$9&0&2&0&0&0&0&$-$3&0&0&0\\
$c'_{7}$& & & & & & & & & & & & & & & & & & & & & &$-$13&1&1&2&0&0&$-$1&$-$4&1&1\\
$c'_{8}$& & & & & & & & & & & & & & & & & & & & & & &$-$19&3&3&1&$-$1&0&$-$1&0&3\\
$c'_{9}$& & & & & & & & & & & & & & & & & & & & & & & &1&0&0&$-$1&$-$1&$-$1&0&1\\
$s'_{1}$& & & & & & & & & & & & & & & & & & & & & & & & &3&4&$-$1&8&3&8&$-$11\\
$s'_{2}$& & & & & & & & & & & & & & & & & & & & & & & & & &$-$13&$-$2&12&8&9&$-$2\\
$s'_{3}$& & & & & & & & & & & & & & & & & & & & & & & & & & &$-$17&$-$1&12&8&4\\
$s'_{4}$& & & & & & & & & & & & & & & & & & & & & & & & & & & &$-$7&0&14&12\\
$s'_{5}$& & & & & & & & & & & & & & & & & & & & & & & & & & & & &$-$9&1&$-$5\\
$s'_{6}$& & & & & & & & & & & & & & & & & & & & & & & & & & & & & &$-$19&$-$5\\
$s'_{7}$& & & & & & & & & & & & & & & & & & & & & & & & & & & & & & &$-$20\\
 \hline
    \end{tabular}
    \end{sideways}
    \caption{Correlation matrix ($\%$) of statistical uncertainties for the equal binning scheme with $\Delta c_{i}$ and $\Delta s_{i}$ constraints.}
    \label{tab:eqstatdcs}
\end{table}

 \begin{table}[!htbp]
\begin{sideways}
    \scriptsize       
    \setlength\tabcolsep{3pt}
    \centering
    \begin{tabular}{cccccccccccccccccccccccccccccccc}
       \hline    
         &$c_{2}$&$c_{3}$&$c_{4}$&$c_{5}$&$c_{6}$&$c_{7}$&$c_{8}$&$s_{1}$&$s_{2}$&$s_{3}$&$s_{4}$&$s_{5}$&$s_{6}$&$s_{7}$&$s_{8}$&$c'_{1}$& $c'_{2}$&$c'_{3}$&$c'_{4}$&$c'_{5}$&$c'_{6}$&$c'_{7}$&$c'_{8}$&$s'_{1}$&$s'_{2}$&$s'_{3}$&$s'_{4}$&$s'_{5}$&$s'_{6}$&$s'_{7}$&$s'_{8}$\\
        \hline
        $c_{1}$&$-$2&$-$6&$-$3&$-$1&$-$1&$-$1&$-$2&$-$2&4&2&$-$3&$-$2&2&4&$-$2&68&$-$2&$-$8&$-$3&$-$1&$-$1&$-$1&$-$1&$-$2&4&2&$-$3&$-$2&2&4&$-$3\\
$c_{2}$& &$-$10&2&0&$-$6&$-$7&0&0&$-$3&1&0&0&0&0&0&$-$2&96&$-$18&2&0&$-$7&$-$8&1&0&$-$3&1&0&0&0&0&0\\
$c_{3}$& & &$-$1&$-$1&$-$2&0&$-$1&1&5&0&$-$1&$-$1&2&1&$-$2&$-$5&$-$10&14&$-$1&$-$1&$-$2&0&1&1&5&0&$-$1&$-$1&2&1&$-$2\\
$c_{4}$& & & &$-$3&0&6&1&0&1&0&$-$5&$-$1&0&2&$-$1&$-$3&2&0&85&$-$2&0&6&2&0&1&0&$-$5&$-$1&0&1&$-$1\\
$c_{5}$& & & & &$-$4&$-$1&0&$-$1&$-$2&$-$1&1&$-$1&$-$1&$-$1&2&$-$1&1&0&$-$3&43&$-$4&0&0&$-$1&$-$2&$-$1&1&$-$1&$-$1&$-$1&2\\
$c_{6}$& & & & & &0&0&$-$1&$-$4&$-$2&2&1&$-$1&$-$2&1&$-$1&$-$6&0&0&$-$3&82&0&1&$-$1&$-$4&$-$2&2&1&$-$1&$-$2&1\\
$c_{7}$& & & & & & &0&0&0&$-$1&2&0&0&$-$4&1&$-$1&$-$7&0&6&$-$1&0&65&$-$1&0&0&$-$1&2&0&0&$-$3&0\\
$c_{8}$& & & & & & & &0&0&0&1&0&0&$-$1&3&$-$1&0&0&1&0&0&0&4&0&0&0&1&0&0&0&1\\
$s_{1}$& & & & & & & & &$-$1&$-$12&$-$1&17&13&4&$-$5&$-$2&0&0&0&$-$1&0&0&0&98&$-$2&$-$12&$-$1&17&13&6&$-$9\\
$s_{2}$& & & & & & & & & &$-$16&2&$-$1&$-$5&$-$7&$-$2&3&$-$3&1&1&$-$2&$-$3&0&0&$-$1&86&$-$16&2&$-$1&$-$5&$-$8&$-$2\\
$s_{3}$& & & & & & & & & & &$-$7&3&1&8&4&2&1&$-$1&0&$-$1&$-$1&0&0&$-$12&$-$16&97&$-$7&3&1&5&$-$1\\
$s_{4}$& & & & & & & & & & & &11&0&9&5&$-$3&0&0&$-$4&2&2&1&0&$-$1&2&$-$7&97&10&$-$1&6&10\\
$s_{5}$& & & & & & & & & & & & &$-$10&$-$4&5&$-$2&0&0&$-$1&$-$2&1&0&0&17&$-$1&3&11&86&$-$10&$-$2&1\\
$s_{6}$& & & & & & & & & & & & & &$-$1&7&2&0&0&0&0&$-$1&0&0&13&$-$4&1&0&$-$10&99&1&3\\
$s_{7}$& & & & & & & & & & & & & & &3&3&0&0&2&0&$-$1&$-$3&0&4&$-$9&8&9&$-$4&$-$2&66&2\\
$s_{8}$& & & & & & & & & & & & & & & &$-$2&0&0&$-$1&1&1&0&1&$-$5&$-$1&4&5&5&7&1&29\\
$c'_{2}$& & & & & & & & & & & & & & & & &$-$2&$-$12&$-$4&$-$1&$-$1&$-$2&$-$2&$-$2&3&2&$-$2&$-$2&2&3&$-$2\\
$c'_{3}$& & & & & & & & & & & & & & & & & &$-$19&2&0&$-$7&$-$9&1&0&$-$3&1&0&0&0&0&0\\
$c'_{4}$& & & & & & & & & & & & & & & & & & &0&0&0&1&0&0&1&$-$1&0&0&0&0&0\\
$c'_{5}$& & & & & & & & & & & & & & & & & & & &$-$2&0&5&2&0&0&0&$-$4&$-$1&0&1&$-$1\\
$c'_{6}$& & & & & & & & & & & & & & & & & & & & &$-$4&$-$1&0&$-$1&$-$2&$-$1&2&$-$2&0&0&1\\
$c'_{7}$& & & & & & & & & & & & & & & & & & & & & &0&1&0&$-$3&$-$1&2&1&$-$1&$-$1&1\\
$c'_{8}$& & & & & & & & & & & & & & & & & & & & & & &$-$2&0&0&0&1&0&0&$-$3&0\\
$c'_{9}$& & & & & & & & & & & & & & & & & & & & & & & &0&0&0&0&0&0&0&1\\
$s'_{1}$& & & & & & & & & & & & & & & & & & & & & & & & &$-$1&$-$12&$-$1&16&13&6&$-$9\\
$s'_{2}$& & & & & & & & & & & & & & & & & & & & & & & & & &$-$16&2&0&$-$4&$-$10&$-$1\\
$s'_{3}$& & & & & & & & & & & & & & & & & & & & & & & & & & &$-$7&3&1&5&$-$1\\
$s'_{4}$& & & & & & & & & & & & & & & & & & & & & & & & & & & &10&0&6&9\\
$s'_{5}$& & & & & & & & & & & & & & & & & & & & & & & & & & & & &$-$9&$-$2&1\\
$s'_{6}$& & & & & & & & & & & & & & & & & & & & & & & & & & & & & &1&3\\
$s'_{7}$& & & & & & & & & & & & & & & & & & & & & & & & & & & & & & &0\\
 \hline
    \end{tabular}
    \end{sideways}
    \caption{Correlation matrix ($\%$) of statistical uncertainties for the optimal binning scheme with $\Delta c_{i}$ and $\Delta s_{i}$ constraints.}
    \label{tab:opstatdcs}
\end{table}

 \begin{table}[!htbp]
\begin{sideways}
    \scriptsize       
    \setlength\tabcolsep{3pt}
    \centering
    \begin{tabular}{cccccccccccccccccccccccccccccccc}
          \hline     
         &$c_{2}$&$c_{3}$&$c_{4}$&$c_{5}$&$c_{6}$&$c_{7}$&$c_{8}$&$s_{1}$&$s_{2}$&$s_{3}$&$s_{4}$&$s_{5}$&$s_{6}$&$s_{7}$&$s_{8}$&$c'_{1}$& $c'_{2}$&$c'_{3}$&$c'_{4}$&$c'_{5}$&$c'_{6}$&$c'_{7}$&$c'_{8}$&$s'_{1}$&$s'_{2}$&$s'_{3}$&$s'_{4}$&$s'_{5}$&$s'_{6}$&$s'_{7}$&$s'_{8}$\\
        \hline
        $c_{1}$&0&$-$2&$-$3&$-$1&0&$-$3&0&$-$1&1&0&0&0&0&0&0&17&0&$-$2&$-$3&$-$1&0&$-$1&0&$-$1&1&0&0&0&0&0&0\\
$c_{2}$& &$-$8&3&1&$-$1&$-$1&$-$1&0&$-$3&0&0&0&0&0&0&1&66&$-$9&3&1&$-$1&$-$2&$-$1&0&$-$2&0&0&0&0&0&0\\
$c_{3}$& & &$-$1&$-$1&$-$2&$-$4&$-$1&$-$2&7&1&$-$2&$-$2&3&3&$-$2&$-$1&$-$9&61&$-$2&$-$1&$-$2&$-$2&0&$-$2&7&1&$-$2&$-$2&3&3&$-$3\\
$c_{4}$& & & &$-$3&0&4&3&0&1&1&$-$4&0&0&1&$-$1&$-$3&3&$-$1&90&$-$2&1&4&2&0&1&1&$-$4&0&0&1&$-$1\\
$c_{5}$& & & & &$-$3&0&1&1&$-$2&$-$1&$-$1&$-$1&0&$-$1&0&$-$1&2&0&$-$2&41&$-$2&2&1&1&$-$2&$-$1&$-$1&$-$1&0&$-$1&0\\
$c_{6}$& & & & & &$-$4&0&2&$-$6&$-$3&3&1&$-$4&$-$2&2&1&$-$1&$-$1&0&$-$3&68&$-$4&1&1&$-$6&$-$3&3&1&$-$4&$-$2&3\\
$c_{7}$& & & & & & &1&0&0&$-$1&1&0&0&$-$3&0&$-$1&$-$1&$-$3&4&0&$-$4&21&$-$1&0&0&$-$1&2&0&0&$-$3&0\\
$c_{8}$& & & & & & & &0&$-$1&0&1&0&0&0&4&$-$1&$-$1&$-$1&4&1&0&0&5&0&$-$1&0&1&0&0&0&1\\
$s_{1}$& & & & & & & & &$-$2&$-$14&2&0&2&$-$7&$-$5&$-$2&0&$-$1&0&1&1&0&0&96&$-$3&$-$14&3&0&2&$-$9&9\\
$s_{2}$& & & & & & & & & &$-$2&2&$-$5&9&0&$-$6&1&$-$3&6&1&$-$1&$-$5&0&0&$-$2&94&$-$2&2&$-$5&9&$-$1&$-$8\\
$s_{3}$& & & & & & & & & & &$-$4&16&10&13&1&0&0&0&1&$-$1&$-$2&0&0&$-$13&$-$2&96&$-$4&16&10&14&$-$4\\
$s_{4}$& & & & & & & & & & & &1&5&18&0&0&0&$-$2&$-$4&0&3&0&0&3&2&$-$4&95&1&5&16&6\\
$s_{5}$& & & & & & & & & & & & &$-$17&$-$2&7&0&0&$-$2&0&$-$3&1&0&0&1&$-$5&16&1&82&$-$17&$-$2&9\\
$s_{6}$& & & & & & & & & & & & & &$-$4&$-$3&0&0&2&0&1&$-$3&0&0&3&9&10&6&$-$17&98&$-$2&$-$5\\
$s_{7}$& & & & & & & & & & & & & & &8&1&0&3&1&$-$1&$-$2&$-$3&0&$-$6&$-$1&13&18&$-$3&$-$4&84&5\\
$s_{8}$& & & & & & & & & & & & & & & &0&0&$-$2&$-$1&$-$1&3&0&0&$-$5&$-$6&1&0&7&$-$2&6&19\\
$c'_{2}$& & & & & & & & & & & & & & & & &1&$-$2&$-$3&$-$1&1&$-$4&$-$1&$-$2&1&0&0&0&0&1&0\\
$c'_{3}$& & & & & & & & & & & & & & & & & &$-$11&3&1&$-$1&$-$3&$-$2&0&$-$3&0&0&0&0&0&0\\
$c'_{4}$& & & & & & & & & & & & & & & & & & &$-$1&$-$1&$-$1&$-$3&$-$1&$-$1&6&0&$-$1&$-$2&2&3&$-$2\\
$c'_{5}$& & & & & & & & & & & & & & & & & & & &$-$2&0&3&2&0&1&1&$-$4&0&0&1&$-$1\\
$c'_{6}$& & & & & & & & & & & & & & & & & & & & &$-$4&0&0&1&$-$1&$-$1&0&$-$3&1&$-$1&0\\
$c'_{7}$& & & & & & & & & & & & & & & & & & & & & &$-$5&1&1&$-$5&$-$3&3&1&$-$3&$-$2&2\\
$c'_{8}$& & & & & & & & & & & & & & & & & & & & & & &$-$2&0&0&0&0&0&0&$-$3&0\\
$c'_{9}$& & & & & & & & & & & & & & & & & & & & & & & &0&0&0&0&0&0&0&1\\
$s'_{1}$& & & & & & & & & & & & & & & & & & & & & & & & &$-$2&$-$13&3&1&3&$-$8&9\\
$s'_{2}$& & & & & & & & & & & & & & & & & & & & & & & & & &$-$2&2&$-$5&9&$-$1&$-$8\\
$s'_{3}$& & & & & & & & & & & & & & & & & & & & & & & & & & &$-$4&15&10&13&$-$4\\
$s'_{4}$& & & & & & & & & & & & & & & & & & & & & & & & & & & &0&5&16&6\\
$s'_{5}$& & & & & & & & & & & & & & & & & & & & & & & & & & & & &$-$17&$-$2&7\\
$s'_{6}$& & & & & & & & & & & & & & & & & & & & & & & & & & & & & &$-$2&$-$5\\
$s'_{7}$& & & & & & & & & & & & & & & & & & & & & & & & & & & & & & &3\\
 \hline
    \end{tabular}
    \end{sideways}
    \caption{Correlation matrix ($\%$) of statistical uncertainties for the modified optimal binning scheme with $\Delta c_{i}$ and $\Delta s_{i}$ constraints.}
    \label{tab:modistatdcs}
\end{table}

 \begin{table}[!htbp]
\begin{sideways}
    \scriptsize       
    \setlength\tabcolsep{3pt}
    \centering
    \begin{tabular}{cccccccccccccccccccccccccccccccc}
         \hline       
         &$c_{2}$&$c_{3}$&$c_{4}$&$c_{5}$&$c_{6}$&$c_{7}$&$c_{8}$&$s_{1}$&$s_{2}$&$s_{3}$&$s_{4}$&$s_{5}$&$s_{6}$&$s_{7}$&$s_{8}$&$c'_{1}$& $c'_{2}$&$c'_{3}$&$c'_{4}$&$c'_{5}$&$c'_{6}$&$c'_{7}$&$c'_{8}$&$s'_{1}$&$s'_{2}$&$s'_{3}$&$s'_{4}$&$s'_{5}$&$s'_{6}$&$s'_{7}$&$s'_{8}$\\
  \hline
        $c_{1}$&80&74&22&$-$28&14&62&79&9&28&42&29&16&$-$20&$-$46&$-$46&76&67&69&5&$-$32&$-$1&56&69&$-$7&38&46&27&9&$-$20&$-$44&$-$48\\
$c_{2}$& &71&23&$-$18&19&61&72&4&31&39&29&11&$-$15&$-$41&$-$38&55&87&63&11&$-$22&6&52&60&$-$2&39&43&27&4&$-$15&$-$39&$-$40\\
$c_{3}$& & &47&9&49&80&76&8&24&34&24&$-$4&$-$25&$-$31&$-$25&42&51&86&35&2&31&60&54&$-$1&40&43&20&$-$14&$-$25&$-$31&$-$35\\
$c_{4}$& & & &66&68&47&33&1&$-$30&$-$46&$-$60&$-$16&36&26&12&$-$7&5&27&88&61&60&29&8&$-$23&$-$23&$-$40&$-$63&$-$22&36&27&4\\
$c_{5}$& & & & &61&15&$-$8&4&$-$37&$-$55&$-$54&$-$31&34&46&39&$-$53&$-$30&$-$7&70&98&64&2&$-$31&$-$10&$-$36&$-$51&$-$58&$-$33&34&47&36\\
$c_{6}$& & & & & &52&30&3&$-$11&$-$21&$-$25&$-$23&16&20&22&$-$16&0&28&65&56&90&28&2&$-$2&0&$-$13&$-$30&$-$30&17&20&14\\
$c_{7}$& & & & & & &66&2&13&25&15&$-$14&$-$22&$-$31&$-$14&35&42&62&37&7&34&77&44&$-$5&32&34&12&$-$25&$-$21&$-$32&$-$28\\
$c_{8}$& & & & & & & &11&22&32&20&7&$-$15&$-$33&$-$38&53&58&65&18&$-$14&13&51&83&$-$5&34&39&17&$-$2&$-$15&$-$32&$-$50\\
$s_{1}$& & & & & & & & &17&5&0&19&6&11&6&2&10&13&$-$3&5&5&1&7&17&9&5&$-$1&19&7&12&1\\
$s_{2}$& & & & & & & & & &50&49&19&$-$40&$-$29&$-$15&26&24&31&$-$32&$-$37&$-$14&11&23&21&78&51&49&17&$-$40&$-$29&$-$17\\
$s_{3}$& & & & & & & & & & &84&$-$9&$-$73&$-$58&$-$39&46&43&47&$-$38&$-$55&$-$20&33&42&10&54&98&84&$-$11&$-$73&$-$57&$-$41\\
$s_{4}$& & & & & & & & & & & &$-$7&$-$73&$-$55&$-$29&36&36&37&$-$47&$-$53&$-$23&23&31&12&52&82&99&$-$8&$-$73&$-$54&$-$32\\
$s_{5}$& & & & & & & & & & & & &35&11&8&14&9&$-$8&$-$41&$-$34&$-$40&$-$26&2&37&23&$-$6&$-$3&83&35&9&8\\
$s_{6}$& & & & & & & & & & & & & &54&35&$-$29&$-$19&$-$38&18&33&9&$-$33&$-$28&13&$-$43&$-$70&$-$72&36&98&54&42\\
$s_{7}$& & & & & & & & & & & & & & &56&$-$51&$-$45&$-$44&19&46&17&$-$49&$-$48&27&$-$30&$-$55&$-$55&12&55&95&60\\
$s_{8}$& & & & & & & & & & & & & & & &$-$49&$-$42&$-$42&4&36&14&$-$36&$-$57&39&$-$9&$-$33&$-$28&6&36&52&85\\
$c'_{2}$& & & & & & & & & & & & & & & & &64&56&$-$11&$-$56&$-$19&46&69&$-$17&30&46&36&11&$-$30&$-$50&$-$49\\
$c'_{3}$& & & & & & & & & & & & & & & & & &59&3&$-$32&$-$2&48&62&$-$9&25&44&35&6&$-$20&$-$43&$-$41\\
$c'_{4}$& & & & & & & & & & & & & & & & & & &31&$-$9&26&65&63&$-$14&35&48&33&$-$13&$-$38&$-$41&$-$47\\
$c'_{5}$& & & & & & & & & & & & & & & & & & & &69&72&34&5&$-$42&$-$33&$-$37&$-$52&$-$43&18&21&$-$0\\
$c'_{6}$& & & & & & & & & & & & & & & & & & & & &63&$-$0&$-$33&$-$14&$-$42&$-$54&$-$57&$-$34&34&48&36\\
$c'_{7}$& & & & & & & & & & & & & & & & & & & & & &30&$-$0&$-$19&$-$16&$-$18&$-$28&$-$42&9&19&11\\
$c'_{8}$& & & & & & & & & & & & & & & & & & & & & & &50&$-$26&16&33&20&$-$30&$-$33&$-$47&$-$41\\
$c'_{9}$& & & & & & & & & & & & & & & & & & & & & & & &$-$16&26&43&30&$-$1&$-$29&$-$46&$-$65\\
$s'_{1}$& & & & & & & & & & & & & & & & & & & & & & & & &32&16&14&34&14&24&44\\
$s'_{2}$& & & & & & & & & & & & & & & & & & & & & & & & & &60&51&16&$-$42&$-$32&$-$17\\
$s'_{3}$& & & & & & & & & & & & & & & & & & & & & & & & & & &82&$-$10&$-$70&$-$55&$-$38\\
$s'_{4}$& & & & & & & & & & & & & & & & & & & & & & & & & & & &$-$4&$-$71&$-$55&$-$31\\
$s'_{5}$& & & & & & & & & & & & & & & & & & & & & & & & & & & & &35&11&10\\
$s'_{6}$& & & & & & & & & & & & & & & & & & & & & & & & & & & & & &54&43\\
$s'_{7}$& & & & & & & & & & & & & & & & & & & & & & & & & & & & & & &58\\
 \hline
    \end{tabular}
    \end{sideways}
    \caption{Correlation matrix ($\%$) of systematic uncertainties for the equal binning scheme with $\Delta c_{i}$ and $\Delta s_{i}$ constraints.}
    \label{tab:eqsystdcs}
\end{table}

 \begin{table}[!htbp]
\begin{sideways}
    \scriptsize       
    \setlength\tabcolsep{3pt}
    \centering
    \begin{tabular}{cccccccccccccccccccccccccccccccc}
         \hline        
         &$c_{2}$&$c_{3}$&$c_{4}$&$c_{5}$&$c_{6}$&$c_{7}$&$c_{8}$&$s_{1}$&$s_{2}$&$s_{3}$&$s_{4}$&$s_{5}$&$s_{6}$&$s_{7}$&$s_{8}$&$c'_{1}$& $c'_{2}$&$c'_{3}$&$c'_{4}$&$c'_{5}$&$c'_{6}$&$c'_{7}$&$c'_{8}$&$s'_{1}$&$s'_{2}$&$s'_{3}$&$s'_{4}$&$s'_{5}$&$s'_{6}$&$s'_{7}$&$s'_{8}$\\
         \hline
        $c_{1}$& 53&83&28&82&60&51&47&$-$13&4&$-$10&$-$11&22&9&32&$-$5&81&35&23&21&66&39&32&14&$-$14&3&$-$11&$-$14&19&15&36&$-$14\\
$c_{2}$& &44&$-$10&36&44&52&37&$-$25&10&$-$12&$-$13&29&12&10&$-$23&42&88&21&$-$16&28&30&42&32&$-$26&7&$-$13&$-$15&28&17&17&$-$29\\
$c_{3}$& & &32&81&55&39&41&$-$7&11&$-$20&$-$3&16&6&33&2&60&26&26&26&56&34&17&1&$-$8&9&$-$21&$-$5&13&12&36&$-$5\\
$c_{4}$& & & &58&6&$-$45&$-$16&53&$-$12&30&31&$-$40&$-$13&43&65&13&$-$19&$-$8&98&53&$-$6&$-$57&$-$58&53&$-$12&31&32&$-$43&$-$10&39&61\\
$c_{5}$& & & & &48&25&34&11&$-$6&6&5&$-$12&$-$2&45&26&62&20&15&53&82&28&4&$-$13&11&$-$8&5&4&$-$16&3&43&20\\
$c_{6}$& & & & & &46&42&$-$18&11&$-$6&$-$6&28&33&25&$-$11&46&30&24&0&40&91&32&20&$-$19&10&$-$7&$-$8&26&38&29&$-$20\\
$c_{7}$& & & & & & &61&$-$49&7&$-$27&$-$24&49&19&2&$-$42&54&49&33&$-$47&25&43&94&58&$-$50&5&$-$29&$-$25&49&21&3&$-$49\\
$c_{8}$& & & & & & & &$-$33&4&$-$13&$-$12&39&12&18&$-$25&46&33&20&$-$19&29&38&50&34&$-$33&4&$-$15&$-$13&38&14&20&$-$29\\
$s_{1}$& & & & & & & & &19&58&59&$-$61&$-$19&29&59&$-$32&$-$31&$-$19&51&$-$2&$-$26&$-$51&$-$45&98&23&59&61&$-$60&$-$18&35&60\\
$s_{2}$& & & & & & & & & &10&7&12&7&6&$-$24&$-$7&4&$-$7&$-$14&$-$19&6&6&8&19&86&11&7&14&8&22&$-$25\\
$s_{3}$& & & & & & & & & & &45&$-$42&$-$21&21&45&$-$22&$-$16&$-$32&27&$-$9&$-$11&$-$30&$-$25&58&17&99&47&$-$42&$-$20&29&43\\
$s_{4}$& & & & & & & & & & & &$-$39&$-$33&32&49&$-$32&$-$24&$-$16&24&$-$13&$-$16&$-$30&$-$31&59&10&47&89&$-$39&$-$30&34&63\\
$s_{5}$& & & & & & & & & & & & &39&$-$4&$-$56&31&30&21&$-$40&1&31&47&35&$-$60&11&$-$43&$-$40&99&39&0&$-$63\\
$s_{6}$& & & & & & & & & & & & & &$-$12&$-$23&11&12&7&$-$12&2&29&18&14&$-$19&7&$-$21&$-$35&39&98&$-$10&$-$32\\
$s_{7}$& & & & & & & & & & & & & & &28&23&3&8&40&42&18&$-$13&$-$23&29&$-$1&18&34&$-$6&$-$10&78&35\\
$s_{8}$& & & & & & & & & & & & & & & &$-$18&$-$28&$-$14&62&19&$-$18&$-$48&$-$41&59&$-$16&46&51&$-$58&$-$22&28&80\\
$c'_{2}$& & & & & & & & & & & & & & & & &41&30&14&70&43&46&27&$-$33&$-$13&$-$26&$-$33&29&12&16&$-$26\\
$c'_{3}$& & & & & & & & & & & & & & & & & &21&$-$19&26&30&47&35&$-$31&$-$1&$-$18&$-$25&29&13&2&$-$33\\
$c'_{4}$& & & & & & & & & & & & & & & & & & &$-$7&28&24&31&22&$-$19&$-$8&$-$34&$-$17&20&8&7&$-$19\\
$c'_{5}$& & & & & & & & & & & & & & & & & & & &54&$-$6&$-$56&$-$57&51&$-$16&27&26&$-$42&$-$11&32&58\\
$c'_{6}$& & & & & & & & & & & & & & & & & & & & &36&14&$-$2&$-$3&$-$27&$-$12&$-$13&$-$2&3&30&12\\
$c'_{7}$& & & & & & & & & & & & & & & & & & & & & &38&26&$-$26&2&$-$13&$-$16&30&29&14&$-$25\\
$c'_{8}$& & & & & & & & & & & & & & & & & & & & & & &66&$-$51&3&$-$32&$-$32&47&18&$-$15&$-$53\\
$c'_{9}$& & & & & & & & & & & & & & & & & & & & & & & &$-$45&6&$-$26&$-$32&36&14&$-$20&$-$45\\
$s'_{1}$& & & & & & & & & & & & & & & & & & & & & & & & &23&60&61&$-$60&$-$18&34&60\\
$s'_{2}$& & & & & & & & & & & & & & & & & & & & & & & & & &19&10&14&10&28&$-$27\\
$s'_{3}$& & & & & & & & & & & & & & & & & & & & & & & & & & &49&$-$43&$-$19&29&44\\
$s'_{4}$& & & & & & & & & & & & & & & & & & & & & & & & & & & &$-$40&$-$32&35&66\\
$s'_{5}$& & & & & & & & & & & & & & & & & & & & & & & & & & & & &39&$-$0&$-$65\\
$s'_{6}$& & & & & & & & & & & & & & & & & & & & & & & & & & & & & &$-$5&$-$32\\
$s'_{7}$& & & & & & & & & & & & & & & & & & & & & & & & & & & & & & &22\\
 \hline
    \end{tabular}
    \end{sideways}
    \caption{Correlation matrix ($\%$) of systematic uncertainties for the optimal binning scheme with $\Delta c_{i}$ and $\Delta s_{i}$ constraints.}
    \label{tab:opsystdcs}
\end{table}

 \begin{table}[!htbp]
\begin{sideways}
    \scriptsize       
    \setlength\tabcolsep{3pt}
    \centering
    \begin{tabular}{cccccccccccccccccccccccccccccccc}
        \hline     
         &$c_{2}$&$c_{3}$&$c_{4}$&$c_{5}$&$c_{6}$&$c_{7}$&$c_{8}$&$s_{1}$&$s_{2}$&$s_{3}$&$s_{4}$&$s_{5}$&$s_{6}$&$s_{7}$&$s_{8}$&$c'_{1}$& $c'_{2}$&$c'_{3}$&$c'_{4}$&$c'_{5}$&$c'_{6}$&$c'_{7}$&$c'_{8}$&$s'_{1}$&$s'_{2}$&$s'_{3}$&$s'_{4}$&$s'_{5}$&$s'_{6}$&$s'_{7}$&$s'_{8}$\\
          \hline
        $c_{1}$&36&65&55&80&51&27&35&18&8&4&7&$-$7&$-$15&7&19&51&10&42&54&62&20&$-$2&$-$17&23&16&2&8&$-$11&$-$6&1&13\\
$c_{2}$& &70&$-$15&42&66&70&60&$-$18&13&$-$29&$-$10&26&30&1&$-$23&40&80&47&$-$17&17&45&52&39&$-$14&27&$-$31&$-$6&27&39&7&$-$31\\
$c_{3}$& & &21&72&76&58&57&$-$8&11&$-$20&$-$4&15&16&6&$-$6&49&32&62&18&44&41&27&14&$-$1&29&$-$22&$-$1&16&27&11&$-$14\\
$c_{4}$& & & &59&3&$-$35&$-$14&31&6&38&17&$-$30&$-$41&8&47&27&$-$32&10&99&67&$-$15&$-$51&$-$57&35&6&37&13&$-$36&$-$38&$-$3&47\\
$c_{5}$& & & & &55&31&36&22&$-$7&$-$4&14&$-$27&$-$22&2&23&47&13&46&57&79&21&$-$1&$-$15&28&2&$-$7&16&$-$30&$-$12&$-$7&20\\
$c_{6}$& & & & & &64&64&$-$14&25&$-$19&$-$8&34&42&16&$-$27&50&46&59&3&37&86&42&26&$-$7&36&$-$22&$-$3&34&51&20&$-$33\\
$c_{7}$& & & & & & &70&$-$15&3&$-$47&$-$6&31&33&$-$1&$-$32&30&59&48&$-$36&1&48&70&49&$-$11&12&$-$48&0&33&41&2&$-$39\\
$c_{8}$& & & & & & & &$-$23&15&$-$31&$-$13&39&34&6&$-$33&37&49&48&$-$14&14&49&51&37&$-$17&23&$-$33&$-$9&39&42&7&$-$33\\
$s_{1}$& & & & & & & & &$-$17&17&69&$-$63&$-$52&11&33&$-$18&$-$27&$-$19&30&8&$-$17&$-$38&$-$36&97&$-$19&19&70&$-$64&$-$52&3&49\\
$s_{2}$& & & & & & & & & &33&$-$11&37&36&38&$-$34&22&7&12&7&$-$4&32&7&4&$-$12&90&31&$-$11&37&36&48&$-$31\\
$s_{3}$& & & & & & & & & & &29&$-$4&$-$16&21&32&$-$14&$-$38&$-$40&36&1&$-$15&$-$43&$-$33&16&32&93&24&$-$4&$-$21&27&37\\
$s_{4}$& & & & & & & & & & & &$-$40&$-$45&26&42&$-$29&$-$25&$-$25&14&$-$8&$-$10&$-$30&$-$22&67&$-$10&30&86&$-$39&$-$46&24&55\\
$s_{5}$& & & & & & & & & & & & &60&14&$-$36&17&30&22&$-$29&$-$23&40&40&33&$-$59&39&$-$5&$-$39&98&62&23&$-$40\\
$s_{6}$& & & & & & & & & & & & & &3&$-$63&21&35&22&$-$40&$-$21&47&45&39&$-$50&39&$-$17&$-$42&63&99&14&$-$67\\
$s_{7}$& & & & & & & & & & & & & & &$-$6&5&$-$4&1&8&$-$0&21&$-$10&$-$7&14&36&20&25&14&5&65&0\\
$s_{8}$& & & & & & & & & & & & & & & &$-$23&$-$36&$-$24&45&19&$-$40&$-$43&$-$39&31&$-$33&34&38&$-$37&$-$65&$-$14&84\\
$c'_{2}$& & & & & & & & & & & & & & & & &43&64&29&67&45&43&22&$-$11&24&$-$18&$-$27&13&30&$-$1&$-$27\\
$c'_{3}$& & & & & & & & & & & & & & & & & &47&$-$30&10&46&67&54&$-$24&9&$-$39&$-$20&30&40&$-$3&$-$42\\
$c'_{4}$& & & & & & & & & & & & & & & & & & &11&48&50&45&29&$-$13&16&$-$43&$-$21&19&31&$-$4&$-$28\\
$c'_{5}$& & & & & & & & & & & & & & & & & & & &68&$-$14&$-$50&$-$57&33&5&35&10&$-$34&$-$37&$-$5&46\\
$c'_{6}$& & & & & & & & & & & & & & & & & & & & &22&$-$2&$-$15&13&$-$3&$-$2&$-$7&$-$28&$-$13&$-$13&16\\
$c'_{7}$& & & & & & & & & & & & & & & & & & & & & &48&36&$-$13&33&$-$18&$-$5&39&52&25&$-$40\\
$c'_{8}$& & & & & & & & & & & & & & & & & & & & & & &71&$-$36&9&$-$44&$-$25&41&49&$-$8&$-$50\\
$c'_{9}$& & & & & & & & & & & & & & & & & & & & & & & &$-$37&5&$-$34&$-$18&36&41&2&$-$41\\
$s'_{1}$& & & & & & & & & & & & & & & & & & & & & & & & &$-$14&17&68&$-$61&$-$49&5&47\\
$s'_{2}$& & & & & & & & & & & & & & & & & & & & & & & & & &30&$-$10&40&41&51&$-$33\\
$s'_{3}$& & & & & & & & & & & & & & & & & & & & & & & & & & &25&$-$6&$-$23&26&38\\
$s'_{4}$& & & & & & & & & & & & & & & & & & & & & & & & & & & &$-$38&$-$43&24&50\\
$s'_{5}$& & & & & & & & & & & & & & & & & & & & & & & & & & & & &64&25&$-$43\\
$s'_{6}$& & & & & & & & & & & & & & & & & & & & & & & & & & & & & &15&$-$69\\
$s'_{7}$& & & & & & & & & & & & & & & & & & & & & & & & & & & & & & &$-$9\\
 \hline
    \end{tabular}
    \end{sideways}
    \caption{Correlation matrix ($\%$) of systematic uncertainties for the modified optimal binning scheme with $\Delta c_{i}$ and $\Delta s_{i}$ constraints.}
    \label{tab:modisystdcs}
\end{table}

\end{document}